\documentclass[12pt, a4paper]{article}
\pdfoutput=1

\usepackage[table]{xcolor}
\usepackage{amsmath}
\usepackage{amsfonts}
\usepackage{amssymb}
\usepackage{graphicx, rotating}
\usepackage{epstopdf}
\usepackage{epsfig}
\usepackage{latexsym}
\usepackage{graphicx}
\usepackage{color}
\usepackage{amsmath,amssymb}
\usepackage{cite}
\usepackage{bbold}
\usepackage{yfonts}
\usepackage{slashed}
\usepackage{hyperref}
\usepackage{ntheorem}
\usepackage{enumitem}
\usepackage{soul}

\hypersetup{colorlinks, citecolor=bluscuro, linkcolor=black, urlcolor=bluscuro}
\definecolor{rossos}{cmyk}{0,1,1,0.55}
\definecolor{bluscuro}{rgb}{0.15, 0.2, .85}
\definecolor{bluchiaro}{cmyk}{1,.3,0.,0.1}
\usepackage{cancel}
\definecolor{oucrimsonred}{rgb}{0.6, 0.0, 0.0}
\definecolor{persianblue}{rgb}{0.11, 0.22, 0.73}
\definecolor{forestgreen}{rgb}{0.13,0.35,0.13}
 \hypersetup{colorlinks, citecolor=oucrimsonred, linkcolor=persianblue, urlcolor=oucrimsonred}
 \hypersetup{colorlinks, citecolor=oucrimsonred, linkcolor=persianblue, urlcolor=oucrimsonred}

\usepackage{lmodern}

\makeatletter
\ifcase \@ptsize \relax
  \newcommand{\miniscule}{\@setfontsize\miniscule{4}{5}}
\or
  \newcommand{\miniscule}{\@setfontsize\miniscule{5}{6}}
\or
  \newcommand{\miniscule}{\@setfontsize\miniscule{5}{6}}
\fi
\makeatother

\usepackage{lipsum}

\newcommand{\be}{\begin{equation}}
\newcommand{\ee}{\end{equation}}
\newcommand{\bea}{\be\begin{aligned}}
\newcommand{\eea}{\end{aligned}\ee}
\newcommand{\bc}{\begin{center}}
\newcommand{\ec}{\end{center}}

\usepackage{adjustbox}
\usepackage{array}
\usepackage{booktabs}
\usepackage{multirow}

\newcolumntype{R}[2]{%
    >{\adjustbox{angle=#1,lap=\width-(#2)}\bgroup}%
    l%
    <{\egroup}%
}

\definecolor{Gray}{gray}{0.95}
\newcommand{\bbox}[1]{\fcolorbox{gray}{Gray}{~$\displaystyle #1$~}}

\def\td{{\rm d}}
\def\diag{{\rm diag}}

\def\eps{\epsilon}

\newtheorem*{theorem}{}

\begin{document}


\vspace*{-2cm}
\begin{flushright}
CERN-TH-2018-224 \\
\vspace*{2mm}
\today
\end{flushright}

\begin{center}
\vspace*{15mm}

\vspace{1cm}
{\Large \bf 
On gravitational echoes from ultracompact exotic stars
} \\
\vspace{1.4cm}

\bigskip

{\bf Alfredo Urbano$^{a}$ and Hardi Veerm\"ae$^{b,c}$}
 \\[5mm]

{\it $^a$ INFN, sezione di Trieste, SISSA, via Bonomea 265, 34136 Trieste, Italy.}\\[1mm]
{\it $^b$ CERN, Theoretical Physics Department, Geneva, Switzerland.}\\[1mm]
{\it $^c$ NICPB, R\"avala 10, 10143 Tallinn, Estonia.}

\end{center}
\vspace*{10mm} 
\begin{abstract}\noindent\normalsize

At the dawn of a golden age for gravitational wave astronomy, we must leave no stone unturned in our quest for new phenomena beyond our current understanding of General Relativity (GR), particle physics and nuclear physics. 
In this paper we discuss gravitational echoes from ultracompact stars.
We restrict our analysis to exact solutions of Einstein field equations in GR that are supported by physically motivated equations of state (EoS), and in particular  we impose the constraint of causality.
Our main conclusion is that ultracompact objects supported by physical EoS are not able to generate gravitational echoes like those that characterize the relaxation phase of a putative black hole mimicker.
Nevertheless, we identify a class of physical exotic objects that are compact enough to accommodate the presence of an external unstable light ring, thus opening the possibility of trapping gravitational radiation and affecting the ringdown phase of a merger event.
Most importantly, we show that once rotation is included these stars -- contrary to what usually expected for ultracompact objects -- are not plagued by any ergoregion instability. 
We extend our analysis for arbitrary values of angular velocity up to the Keplerian limit, and we comment about potential signals relevant for gravitational wave interferometers.

\end{abstract}

\vspace*{3mm}



\tableofcontents

\pagebreak

\section{Introduction and motivations} 
\label{sec:intro}

\begin{figure}[!htb!]
\begin{center}
	\includegraphics[width=.45\textwidth]{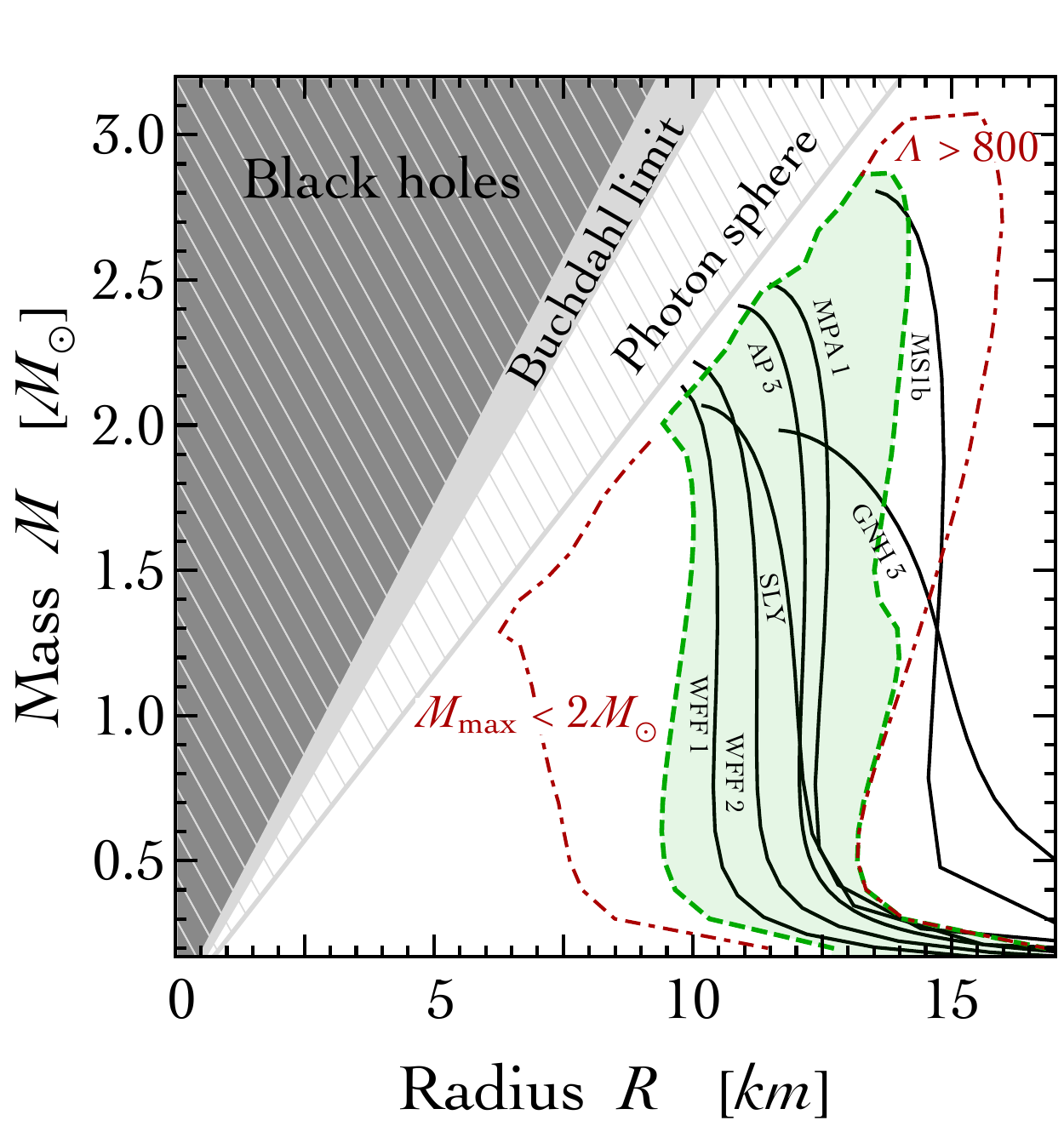}
	\caption{\em \label{fig:MassRadiusEOS} 
M-R relation for ordinary NSs in the static limit for various EoS (solid black lines). The green region (from~\cite{Annala:2017llu}) is compatible with the observations of NSs with mass $M \simeq 2 M_\odot$ and with the bound on the tidal deformability parameter $\Lambda < 800$~\cite{TheLIGOScientific:2017qsa}.
 }
\end{center}
\end{figure}

Neutron stars (NSs) comprise one of the possible evolutionary end-points of high mass stars. NSs are a degenerate state of matter that is formed after the core collapse in a supernova event, where the electrons fall into nuclear matter and get captured by protons forming neutrons~\cite{Ozel:2016oaf}. The actual composition of NSs -- especially in the inner core --  is still unknown, and  their extreme conditions cannot be reproduced in a laboratory.  In   the   era   of   Gravitational   Wave (GW) physics the   characterization of the GW signal emitted by compact binary sources will play a forefront role in understanding the NS properties~\cite{Rezzolla:2016nxn}.  In this respect, the first observation (GW170817) of GWs from a binary neutron star inspiral~\cite{TheLIGOScientific:2017qsa} represented an excellent laboratory to study high-density nuclear physics, and -- very remarkably -- it allowed putting non-trivial constraints on a generic family of neutron-star-matter Equations of State (EoS) that interpolate between state-of-the-art theoretical results at low and high baryon density~\cite{Annala:2017llu,Most:2018hfd}. The situation is summarized in Fig.~\ref{fig:MassRadiusEOS}. We show the mass-radius (M-R hereafter) relation -- obtained by solving with standard methods the Einstein's equations for a non-rotating spherically symmetric fluid star --  for various EoS (extracted from the tabulated results of~\cite{Ozel:2015fia}).
The region shaded in green (taken from~\cite{Annala:2017llu}) is compatible with the observations of NSs with mass $M \simeq 2 M_\odot$~\cite{Demorest:2010bx,Antoniadis:2013pzd} and with the bound on the so-called tidal deformability parameter $\Lambda < 800$ extracted from the GW signal in~\cite{TheLIGOScientific:2017qsa}.  

On general grounds, there are three important regions in the M-R plane describing compact astrophysical objects. The black hole (BH) region (black-hatched in Fig.~\ref{fig:MassRadiusEOS}) is limited by the condition $R= 2M$ 
while compact objects with $4/9 < M/R < 1/2$ (region shaded in gray in Fig.~\ref{fig:MassRadiusEOS}) violate the so-called Buchdahl limit - an upper limit on compactness of fluid stars~\cite{Buchdahl:1959zz}.\footnote{Throughout this paper, we shall use a geometrized unit system with $c = G_N = 1$.} This limit describes the maximum amount of mass that can exist in a sphere before it must undergo gravitational collapse. Finally, compact objects with $1/3 < M/R < 4/9$  (white-hatched region in Fig.~\ref{fig:MassRadiusEOS}) do not violate the Buchdahl limit and possess a photon sphere, that is the unstable circular null geodesic of the external Schwarzschild spacetime metric. The latter property may play an important role for the characterization of the GW signal emitted in the final state of a compact binary merger. If the object originating from the merger process is sufficiently compact to have a photon sphere, the post-merger ringdown waveform  is believed to be initially identical to that of a BH with modifications encoded in subsequent pulses of gravitational radiation (dubbed ``echoes'') ~\cite{Cardoso:2017cqb}. 

The possibility of observing GW echoes seems to be particularly true in the case of BH mimickers~\cite{Cardoso:2016rao,Cardoso:2016oxy,Cardoso:2017njb}, objects that for an external observer are in all respects identical to BHs except for the presence of putative deviations from the prediction of general relativity (GR) at the position of the event horizon. These objects can evade the Buchdahl limit, and, as a consequence, their radius can be arbitrarely close to the BH limit. Objects that fall into this category are, to name a few, gravitational vacuum stars (a.k.a. gravastars)~\cite{Mazur:2004fk,Mazur:2001fv}, Lorentzian traversable wormholes~\cite{Morris:1988tu}, matter-bumpy BHs~\cite{Cardoso:2016oxy,Barausse:2014tra}, Kerr-like wormholes~\cite{Bueno:2017hyj} and Fuzzballs~\cite{Mathur:2005zp}. The possible presence of BH mimickers -- and the consequent detectability of their gravitational echoes -- is a tantalizing hypothesis but it is fair to remember that the existence of none of the aforementioned objects has solid theoretical ground.\footnote{However, see~\cite{Carballo-Rubio:2017tlh} for a recent discussion about 
a possible model for gravastars.}
 Nevertheless, following a pure phenomenological approach, a rich search program for echoes in current and future GW interferometers is undergoing~\cite{Mark:2017dnq,Testa:2018bzd,Tsang:2018uie,Wang:2018gin,Glampedakis:2017cgd}. Furthermore, a tentative evidence (with significance of $2.5\,\sigma$) of gravitational echoes in the three BH merger events GW150914, GW151226, and LVT151012 was proposed in~\cite{Abedi:2016hgu} (see also~\cite{Ashton:2016xff,Abedi:2017isz,Westerweck:2017hus,Abedi:2018pst} for further discussions), and, even more interestingly, in the NS merger event GW170817~\cite{Abedi:2018npz} (with significance of $4.2\,\sigma$). Even if the statistical significance of these tentative analysis is far from being convincing, the relevance and the implications of a possible future detection are, indisputably, of major importance.
  
Recently~\cite{Pani:2018flj}, it has been pointed out that gravitational echoes are not a unique prerogative of deviations from GR at the horizon scale but similar signals may arise in the presence of compact exotic fluid stars with $1/3 < M/R < 4/9$, thus featuring a photon sphere (see also~\cite{Ferrari:2000sr,Kokkotas:2000is} for earlier theoretical studies). As is clear from  Fig.~\ref{fig:MassRadiusEOS}, this possibility cannot be realized by viable neutron-star-matter EoS. Thus, as a baryonic realization seems implausible, Ref.~\cite{Pani:2018flj} considered constant-density stars (CDS) as a toy model for exotic stars capable of producing gravitational echoes -- in this case stable configurations arbitrarily close to the Buchdahl limit exist. However -- as already  noticed in~\cite{Pani:2018flj}, and reiterated in~\cite{Mannarelli:2018pjb} -- a CDS represents an unphysical system because an incompressible fluid violates causality. The aim of this article is to investigate under which physical conditions, if any, gravitational echoes could be produced by compact exotic objects and what is the microphysics that could support the required high compactness.

This article is outlined as follows: In section \ref{sec:stars} we consider static ultracompact exotic objects with a well defined microphysics. Section \ref{sec:echoes} deals with the quasinormal modes and the viability of GW echoes from such objects. In section \ref{sec:rotation} we discuss the effect of rotation on light rings and gravitational echoes. We conclude in section \ref{sec:conclusions}. Some technical details related to the Buchdahl limit, boson stars and the Hartle-Thorne approximation for rotating stars are collected into the appendix~\ref{app:Buchdahl},~\ref{app:BosonStars} and \ref{app:HartleThorne}, respectively.

\section{Ultracompact exotic stars} 
\label{sec:stars}

The dynamics of compact objects depends on both their matter content and the underlying gravitational theory. To limit the theoretical possibilities we will work within the limits of GR and study matter satisfying reasonable physical assumptions. In this section we will consider solutions of the Einstein's  equations for the gravitational field  produced by a static, spherically symmetric mass distribution. As a consequence of the Birkhoff's theorem~\cite{Birkhoff} the exterior of such stars are described by the Schwarzschild solution.\footnote{This is Birkhoff's theorem~\cite{Birkhoff}. However, notice~\cite{Deser:2004gi} that the result was first derived and published by Jebsen~\cite{Jebsen}.} 

The general spherically symmetric static spacetime in Boyer-Lindquist type coordinates $(t,r,\theta,\phi)$ reads
\be\label{eq:LineElement}
	\td s^2 = -e^{\nu(r)}\td t^2 + e^{\lambda(r)}\td r^2 + r^2\left(
	\td \theta^2 + \sin^2\theta \td \phi^2
\right)~,
\ee
In the most general case the matter inside a static spherically symmetric star is described by the stress energy tensor 
\be
	T_{\mu\nu} = \diag( \rho, P_{\rm r}, P_{\rm t}, P_{\rm t} )~,
\ee
where $\rho$ is the proper energy density, $P_{\rm r}$ denotes the radial pressure and $P_{\rm t}$ the tangential pressure. The $(t,t)$ and $(r,r)$ components of the Einstein's equations $G_{\mu\nu} = 8\pi  T_{\mu\nu}$, together with the $r$ component of the continuity equation $\nabla^{\mu}T_{\mu\nu} = 0$, then yield the system
\begin{eqnarray}
	m^{\prime} &=& 4\pi r^2 \rho ~,
	\label{eq:tt} \\
	\frac{\nu^{\prime} }{2r e^{\lambda}}  &=& \frac{m}{r^3} + 4\pi P_{\rm r}~,
	\label{eq:rr} \\
	P_{\rm r}^{\prime} &=& - \frac{1}{2}(P_{\rm r}+\rho) \nu^{\prime} - \frac{2\Delta}{r}~,
	\label{eq:cont}
\end{eqnarray}
where $\Delta \equiv P_{\rm r} - P_{t}$ measures pressure anisotropy. The function $m(r)$ is related to the metric coefficient $\lambda(r)$ by means of $e^{-\lambda(r)} = 1-2 m(r)/r$, and can be  interpreted as  the mass-energy enclosed within the radius $r$. For the sake of compactness we omitted the functional dependence from the radial variable and denoted the derivative with respect to the radial coordinate by the prime. Eliminating $\nu'$ from the continuity equation yields the anisotropic generalization of the Tolman-Oppenheimer-Volkoff (TOV) equation,
\begin{eqnarray}\label{eq:TOV}
	P_{\rm r}^{\prime} &=& - \frac{(P_{\rm r}+\rho) \left(m + 4\pi r^3 P_{\rm r}\right)}{r^2(1-2 m/r)} - \frac{2\Delta}{r} ~.
\end{eqnarray}
Eq.s~(\ref{eq:tt}-\ref{eq:cont}) must be supplemented by additional identities relating $\rho$, $P_{\rm r}$ and $\Delta$ to close the system. These may include field equations of matter comprising the star and may replace Eq.~\eqref{eq:cont} in practical computations. The solutions must satisfy the boundary condition $e^{-\lambda}, e^{\nu} \to 1-2 M/r$ when $r \to \infty$, where $M$ is the mass of the star. The radius of the star can be defined as the value of the radial coordinate $R$ at the point where pressure drops to zero, so $M = M(R)$, e.g. for perfect fluid stars, or, if the star lacks a sharp boundary, e.g. in case of boson stars, one can define an effective radius that contains a fixed fraction, usually $99\%$, of the total mass, so $m(R) \equiv 0.99 M$. 

Production of GW echoes requires the existence of a photon sphere, implying that the compactness $\mathcal{C} \equiv M/R$ must be larger than 1/3 -- such stars are dubbed ultracompact. It is thus important to know what kind of matter may support that kind of compactness. Irrespectively of their composition, stars with a monotonously decreasing energy density profile and a positive pressure anisotropy,
\be\label{eq:Buch_cond}
	\rho' \leqslant 0~, \qquad \Delta \geqslant 0~,
\ee
must satisfy~\cite{Guven:1999wm} (for a proof see appendix~\ref{app:Buchdahl})
\be\label{eq:Buch}
\bbox{
	{\rm Buchdahl\,\,limit}~~~~~~\mathcal{C} \leqslant \frac{4}{9}}
\ee
as long as $\nu'$ is continuous. This bound is slightly lower than the compactness of BHs, $\mathcal{C} = 1/2$. 
We remark that if Eq.~\eqref{eq:Buch_cond} is satisfied then it must be saturated in order to saturate Eq.~\eqref{eq:Buch}. Such stars have isotropic pressure and constant density. These extremal conditions make CDS an important toy model for compact objects.
 The inequalities in \eqref{eq:Buch_cond} are relatively mild as we will argue based on the examples of fluid stars and boson stars.


Isotropic perfect fluids are described by $\Delta = 0$ and an EoS $\rho = \rho(P)$. We denoted $P\equiv P_{\rm r} = P_{\rm t}$. The EoS together with Eq.s~(\ref{eq:tt},\,\ref{eq:TOV}) form a closed system that can be integrated from the origin outward with initial condition $m(r = 0) =0$ and an arbitrary choice for the central density until the pressure becomes zero. The latter condition defines the surface of the star $r=R$, with the mass of the star given by $M \equiv m(R)$. Outside the mass distribution, which terminates at the radius $R$, there is vacuum with $\rho(r) = P(r) = 0$ and so $\nu_{\rm ext}(r) = -\lambda_{\rm ext}(r) = \log(1-2M /r)$. We remark that for perfect fluids the continuity equation $P^{\prime} + \nu^{\prime} (\rho + P)/2 = 0$ can be integrated without explicitly knowing the radial dependence of $\rho$ and $P$, yielding
\be\label{eq:nu_int}
	\nu_{\rm int}(r) = -2H[P(r)] + \log\left(1 - 2M/R\right)~,
\ee 
where $H(P) \equiv \int^P_0 \td P/(P+\rho(P) )$ is the pseudoenthalpy. The constant of integration was fixed by matching the interior and the exterior metric at the surface of the star. Notably, perfect fluid stars that satisfy the weak energy condition and are microscopically stable obey both conditions \eqref{eq:Buch_cond} and thus also the Buchdahl limit \eqref{eq:Buch}~\cite{Buchdahl:1959zz}. Under these assumptions, $\rho' \leqslant 0$ follows from the TOV equation \eqref{eq:TOV}, while $\Delta \geqslant 0$ is satisfied trivially by definition.
If, in addition, one also assumes causality, the maximal compactness of fluid stars is constrained even further, with $\mathcal{C}_{\rm max} = 0.354$~\cite{1984ApJ...278..364L,PhysRevD.46.4161,Lattimer:2006xb}.

A phenomenologically well motivated example of exotic stars with anisotropic pressure are boson stars (for reviews see e.g. \cite{Schunck:2003kk,Liebling:2012fv} and appendix~\ref{app:BosonStars} for field equations). Since static non-topological scalar field solutions can not be stable~\cite{Derrick:1964ww} boson stars are usually constructed from a complex scalar field $\Phi$ with the stationary field configuration $\Phi(r,t) = \phi(r) \exp(-i\omega t)$. In this case the spacetime remains static in spite of a time dependent field. The dynamics of the scalar field is determined by its potential $V(|\Phi|)$. Common examples considered in literature are the mini boson star with $V(\Phi) = \mu^2 |\Phi|^2$ that support a maximal compactness $\mathcal{C}_{\rm max} \approx 0.08$~\cite{Kaup:1968zz}, massive boson star with $V(\Phi) = \mu^2 |\Phi|^2 + \lambda |\Phi|^4/2$ and $\mathcal{C}_{\rm max} \approx 0.16$~\cite{Colpi:1986ye,Guzman:2005bs,AmaroSeoane:2010qx} and solitonic boson stars with $V(\Phi) = \mu^2 |\Phi|^2(1 - |\Phi|^2/\sigma_0^2)^2$ and a maximal compactness $\mathcal{C}_{\rm max} \approx 0.349$~\cite{Lee:1986ts,Friedberg:1986tq}. 

In all cases listed, the Buchdahl bound \eqref{eq:Buch} is satisfied, thus, even if it may be a coincidence, it makes sense to consider their relation to conditions \eqref{eq:Buch_cond}. First, the pressure anisotropy of boson stars is  positive, $\Delta = 2e^{-\lambda}(\phi')^2 \geqslant 0$, as required by the second condition in \eqref{eq:Buch_cond}. So, to evade the Buchdahl limit, the boson star has to contain regions with 
a growing energy density.  This might require exited states or bumpy potentials. Namely, if we limit the discussion to ground state configurations where the field is monotonous in the radial coordinate, the condition $\rho' \leqslant 0$ holds if the pressure is positive and the potential decreases monotonously within the star, $V' \leqslant 0$. This is so because for boson stars $\rho - P_{\rm r} = 2 V$, so $V'\leqslant 0$ implies $\rho' \leqslant P_{\rm r}'$. Since boson stars satisfy the weak energy condition, then $P_{\rm r}\geqslant 0$ implies $P_{\rm r}' \leqslant 0$ by the TOV equation and therefore $\rho' \leqslant 0$. So, at least in some fairly generic cases boson stars will respect the Buchdahl limit.\footnote{Note that in this case $0 \geqslant P_{\rm r}'/\rho' \geqslant 1$, where $P_{\rm r}'/\rho' \equiv \td P_{\rm r}/ \td \rho$ may be interpreted as the square of an effective speed of sound. Of course, this quantity will depend on a given solution as there is generally no unique relation between $P_{\rm r}$ and $\rho$ in boson stars.} 

For monotonously decreasing fields, $V' \leqslant 0$ is implied by $\partial V/\partial\phi \geqslant 0$. The last condition is satisfied in mini boson stars and massive boson stars, but it can be violated in solitonic boson stars (see examples in appendix~\ref{app:BosonStars}).  Namely, the most compact configurations of the solitonic boson star are characterized by an interior region where the field is roughly constant that is surrounded by a shell with a thickness of the order of $\mu^{-1}$. Within this shell, the field transitions to the true vacuum and, as the field rolls over the potential barrier, a sharp peak in the energy density arises. Solitonic boson stars attain their maximal compactness in the limit $\sigma_0 \to 0$, where the shell becomes infinitesimally narrow. In this case the solution can be obtained by first solving the TOV equations with the ansatz $P = \rho \propto \exp(-2\nu)$ inside the star and then matching $\nu$ smoothly to the exterior Schwarzschild spacetime~\cite{Friedberg:1986tq,Kesden:2004qx}. As this calculation is independent of the shape of the potential, this limiting case may be considered universal. In particular, the parameters within the potential do not determine the stars compactness. Instead, they mainly serve to fix the scale of boson stars mass, $m_{\rm Pl}^4/(\mu \sigma_0^2)$. Although the solitonic boson star is an example of a compact object that does not assume exotic matter and can also evade the conditions \eqref{eq:Buch_cond}, its maximal compactness does not surpass the maximal compactness allowed for physical fluid stars.

In the following we shall thus focus on fluid stars. On top of an EoS $P = P(\rho)$, we will require the additional conditions, that should be satisfied for any physically viable fluid star:
\begin{enumerate}

\item Matter satisfies the weak energy condition\footnote{In full generality, the weak energy condition 
stipulates that $\rho \geqslant 0$ and, for each $i$, $\rho + P_i \geqslant 0$.}
\be\label{eq:WEC}
	\rho \geqslant 0~, \qquad
	\rho + P \geqslant 0~.
\ee
In other words, the star is made of non-exotic matter.

\item Matter is microscopically stable. This assumption can be equivalently formulated in terms of the requirements 
\be\label{eq:MS}
	P \geqslant 0~,\qquad
	\frac{\td P}{\td \rho} \geqslant 0~.
\ee
In the presence of negative pressure, collapse is energetically favored. If $P\geqslant 0$ but $\td P/\td \rho <0$, the system is unstable with respect to volume fluctuations since a contraction would imply a decrease of the pressure and, consequently, generate a further contraction. Note that $\rho + P \geqslant 0$ and $P \geqslant 0$ are equivalent when $\td P/\td \rho \geqslant 0$ and $\rho \geqslant 0$ are assumed.

\item The speed of sound of the fluid $c_s$ -- that is the speed at which pressure disturbances travel in the fluid -- respects the causality constraint 
\be\label{eq:causality}
	c_s = \left(\frac{\td P}{\td \rho}\right)^{1/2} \leqslant 1~.
\ee

\end{enumerate}

We remark that the first two conditions exclude traversable wormholes, another class of potential candidates of compact objects, as they necessarily violate the weak energy condition~\cite{Morris:1988tu,Visser:1995cc}.\footnote{Traversable wormholes without exotic matter may be possible within modified gravity (see e.g. ~\cite{Capozziello:2012hr,Harko:2013yb,Duplessis:2015xva,Hohmann:2018shl,Ovgun:2018xys}).} For similar reasons we will also not consider gravastars -- they violate the second condition since they are supported by an inner core with negative pressure confined inside a thin shell of ultra-relativistic stiff matter. Furthermore, gravastars require a negative pressure anisotropy in order to avoid  the presence of a pressure discontinuity at the junction between the inner core and the crust~\cite{Cattoen:2005he}. 

\subsection{Constant density stars}\label{sec:CDS}

The Buchdahl limit in (\ref{eq:Buch}) does not pose any obstruction against the existence of ultracompact stars with 
\be\label{eq:LightRing}
\bbox{
{\rm Ultracompact\,\,stars}~~~~~~\frac{1}{3} \leqslant \mathcal{C} \leqslant \frac{4}{9}
}
\ee
However, we remark that the derivation of the Buchdahl limit did not use the causality condition~3. In this sense, the case of CDS is a representative example for compact objects that respect the Buchdahl limit, have the chance to be ultracompact but violate the causality condition. This is indeed the case explored in~\cite{Pani:2018flj} to discuss gravitational echoes in the context of ultracompact stars. The M-R relation is given by $M=4\pi\rho_0 R^3/3$, where $\rho_0 = 3\mathcal{C}/4\pi R^2$ is the constant density of the star. The TOV equation can be analytically integrated, and the pressure inside the star is given by 
\be\label{eq:P_Cstar}
	P_{\rm CDS}(r) 
	= \frac{3\mathcal{C}}{4\pi R^2}\left[
		\frac{\left(1 - 2\mathcal{C}\right)^{1/2} - \left(1 - 2\mathcal{C}r^2/R^2\right)^{1/2}}
		{\left(1 - 2\mathcal{C}r^2/R^2\right)^{1/2} - 3\left(1 - 2\mathcal{C}\right)^{1/2}}
	\right]~,
\ee
while the metric function $\nu(r)$ is given by
\be\label{eq:nu_Cstar}
e^{\nu_{\rm CDS}(r)} = \left[
	\frac{3}{2}\left(1-2\mathcal{C}\right)^{1/2} 
- 	\frac{1}{2}\left(1-2\mathcal{C}r^2/R^2\right)^{1/2}
\right]^2~.
\ee
The Buchdahl limit $\mathcal{C} \leqslant 4/9$ follows from the stability requirement $P\geqslant 0$ at the center of the star. 
A CDS saturating the Buchdahl limit has infinite central pressure. 

Another relevant quantity related to gravitational echoes is the echo frequency. As we will show shortly, CDS will produce the highest echo frequencies when compared to other fluid stars with the same mass and radius. The echo frequency can be roughly estimated from the inverse of the time $\tau$ it takes a massless test particle to travel from the unstable light ring to the center of the star,
\be\label{eq:EchoFreq}
	\tau_{\rm echo} \equiv \int^{3M}_{0} \td r \, e^{(\lambda(r)-\nu(r))/2}~.
\ee
The corresponding echo frequency is $f \approx \pi/\tau_{\rm echo}$.

The geometry of the fluid stars satisfies the following constraints 
\be
	e^{\nu(r)} \leqslant e^{\nu_{\rm CDS}(r)}~, \qquad
	e^{\lambda(r)} \geqslant e^{\lambda_{\rm CDS}(r)} = (1 - 2M r^2/R^3)^{-1}~,
\ee
where the metric components for the CDS are evaluated for the same mass and radius (for details see appendix~\ref{app:Buchdahl}). Thus for a fluid star with compactness $\mathcal{C}$ and mass $M$ the following limit holds
\be \label{eq:TauBound}
	\frac{\tau_{\rm echo}}{M} \geqslant \left.\frac{\tau_{\rm echo}}{M}\right|_{\rm CDS}
	=  	\frac{ \cot^{-1}\left( \sqrt{4 - \frac{9}{\mathcal{C}}} \right) + \tan^{-1}\left( 3/\sqrt{4 - \frac{9}{\mathcal{C}}} \right) }{ \mathcal{C}^2  \sqrt{4 - 9/\mathcal C } }
	-	2 \log\left( \frac{1}{\mathcal{C}} - 2 \right) + 3 - \frac{1}{\mathcal{C}}~.
\ee
Note that $\tau_{\rm echo} \to \infty$ as $\mathcal{C} \to 4/9$.

Although CDS provide a useful analytic toy model, they are unphysical because, as in any other ideal incompressible fluid or medium, the speed of sound is infinite. It is therefore important to investigate what are the restrictions imposed by causality  on the ultracompact condition in (\ref{eq:LightRing}).

\begin{figure}[!htb!]
\centering
  \includegraphics[width=.8\linewidth]{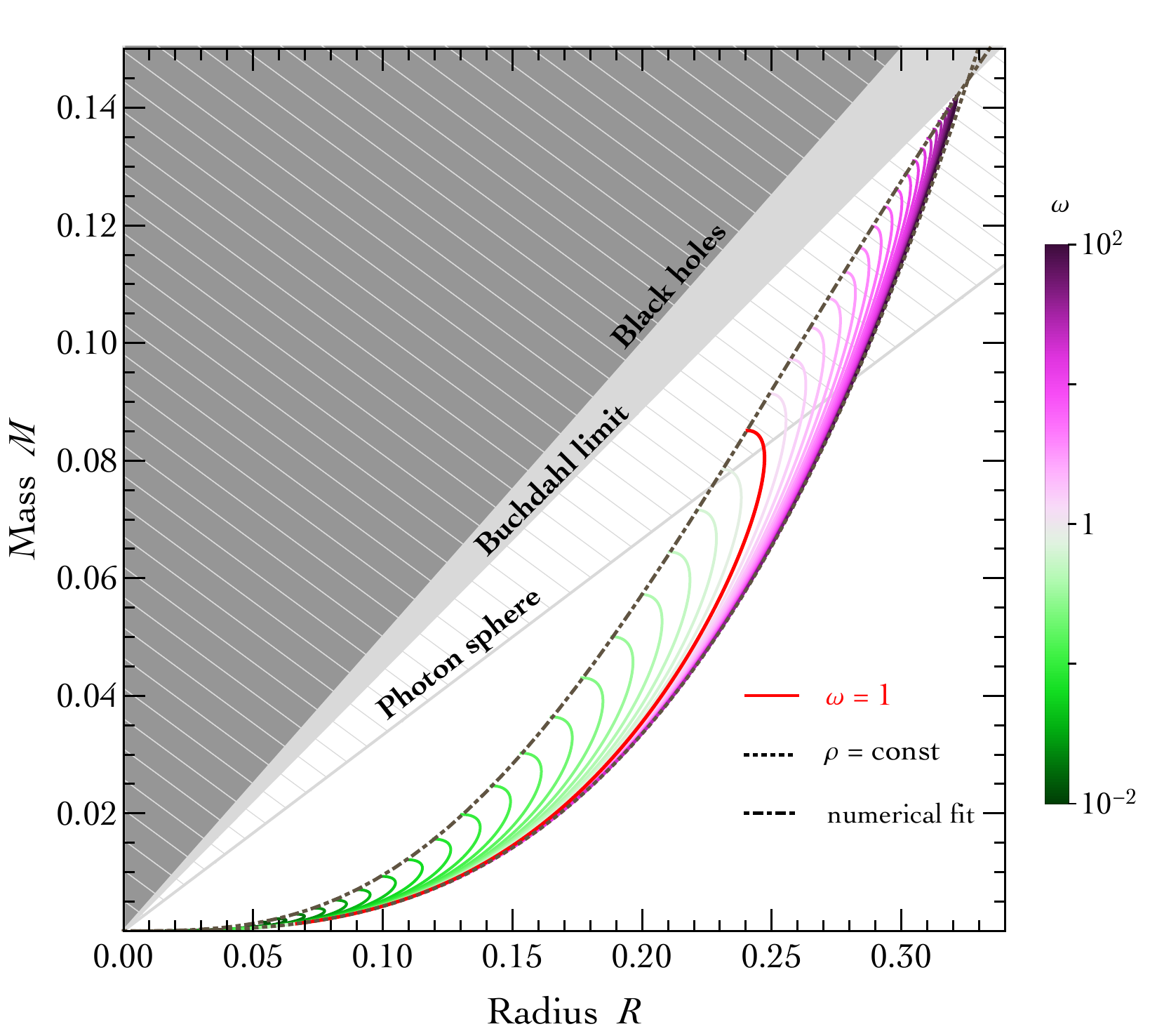}
\vspace{-0 cm}
\caption{\label{fig:MassRadiusPlot}\em The M-R curves for different linear EoS $\rho = \rho_0 + P/\omega$. The red M-R curve saturates the causal limit, the magenta curves violate causality and the green curves obey it. The M-R curves are bounded by the dash-dotted curve obtained numerically, and the dotted curve representing the M-R relation of CDS, $M = \rho_0 (4\pi/3) R^3$, which also corresponds to limiting case $\omega \to \infty$. The mass and radius are given in units of $\rho_0^{-1/2}$.}
\end{figure}


\subsection{Causality and maximal compactness}

Causality \eqref{eq:causality} implies that the speed of sound in the medium should always be smaller than the speed of light. Together with stability they impose the condition $0 \leqslant c_s^2 \equiv \td P/\td \rho \leqslant 1$ on the EoS. For the sake of generality we will consider the constraint $0 \leqslant \td P/ \td \rho \leqslant \omega$ instead, allowing also for non-causality, $\omega > 1$. The integrated equivalent of this condition implies that such EoS should satisfy
\be\label{eq:EoS_cond}
	P/\omega \geqslant \rho(P) \geqslant \rho_0 + P/\omega 
	\qquad \mbox{for all} \qquad P \geqslant 0, 
\ee
where $\rho_0 > 0$ is a constant. Notice the strict inequality: $\rho_0 < 0$ is clearly excluded by the condition $\rho(P) > 0$ for $P>0$, while $\rho_0 = 0$ is excluded because the EoS $\rho(P) = P/\omega$ does not support stable stars.\footnote{Since the TOV equation \eqref{eq:TOV} with an EoS $P = \omega \rho$ (with $\omega > 0$) does not contain any explicit scales, then given a solution $m(r)$ other solutions can be obtained by the rescaling $m(r) \to \alpha  m(r/\alpha)$, where $\alpha$ is a constant. This would imply $M \propto P_c^{-1/2}$, i.e the mass decreases with increasing central pressure indicating instability against spherical perturbations. Moreover, because the exact solution $m(r) = 2r\omega/(1 + 6\omega + \omega^2)$ is an attractor when $r \to \infty$, such stars do not even possess a well defined boundary.}

\begin{figure}[!tb!]
\begin{center}
\includegraphics[width=.55\textwidth]{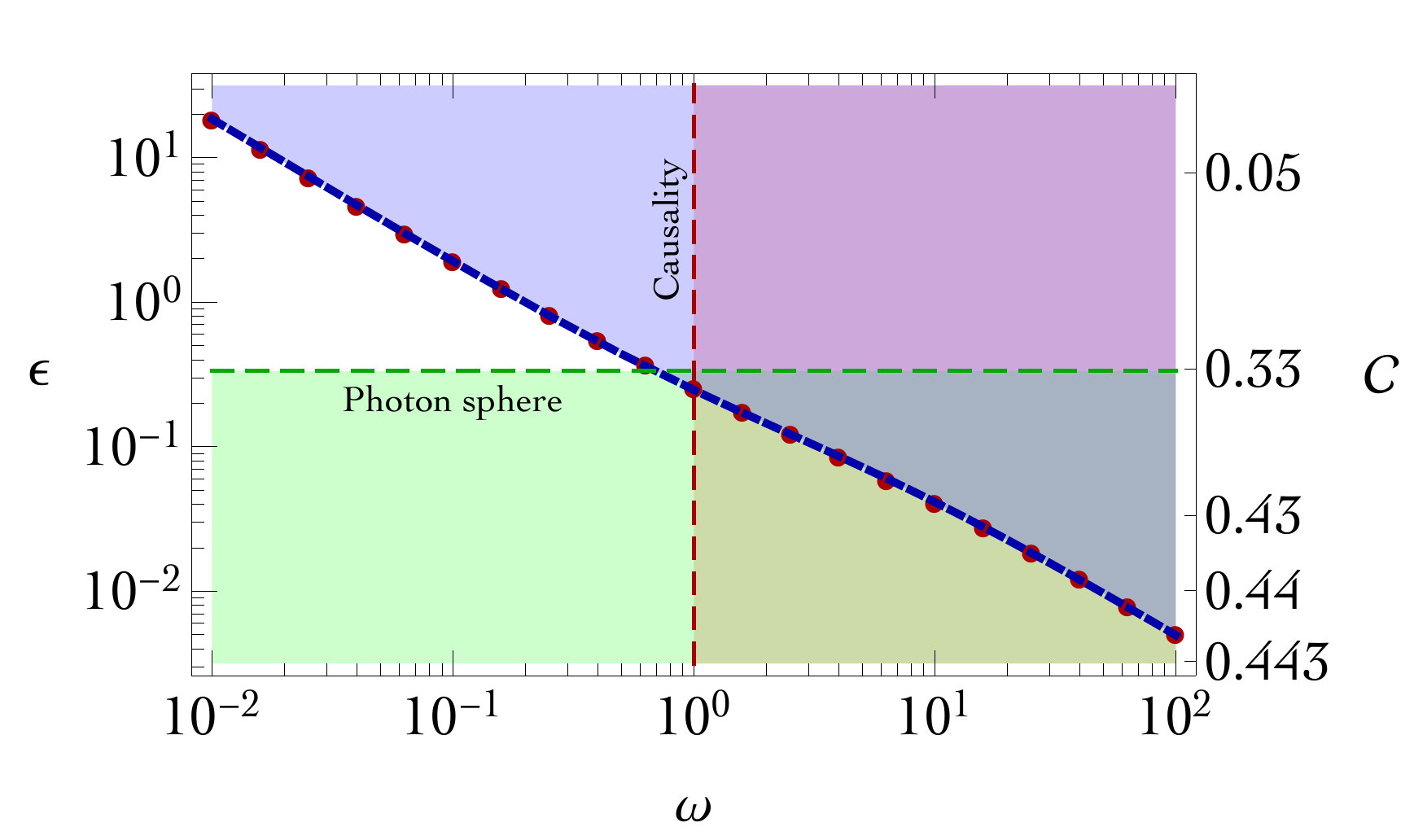}
\caption{\em \label{fig:MaxCPlot} The maximal compactness for the LinEoS. The numerical estimate obtained by solving the TOV equation is denoted by red dots. 
The dashed blue curve corresponds to the analytic interpolation \eqref{eq:lin_eps_min}. 
The red region is excluded  by causality and the green region contains stars with a photon sphere.
The region shaded in blue on the left of the vertical dashed red line, therefore, encompasses physically viable solutions. }
\end{center}
\end{figure}

\begin{figure}[!tb!]
\begin{center}
$$\includegraphics[width=.434\textwidth]{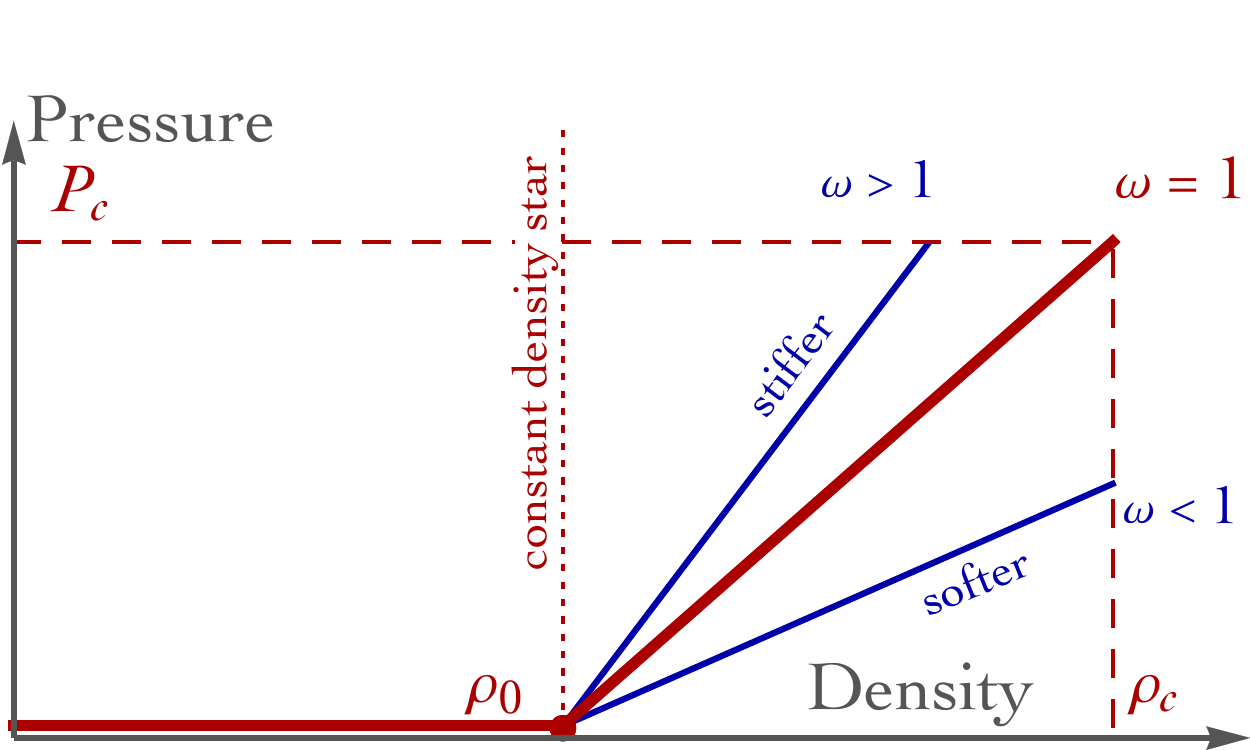}
\qquad\includegraphics[width=.434\textwidth]{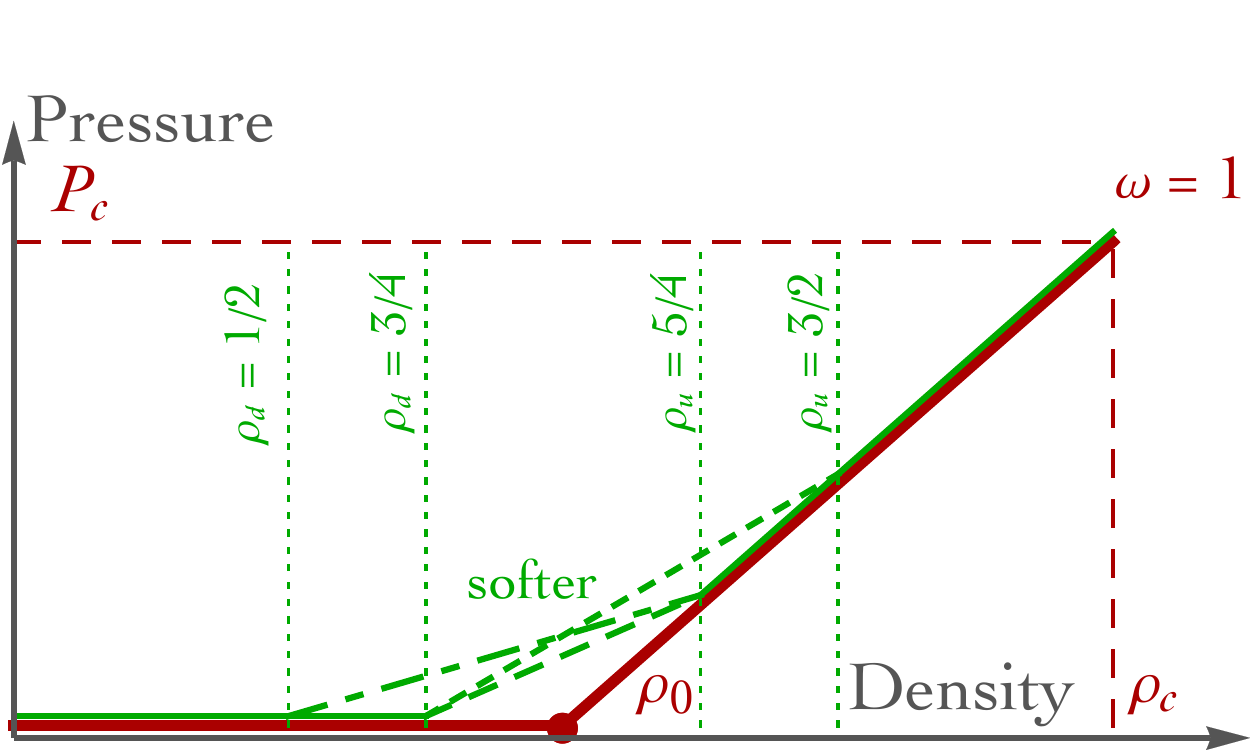}$$
\caption{\em \label{fig:EOS} 
Left panel.
Comparison between the LinEoS (solid red) and the EOS for a CDS (dashed red) for different values of $\omega$. 
Right panel. Modifications of the LinEoS in the outer region of the star according to the parametrization proposed 
in Eq.~(\ref{eq:LinEoSMod}).
 }
\end{center}
\end{figure}

The condition $\td P/ \td \rho \leqslant \omega$, or equivalently \eqref{eq:EoS_cond}, is saturated by the linear EoS (LinEoS hereafter),
\be\label{eq:EoS_lin}
	\rho(P) = \rho_0 + P/\omega~,
\ee
which describes the stiffest possible EoS as the speed of sound, $c_s = \sqrt{\omega}$, takes the maximal value throughout the star. In the left panel of Fig.~\ref{fig:EOS} we highlight the comparison with the EoS of a CDS. It is seen that the latter can be obtained in the limit $\omega \to \infty$. 

The maximal compactness for a given LinEoS \eqref{eq:EoS_lin} is a function of $\omega$ only, since $\rho_0$ be absorbed in the redefinitions $\tilde{P} \equiv P/\rho_0$, $\tilde{\rho}\equiv \rho/\rho_0$.\footnote{This is equivalent to working in units where $\rho_0 = 1$.} The maximal compactness for the LinEoS for a given $\omega$ can be approximated the rational function
\be\label{eq:lin_eps_min}
	\eps_{\rm min}(\omega) \approx  \frac{0.77 + 0.51 \, \omega}{\omega \, (4.18 + \omega)}~,	\qquad \mbox{where} \qquad
	\eps \equiv \frac{4}{9\mathcal C}- 1~,
\ee
quantifies the distance from the Buchdahl bound. The numerical results for the compactness of the LinEoS are summarized in Fig.~\ref{fig:MaxCPlot}, where the red dots represent the numerical estimates while the blue line corresponds to the analytic approximation \eqref{eq:lin_eps_min}. The analytic fit \eqref{eq:lin_eps_min} deviates from the numerical prediction by at most $3.6\%$ .

The M-R curves of the LinEoS \eqref{eq:EoS_lin} are shown in Fig.~\ref{fig:MassRadiusPlot} for $\omega$ ranging from 0.01 to 100. It can be seen that for low masses the M-R curve roughly follows a star with constant density $\rho_0$. The density in such stars is approximately constant, $\rho \approx \rho_0$, so the pressure inside the star must satisfy $P \ll \omega \rho_0$. On the other hand, when $P \gg \omega \rho_0$ in the center of the star, the density profile will be cored and the M-R curve can rise above the M-R curve for the CDS. Unlike the CDS, however, the LinEoS can not support an infinitely high pressure. When the pressure is increased above some critical value, the mass will begin to decrease and the star becomes unstable~\cite{Harrison1965}. The maximal compactness of the CDS and LinEoS are thus determined by fundamentally different mechanisms: for the LinEoS it is set by the stability of the star while, for the CDS it follows from the positivity of the $g_{tt}$ component of the metric, given by Eq.~\eqref{eq:nu_Cstar}.

The LinEoS \eqref{eq:EoS_lin} deserves attention as another extremal toy model since 
it produces the most compact stars from all possible EoS satisfying $\td P/ \td \rho \leqslant \omega$. For all such stars 
\be\label{eps_bound}
	\eps \geqslant \eps_{\rm min}(\omega)~.
\ee
The allowed values for compactness are depicted by the blue region in Fig. \ref{fig:MaxCPlot}. Although a rigorous analytic proof of this statement is missing, it is supported by numerical studies which has so far failed to provide counterexamples. We remark that although it is proved that saturating the bound for causality or microscopic stability, $0 \leqslant \td P/ \td \rho \leqslant 1$, gives the extremal mass for a fixed central density and pressure~\cite{PhysRevLett.32.324}, this does not imply maximal compactness. In fact, it is relatively easy to produce counterexamples, e.g. configurations where a LinEoS with a smaller $\omega$ but gives a more compact star if the central density and pressure are fixed.  As a consistency check, for $\omega =1$ we find that $\mathcal{C}_{max} = 0.354$, which matches the previously obtained upper bounds on compactness from causality~\cite{1984ApJ...278..364L,PhysRevD.46.4161,Lattimer:2006xb}.

\begin{figure}[!htb!]
\begin{center}
$$\includegraphics[width=.45\textwidth]{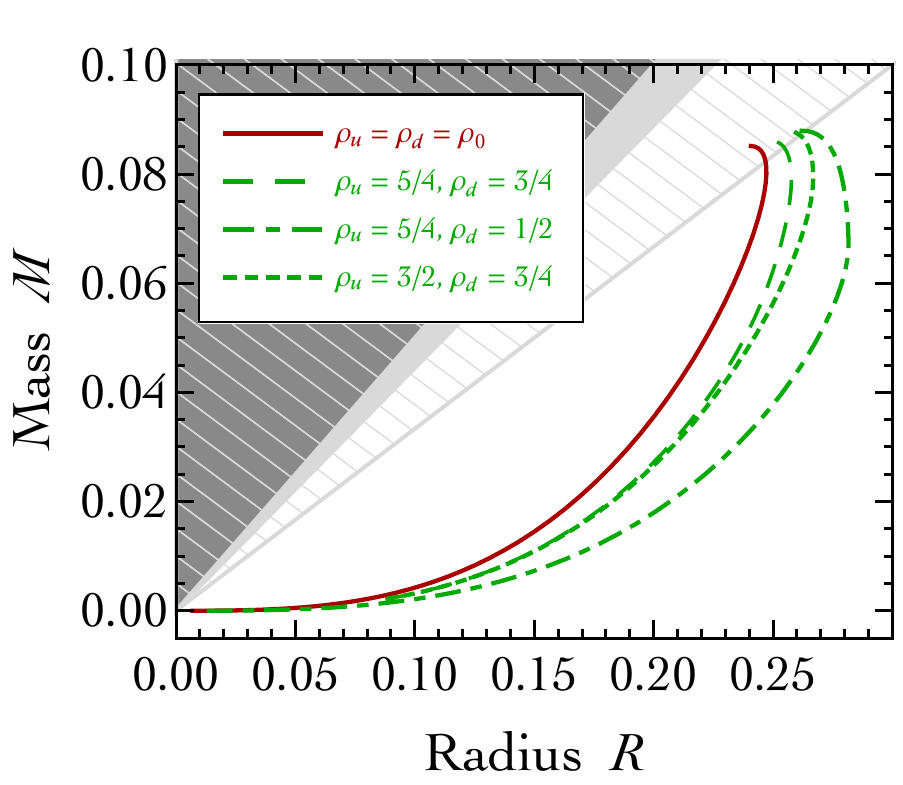}
\qquad\includegraphics[width=.434\textwidth]{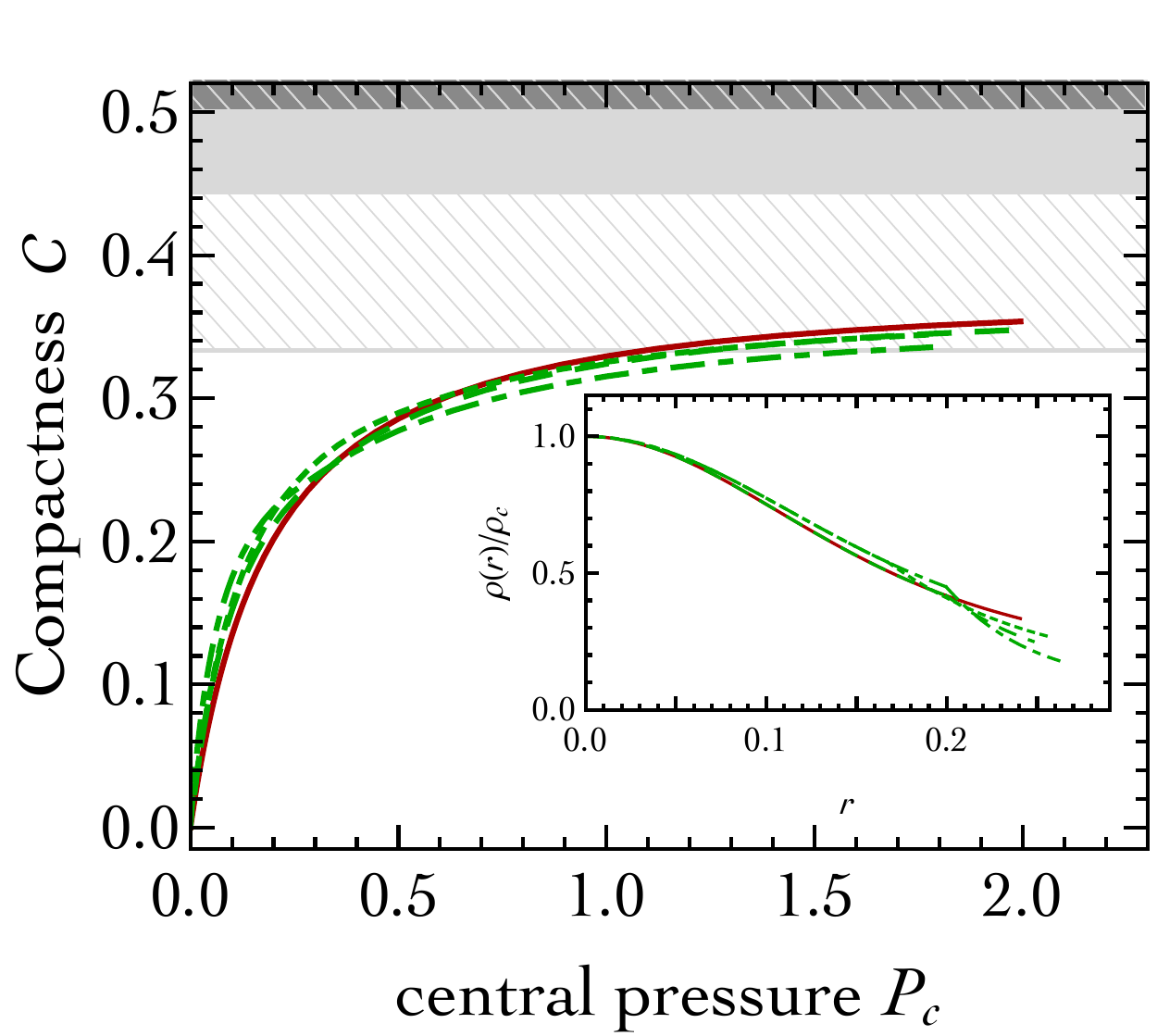}$$
\caption{\em \label{fig:MassRadiusComparison} 
Left panel. M-R relation for the modified LinEoS in the right panel of Fig.~\ref{fig:EOS}.
Right panel. Relation between central pressure and compactness for the same modified LinEoS. In the inset plot
we show the radial profile of the energy density (normalized w.r.t. its central value).
 }
\end{center}
\end{figure} 

To test the hypothesis \eqref{eps_bound} that the LinEoS yields the most compact stars we compared the LinEoS~\eqref{eq:EoS_lin} with other EoS satisfying $\td P/ \td \rho \leqslant \omega$. This is illustrated in Fig.~\ref{fig:EOS} where we test modification of the EoS $\tilde{P} = \tilde{\rho} - 1$ (thus with $\omega = 1$) with a decrease in the speed of sound in the outer region of the star. As an example, consider modifications with a functional form
\be \label{eq:LinEoSMod}
	\tilde{P} = \tilde{\rho}\left(
	\frac{\rho_u/\rho_0 - 1}{\rho_u/\rho_0 - \rho_d/\rho_0}
	\right) - \frac{\rho_d}{\rho_0}\left( 
	\frac{\rho_u/\rho_0 - 1}{\rho_u/\rho_0 - \rho_d/\rho_0}
	\right) \equiv a\tilde{\rho} - b~.
\ee
The M-R curves corresponding to the linear and the modified EoS are shown in the left panel of Fig.~\ref{fig:EOS} while the right panes depicts the compactness $\mathcal{C}$ against the central pressure $P_c$. Fig.~\ref{fig:MassRadiusComparison} also indicates that the LinEoS generally allows higher central pressures.

The upper bound \eqref{eps_bound} also applies to stars composed of different types of matter, each with their own density $\rho_i$ and $P_i$ satisfying $0 \leqslant \td P_i/ \td \rho_i \leqslant \omega_i$. The TOV equation \eqref{eq:TOV} is then expressed in terms of the total density and pressure, $\rho = \sum_i \rho_i$, $P = \sum_i P_i$. Although there is no unique relation between $\rho$ and $P$ in that case, it holds that inside the star
\be
	\frac{\td P}{\td \rho} \leq \sum_i \omega_{i} \frac{\rho_{i}^{\prime}}{\rho^{\prime}} \leq \max \omega_ i~,
\ee
given that $\rho_{i}^{\prime} \leq 0$. The latter condition is implied by the TOV equations for individual matter components if all components are microscopically stable and satisfy the weak energy condition.  If $\rho$ is a decreasing function inside the star, it can substitute the radial coordinate and we may use $\td P/\td \rho \equiv P^{\prime}/\rho^{\prime}$ to define a effective equation of state. Thus, if \eqref{eps_bound} holds for a single fluid star, it will also hold for stars containing multiple fluids with $\max \omega_ i \leqslant \omega$.

In Fig.~\ref{fig:MassRadiusPlot} we used dimensionless values of mass and radius in units of $\rho_0^{-1/2}$. In order to gain some intuition about the typical size of these objects we need to specify its value. For definiteness, let us focus on the case saturating the causality bound $\omega = 1$. The maximal mass corresponds to
\be
\bbox{
	M \simeq 3\times \left(
	\frac{\rho_0}{5\times 10^{14}\,{{\rm g}/{\rm cm}^3}}
	\right)^{-1/2}M_{\odot}~,~~~~~
	R \simeq 12.5\times\left(
	\frac{\rho_0}{5\times 10^{14}\,{{\rm g}/{\rm cm}^3}}
	\right)^{-1/2} {\rm km}
}
\ee
It is thus not implausible to speculate about the possible existence of ultra-high density exotic fluid stars with maximally stiff EoS and mass exceeding the maximal mass expected for conventional NS at the same value of $R$ (cfr. Fig.~\ref{fig:MassRadiusEOS}).

\section{Quasi-normal modes, light rings and ringdown} 
\label{sec:echoes}

As a consequence of a generic perturbation, the metric functions in Eq.~(\ref{eq:LineElement}) and the fluid variables change by a small amount with respect to their unperturbed values. The dynamics of the perturbed system is described by the Einstein's equations coupled to the hydrodynamical equations and the conservation of baryon number. A remarkable result -- obtained after expanding in tensor spherical harmonics -- is that the perturbed equations decouple into two sets  according to the parity of the harmonics:  polar (or even) and axial (or odd) perturbations. Noticeably,  the same distinction is valid for a Schwarzschild BH. In that case, however, polar and axial perturbations are isospectral.  In the case of stars, on the contrary, the polar and axial perturbations are fundamentally different since the former couple metric and hydrodynamical perturbations while the latter describe pure spacetime perturbations. This separation is strictly true for a non-rotating star. In order to enforce the parallel between exotic stars and Schwarzschild BHs, we are only interested in spacetime perturbations, and from now on we shall focus on the axial modes. In this case, perturbations of the line element in Eq.~(\ref{eq:LineElement}) reduce to a single wave equation of the form 
\be\label{eq:PertEquation}
\left[
\frac{\partial^2}{\partial t^2} - \frac{\partial^2}{\partial r_*^2} + V_{s,l}(r)
\right]\Psi_{s,l}(r_*,t) = 0~,
\ee
where we introduced the Regge-Wheeler ``tortoise'' coordinate $dr_* = e^{(\lambda - \nu)/2}dr$, and where the effective radial potential is 
\be\label{eq:Potential}
V_{s,l}(r) = e^{\nu(r)}\left\{
\frac{l(l+1)}{r^2} + \frac{1-s^2}{2re^{\lambda(r)}}\left[
\nu^{\prime}(r) - \lambda^{\prime}(r)
\right] + 8\pi[P(r) - \rho(r)]\delta_{s,2}
\right\}~.
\ee
 \begin{figure}[!tb!]
\begin{center}
$$\includegraphics[width=.434\textwidth]{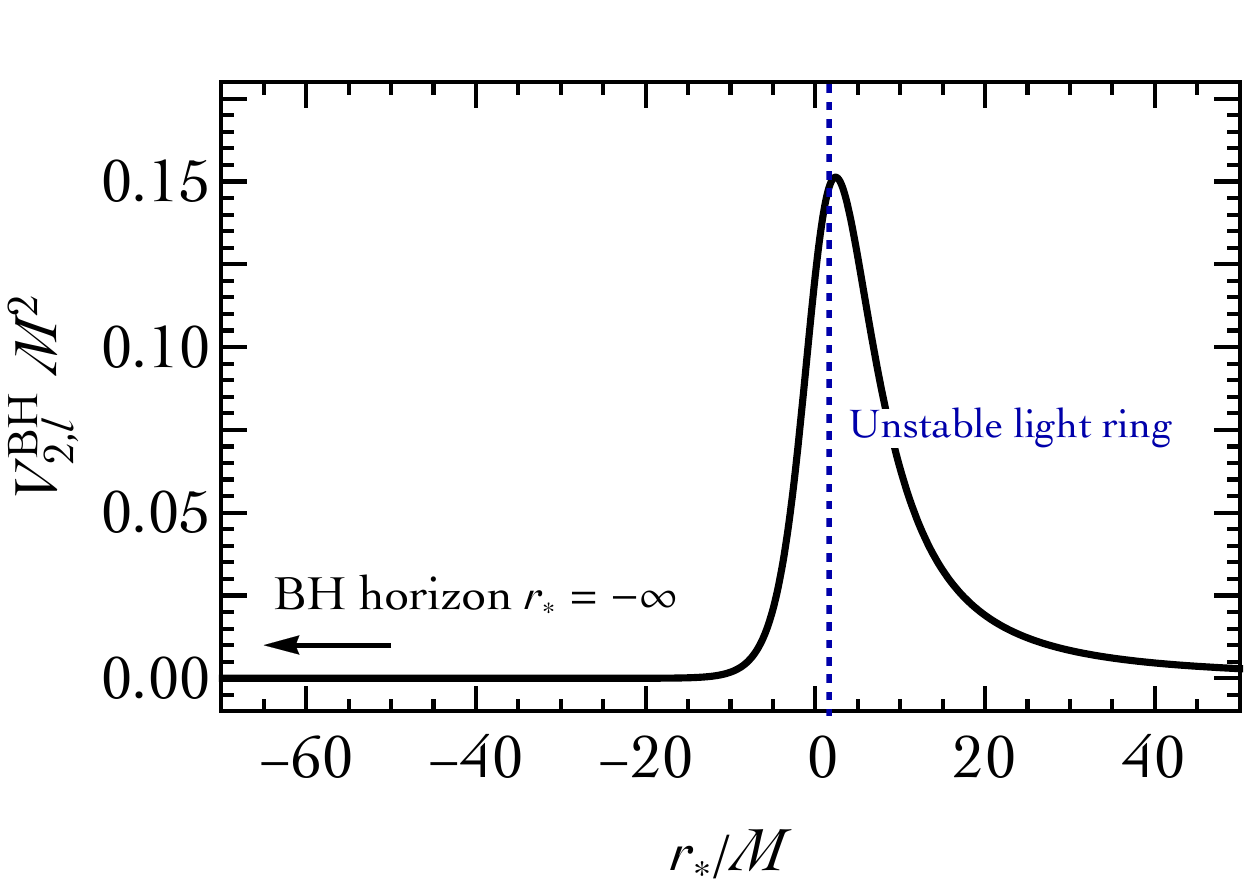}
\qquad\includegraphics[width=.434\textwidth]{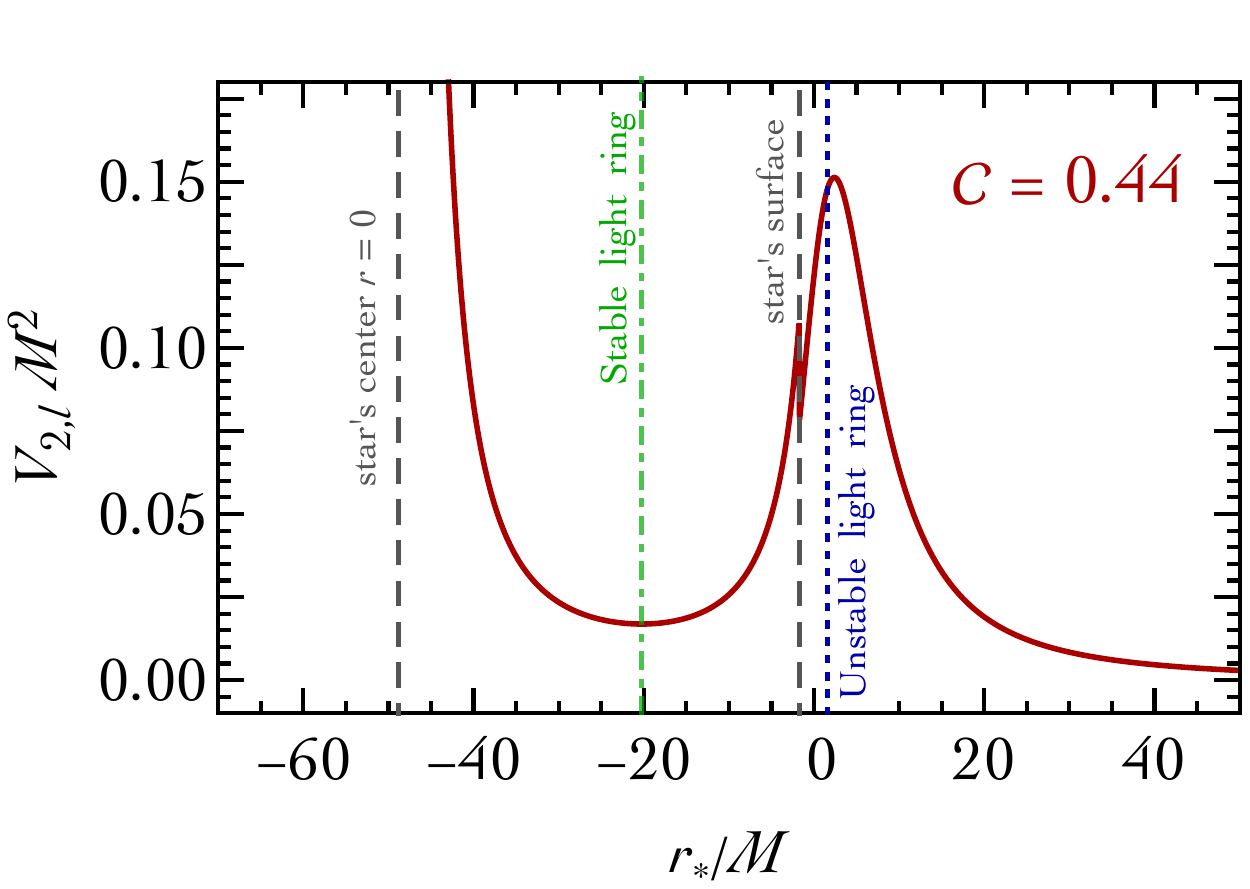}$$
\caption{\em \label{fig:Pot} 
Effective radial potential in Eq.~(\ref{eq:PotConstantDensityStar}) as function of the tortoise coordinate for a
Schwarzschild BH (left panel) and a CDS (right panel).
 }
\end{center}
\end{figure} 
The tortoise coordinate $r_*$ has an important physical meaning. In the metric described by the line element in Eq.~(\ref{eq:LineElement}), radial null geodesics obey $ds^2 = -e^{\nu}dt^2 + e^{\lambda}dr^2 = 0$, and we have $dt/dr = \pm e^{(\lambda - \nu)/2}$. This equation can be integrated and solved for $t = t(r)$,  and, schematically, we find $t = \int dr e^{(\lambda - \nu)/2} = r_*$. This means that the tortoise coordinate measures the coordinate time that elapses along a radial null geodesics. In Eq.~(\ref{eq:Potential}) the azimuthal quantum number satisfies $l \geqslant s$, and 
 $s = 0,\pm 1,\pm 2$ for scalar, vector and tensor modes, respectively. The effective potential or axial gravitational perturbations  -- after using the field equations -- reduces to
 \be\label{eq:PotConstantDensityStar}
 V_{2,l}(r) = e^{\nu(r)}\left[
 \frac{l(l+1)}{r^2} - \frac{6m(r)}{r^3} - 4\pi(P - \rho)
 \right]~.
 \ee
We look for stationary solutions of Eq.~(\ref{eq:PertEquation}) in the form $\Psi_{2,l}(r,t) = e^{-i\omega t}\psi_{2,l}(r)$. The radial mode thus satisfies the time-independent Schr\"odinger-like equation 
\be\label{eq:EigenvalueProblem}
	\frac{\td^2\psi_{2,l}}{\td r_*^2} + \left[\omega^2 - V_{2,l}(r)\right]\psi_{2,l} = 0~.
\ee
 
 \begin{figure}[!tb!]
\centering
  \includegraphics[width=.7\linewidth]{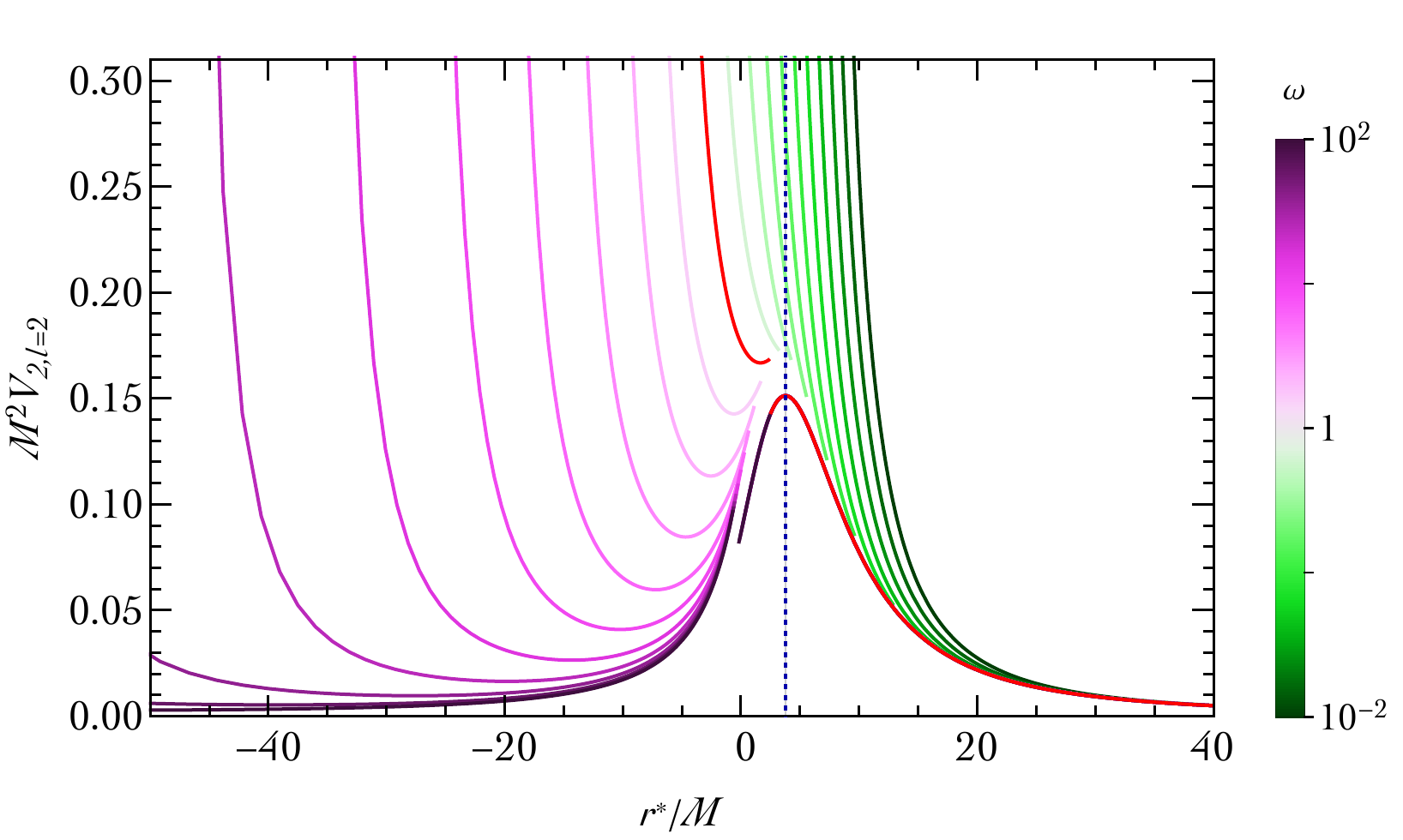}
\vspace{-0 cm}
\caption{\label{fig:PotentialPlot}\em The effective potentials of the maximally compact stars with a LinEoS $\rho = \rho_0 + p/\omega$. The red line corresponds to the maximally compact star saturating the causal limit. The magenta potentials have a non-causal LinEoS with $\omega >1$. The green potentials respect causality yet they do not possess a deep enough potential well to produce gravitational echoes.}
\end{figure}
 
For a Schwarzschild BH one finds
\be\label{eq:PotSch}
 V_{2,l}^{\rm BH}(r) = \frac{1}{r^3}\left(
 1 - \frac{2M}{r}
 \right)\left[
 l(l+1)r - 6M
 \right]~,~~~~~~r_* = r + 2M\log\left(\frac{r}{2M}-1\right)~.
\ee
The potentials in Eq.~(\ref{eq:PotConstantDensityStar}) and Eq.~(\ref{eq:PotSch}) are shown in Fig.~\ref{fig:Pot} (in the right and left panel, respectively). As a benchmark example, we consider a CDS with compactness $\mathcal{C} = 0.44 > 1/3$. The surface of this star is located at $R = M/\mathcal{C} < 3M$, and -- as evident in both the left and right panel of Fig.~\ref{fig:Pot} --  the exterior Schwarzschild potential has a barrier at $r \simeq 3M$: This is the location of the unstable circular null geodesic for a Schwarzschild spacetime, a.k.a. the light ring.\footnote{To be more precise, the light ring at $r = 3M$ corresponds to the position of the maximum of the potential barrier obtained for $s=l=1$ since this is the case describing  massless photons. However, with a slight abuse of language, we shall denote as ``light ring'' also the position of the maximum of the barrier describing gravitational perturbations, $r \simeq 3M$.}
 The important difference is that the Schwarzschild potential vanishes at the BH horizon (corresponding to $r_* \to -\infty$ in tortoise coordinate) while the potential of a perturbed star tends to infinity at $r = 0$ because of the presence of the centrifugal contribution $l(l+1)/r^2$. A trivial consequence of that is that the potential of the ultracompact star enjoys the presence of a potential well in between the light ring at $r = 3M$ and the center of the star. This also means that the star possesses a second (stable) light ring at the minimum of the potential. This observation is, in fact, very general: all horizonless compact objects that are formed of matter satisfying the null energy condition, and that possess an unstable light ring -- which is a necessity for the production of gravitational echoes -- will also have a stable light ring~\cite{Cunha:2017qtt}.

The situation for compact stars governed by the LinEoS is illustrated in Fig.~\ref{fig:PotentialPlot}. Of particular relevance is the $\omega$-dependence. At large $\omega \gg 1$ the potential shows the same qualitative features compared to CDS with a deep well in which radiation can be easily trapped. This situation is, however, unphysical due to violation of causality. For physical values $\omega \leqslant 1$ the potential well rapidly disappears.

An important comment that concerns the unavoidable presence of an internal stable light ring is mandatory. Spacetimes with a stable light ring may be non-linearly unstable because,
 roughly put, a stable light ring is able to trap radiation faster than it can escape the star, thus perturbations can accumulate until non-linear effects become relevant~\cite{Cunha:2017qtt,Keir:2014oka,Stuchlik:2017qiz}. To be more precise, it was shown in~\cite{Keir:2014oka} that metric perturbations of ultracompact stars cannot decay, at the linear level, faster than logarithmically. This is supposed to be problematic because, when proving stability at the full non-linear level, one usually requires faster-than-logarithmic decay of linear perturbations. In this sense, the instability related to the stable light ring would manifest itself only at the non-linear level, and -- given the level of complexity of these computations for non-trivial space-time metrics and also their possible dependence on the matter field equations -- no firm conclusion about its actual presence can be made. In the following, we shall continue our discussion under the tacit assumption that the instability associated with the stable light ring, if any, operates on timescales longer than a Hubble time.

The differential equation in Eq.~(\ref{eq:EigenvalueProblem}) must be solved with appropriate boundary conditions.
First, take the a Schwarzschild BH for which $V_{2,l}^{\rm BH}(r \to 2M) \to 0$, and a general solution in the vicinity of the horizon is a spherical wave $\Psi_{2,l}(r_*,t)\sim e^{-i\omega(t\pm r_*)}$. Since nothing can escape from the BH we must impose an inward-directed spherical boundary condition at the horizon
\be\label{eq:BC1BH}
	\Psi_{2,l}(r_*,t)\sim e^{-i\omega(t + r_*)}~,~~~~~r_* \to -\infty~~(r\to 2M)~.
\ee
In an asymptotically flat spacetime, the potential at spatial infinity tends to zero as well. As second boundary condition we require for the solution to be a spherical outward-directed wave at spatial infinity 
\be\label{eq:BC2BH}
	\Psi_{2,l}(r_*,t)\sim e^{-i\omega(t - r_*)}~,~~~~~r_* \to +\infty~~(r\to +\infty)~.
\ee
The boundary conditions in~(\ref{eq:BC1BH},\,\ref{eq:BC2BH}) define the QNMs for a BH, and their physical meaning can be understood as follows: 
We are interested in the dissipation of metric fluctuations around a Schwarzschild BH. Thus the boundary conditions~(\ref{eq:BC1BH},\,\ref{eq:BC2BH}) impose that perturbations initially localized around the BH can either escape to infinity or fall into the BH, there is no gravitational radiation coming from spatial infinity (or the horizon) that may continue perturbing the BH. The eigenfrequencies obtained by solving Eq.~\eqref{eq:EigenvalueProblem} with the aforementioned boundary conditions have both a real and an imaginary part, the latter giving the inverse of the damping time $\tau_{d}$ of the corresponding mode, $\omega = \mathbb{Re}[\omega] + i \mathbb{Im}[\omega] \equiv \mathbb{Re}[\omega] - i2\pi/\tau_{d}$.
 
For compact stars the situation is slightly different as they lack an event horizon. In order to define the QNMs the boundary condition \eqref{eq:BC1BH} must be replaced by a regularity condition at the center of the star, while the condition~\eqref{eq:BC2BH} at spatial infinity is unaltered. This change in the boundary conditions dramatically affects the spectrum of the QNMs of the compact star with respect to the BH.
 \begin{figure}[!htb!]
\centering
  \includegraphics[width=.5\linewidth]{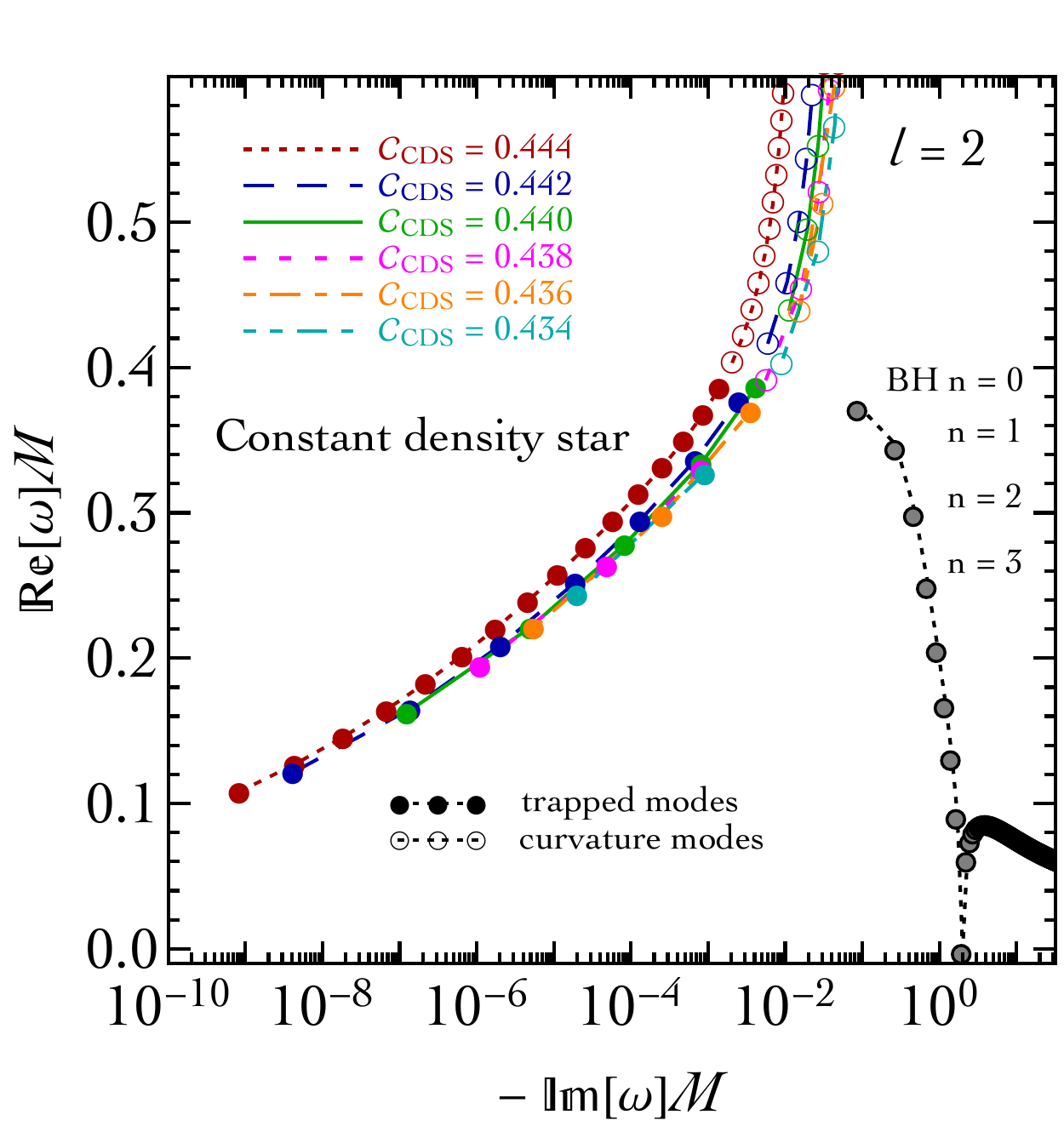}
\vspace{-0 cm}
\caption{\label{fig:QNM}\em QNMs for a CDS for different benchmark values of its compactness. For comparison, we also show (filled gray circles) the QNMs of a Schwarzschild BH.
}
\end{figure}
This is evident from Fig.~\ref{fig:QNM} where we compared the spectrum of the QNMs of a compact star with those of a BH. We focus on the lowest axial gravitational modes with $s=l=2$. We consider the (unphysical) case of a CDS, and show the QNM spectrum for different values of the compactness $\mathcal{C}$ close to the Buchdahl limit $\mathcal{C} = 4/9$. As a general result, already well known in the literature, we see that $\mathbb{Im}[\omega]$ increases with decreasing compactness. It is also interesting to note the distinction between {\it curvature} (empty circles) and {\it trapped} (filled circles)  modes~\cite{Kokkotas:1999bd}.  
Trapped modes exist only for ultracompact stars, i.e. with $\mathcal{C} > 1/3$, and  they correspond to spacetime perturbations that become trapped inside the potential.  There are finitely many trapped modes, although their number grows when compactness is increased because then the potential well becomes wider. 
The rest of the spectrum consists of the curvature modes.
 
QNMs describe, by definition, dissipative properties of spacetime metrics. Consequently, they play an important role in the ringdown phase, the last stage of a binary merger. During the ringdown phase, the compact object created in the final state of the collision relaxes to a stable state, and any distortion in its shape is dissipated in the form of GWs. The ringdown phase of ultracompact objects has a remarkable feature. As the final object possesses an unstable light ring, the primary signal of the ringdown phase results from the slow escape of radiation trapped on the unstable circular orbit. Its frequency and damping time-scale are associated to the orbital frequency and instability time-scale of circular null geodesics. This means that the primary signal in the ringdown phase does not depend on the presence or absence of an event horizon as long as the compact object has an unstable light ring. It is, therefore, possible to distinguish between two master cases:
 
\begin{itemize}

\item[{\it i)}] For BH the ingoing boundary condition at the horizon forces the ringdown waves to cross the event horizon and fall into the BH. This implies that for a BH the QNMs \emph{incidentally} describe also the ringdown phase.

\item[{\it ii)}] In the case of an ultracompact object, the ringdown signal consists in the primary light-ring ringdown modes (which are not QNMs but, as discussed before, radiation trapped at the unstable null geodesic) followed by the proper modes of vibration of the ultracompact objects, i.e. by its QNMs.

\end{itemize}

In the following, we shall elaborate more on the phenomenology of the ringdown phase considering the ultracompact exotic stars discussed in section~\ref{sec:stars}. 
Before proceeding, let us stress that we computed the QNMs by numerically solving 
Eq.~(\ref{eq:EigenvalueProblem}) together with the appropriate boundary data, as discussed before.
However, a better physical intuition can be gained by means of semi-classical methods, along the lines of~\cite{Volkel:2018hwb,Maselli:2017tfq,Volkel:2017kfj,Volkel:2017ofl,Kokkotas:1995av}.

\subsection{Gravitational echoes}\label{sec:GravEchoes}

In order to investigate the existence of gravitational echoes, 
we solve Eq.~\eqref{eq:PertEquation} numerically.\footnote{This means that we analyze gravitational echoes at the linear order. 
General arguments indicate that this approximation may not be appropriate for configurations that are almost as compact as BHs~\cite{Carballo-Rubio:2018jzw}. This is an interesting point that deserves further study.  However, we do not explore this issue any further since our analysis is not focused on BH mimickers.}
This task requires both initial data (at $t=0$) and boundary conditions (at the center of the star and spatial infinity) properly chosen.  We consider
\be\label{eq:In}
{\rm Initial~~data:}~~\left\{
\begin{array}{c}
   \Psi_{2,2}(0,r_*) = 0,    \\
        \\
   \frac{\partial \Psi_{2,2}}{\partial t}(0,r_*) = f(r_*) .
\end{array}
\right.~~~~~
{\rm BCs:}~~\left\{
\begin{array}{c}
   \Psi_{2,2}(t,r_*^c) = 0 ,    \\
      \\
\frac{\partial \Psi_{2,2}}{\partial t}(t,\infty) = - 
\frac{\Psi_{2,2}}{\partial t}(t,\infty)  .     
\end{array}
\right.
\ee
The first boundary condition imposes regularity at the center of the star (with $r_*^c \equiv r_*(r=0)$) while the second one corresponds to only outgoing waves at spatial infinity.\footnote{The potential vanishes at spatial infinity and so $\Psi(t,r)$ obeys the one-dimensional wave equation $\partial^2\Psi/\partial t^2 = \partial^2\Psi/\partial r^2$ solved by $\Psi(t,r) = F_1(r-t) + F_2(r + t)$, where $F_{1}$ ($F_2$) describe waves advancing toward (returning from) the boundary. Pure outgoing waves correspond to $F_2 = 0$, and $F_1$ trivially satisfying $\partial\Psi/\partial t = - \partial\Psi/\partial r$.}
Initial data encodes the physical conditions of a post-merger phase where a perturbed compact object is created in the final state of the collision. As a toy description, we follow the approach of~\cite{Cardoso:2016oxy} and use an initial gaussian pulse centered at $r_* = r_g$ and with spread $\sigma$,  i.e. $f(r_*) =  e^{-(r_* -r_g)^2/\sigma^2}$.
\begin{figure}[!htb!]
\begin{center}
$$\includegraphics[width=.48\textwidth]{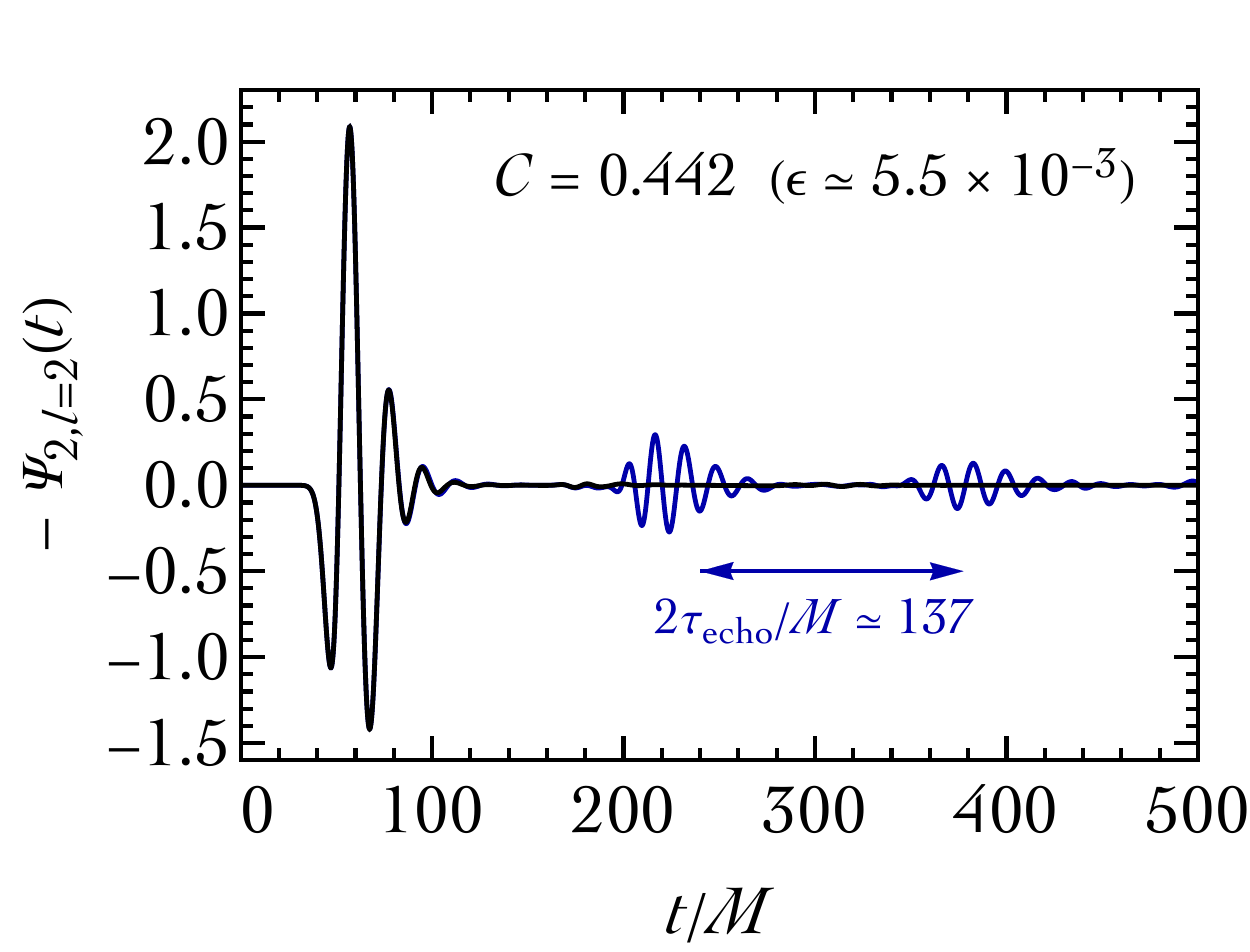}
\qquad\includegraphics[width=.48\textwidth]{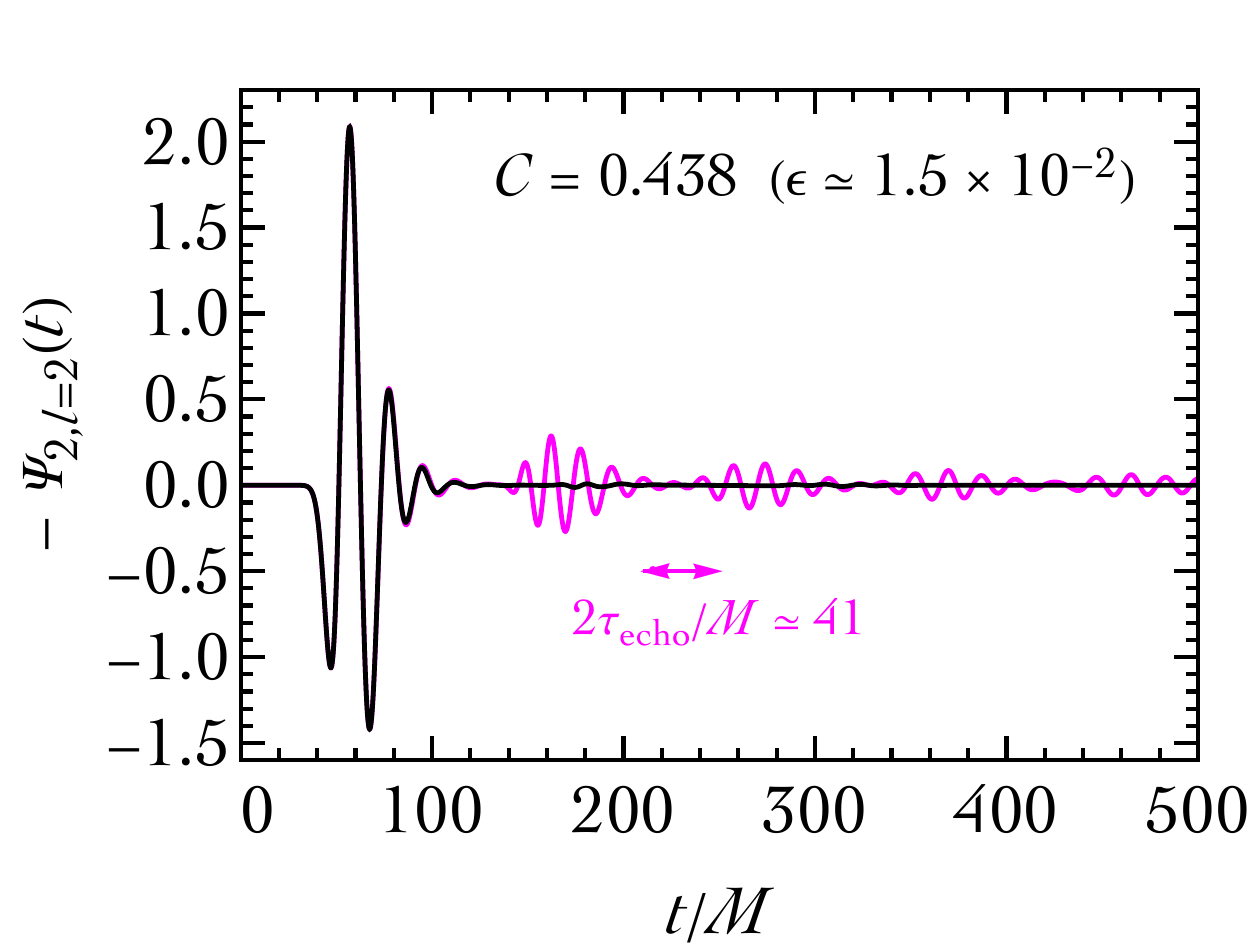}$$
\caption{\label{fig:Echoes}\em  Gravitational echoes for compact stars governed by the LinEoS for different (but uphysical) benchmark values of $\omega$.
}
\end{center}
\end{figure}

\begin{figure}[!htb!]
\begin{center}
$$\includegraphics[width=.48\textwidth]{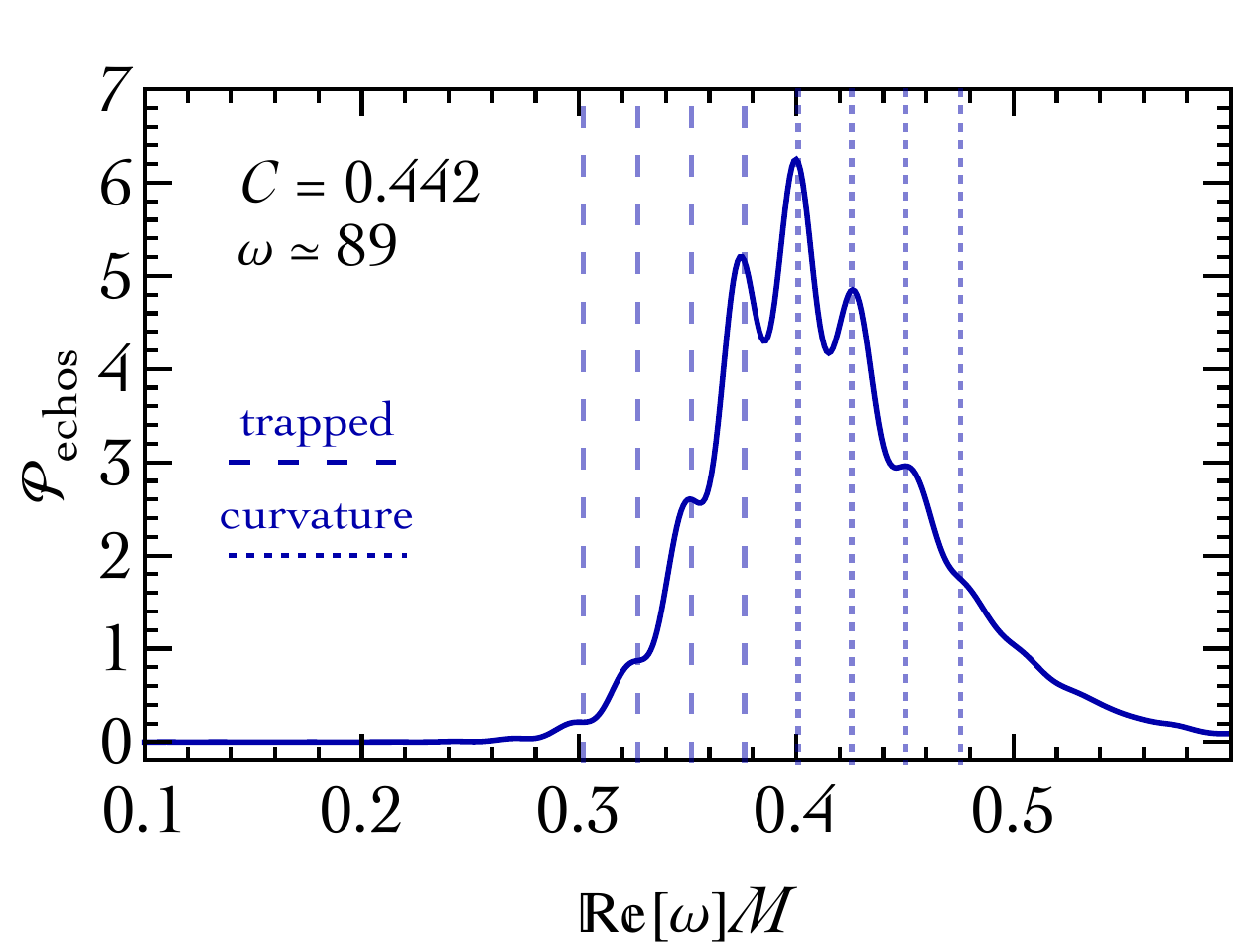}
\qquad\includegraphics[width=.48\textwidth]{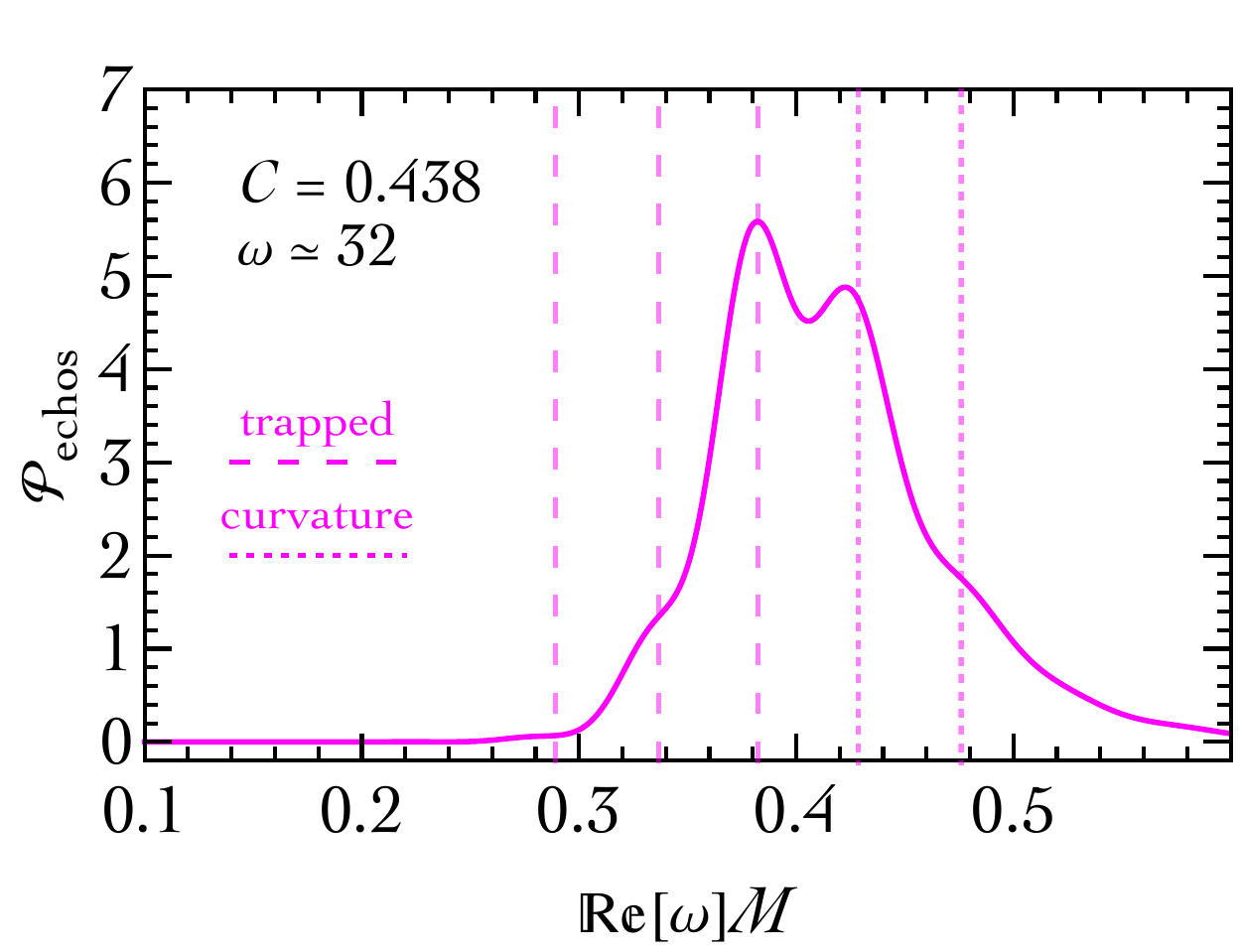}$$
\caption{\label{fig:PowerSpectrum}\em Power spectrum -- after subtraction of the primary ringdown signal -- for the gravitational
echoes displayed in Fig.~\ref{fig:Echoes}. The vertical lines correspond to 
$\mathbb{Re}[\omega]$ for the excited QNMs.
}
\end{center}
\end{figure}

Fig.~\ref{fig:Echoes} depicts the time evolution of the signal 
,$\Psi_{2,l=2}(t) \equiv \Psi_{2,l=2}(t,r_*\to \infty)$, for a highly compact LinEoS. The strain is characterized
by a primary ringdown signal, that coincides with the ringdown signal of a BH, followed by a subsequent train 
of echoes. The latter ring according to the QNMs of the ultracompact object. To better visualize this point, it is instructive to compute the Fourier transform of the waveform, and then subtract the primary BH signal. The resulting power spectrum is shown in Fig.~\ref{fig:PowerSpectrum} for the same benchmark values as used in Fig.~\ref{fig:Echoes}. The peaks in the power spectrum correspond, as numerically verified, to the real parts of the QNMs in Fig.~\ref{fig:QNM} excited by the initial pulse, shown by dashed horizontal lines. The widths of these curves match the imaginary parts of the corresponding QNMs.
 \begin{figure}[!htb!]
\begin{center}
$$\includegraphics[width=.48\textwidth]{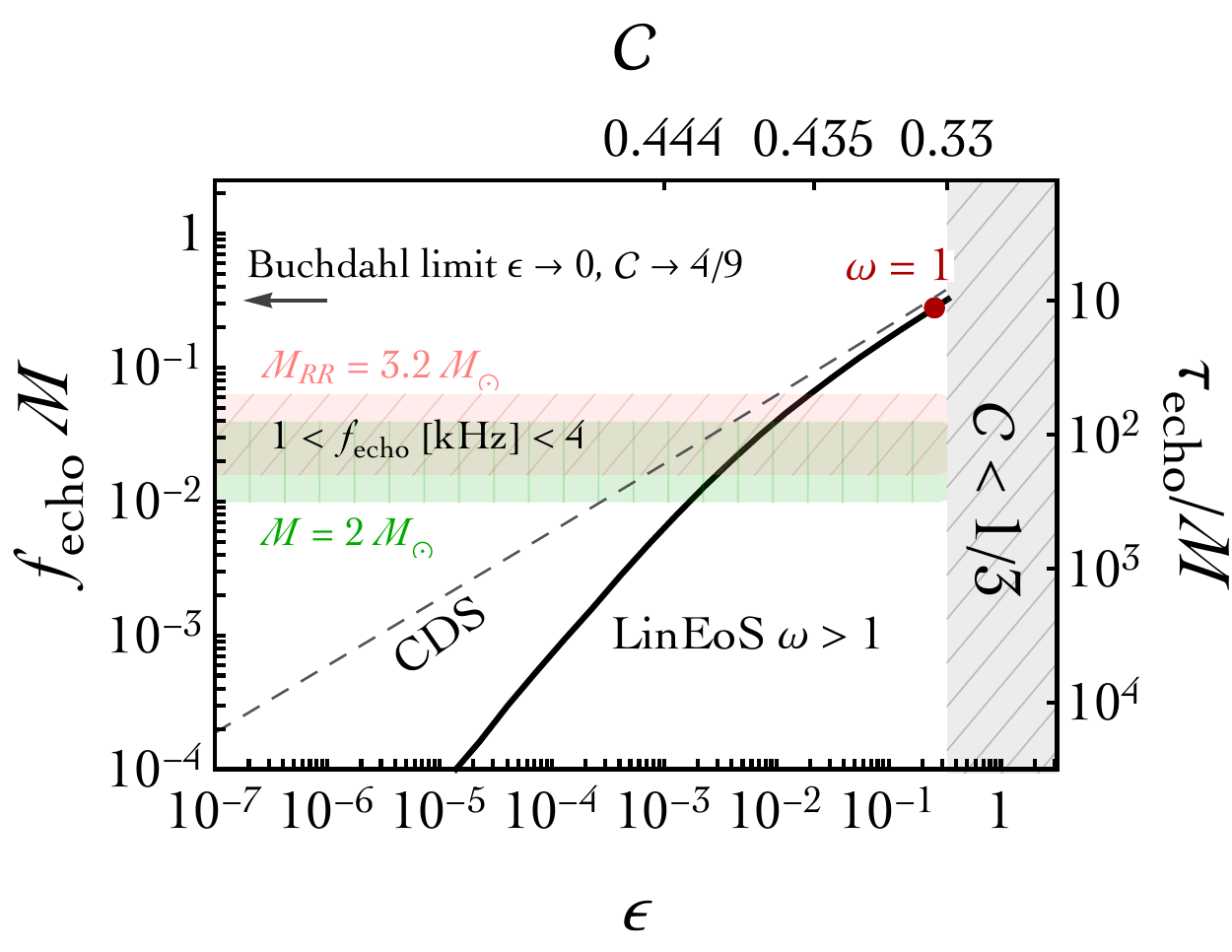}
\qquad\includegraphics[width=.48\textwidth]{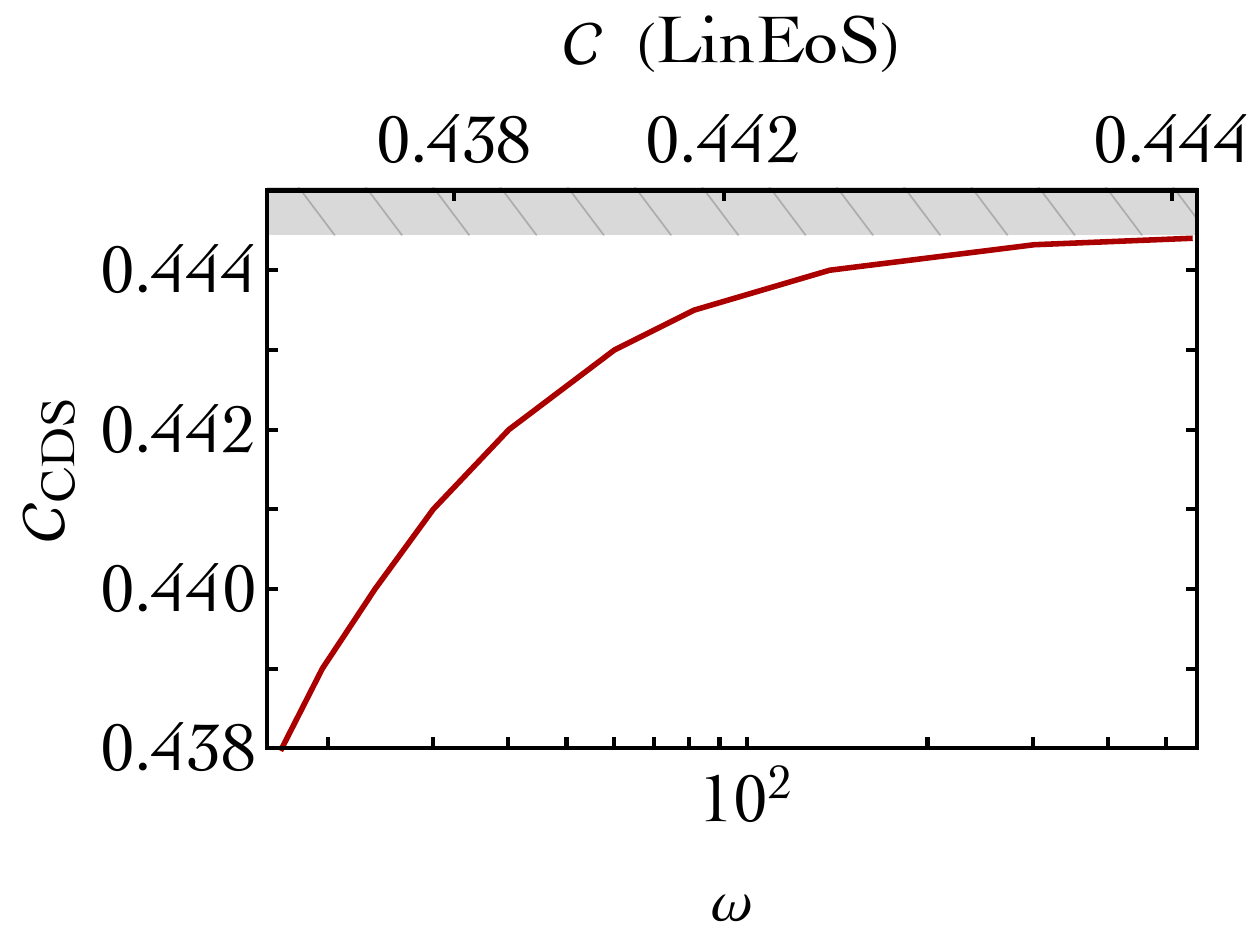}$$
\caption{\em \label{fig:fecho} 
Left panel. Echo frequency as a function of the parameter $\epsilon \equiv 1 -4/9\mathcal{C}$ for 
CDS (black dashed) and maximally compact stars with LinEoS (black solid). The horizontal shaded regions  
highlight the typical aLIGO/Virgo frequency range for the analysis of post-merger signals~\cite{Abbott:2017dke}.
Right panel. The red solid line identifies CDS and maximally compact LinEoS stars 
with the same echo frequency   $f_{\rm echo}M$.
 }
\end{center}
\end{figure} 

We are now in the position to comment about the detectability of GW echoes. The most relevant quantity is the echo timescale defined in Eq.~(\ref{eq:EchoFreq}). In the left panel of Fig.~\ref{fig:fecho} we show this quantity (together with its inverse, the echo frequency $f_{\rm echo} \equiv 2\pi/\tau_{\rm echo}$) as a function of the compactness -- or, equivalently, as a function of the parameter $\epsilon$ defined in Eq.~\eqref{eq:lin_eps_min} -- for the CDS and the LinEoS. As discussed in section~\ref{sec:CDS}, the CDS provides an upper bound for the echo frequency, see Eq.~\eqref{eq:TauBound}. This is consistent with the numerical analysis. The horizontal shaded region corresponds to an echo frequency in the range $f_{\rm echo} = [1,\,4]$ kHz for an ultracompact object with mass $M = 2\,M_{\odot}$ (green) and $M_{\rm RR} = 3.2\,M_{\odot}$ (pink), the latter being known as the Rhoades-Ruffini limit for the maximal mass of NSs~\cite{Rhoades:1974fn}. This frequency range was adopted in Ref.~\cite{Abbott:2017dke} as a region for looking for post-merger signals. We use it as a benchmark value for possible future detections. Fig.~\ref{fig:fecho} shows that, for these light objects, the detectability of GW echoes must rely on an unphysical EoS as the physical branch $\omega \leqslant 1$ cannot sustain visible signals. As already noticed in~\cite{Pani:2018flj} (see also~\cite{Mannarelli:2018pjb}), a compactness extremely close to the Buchdahl limit (at least $\epsilon \lesssim 10^{-2}$) is indeed needed. We remark that this conclusion depends on the value of $M$ chosen to compute the echo frequency $f_{\rm echo}$. Of course, for realistic values close to what we expect for ordinary NSs any speculations about the possible relevance of GW echoes from physically motivated ultracompact objects seems implausible, unless in the presence of a radically new state of matter incompatible with causality. On the other hand, one can still speculate about the existence of exotic -- yet physical -- ultracompact objects with mass $M > M_{\rm RR}$. In such a case, it would be possible to move the shaded region in Fig.~\ref{fig:fecho} towards physical values of the LinEoS. We shall return on this possibility at the end of section~\ref{sec:rotation}.

We close this section with a final comment about the QNMs for a star sustained by the LinEoS. 
In the left panel of Fig.~\ref{fig:fecho} lines with constant $\tau_{\rm echo}$ define 
two ultracompact star configurations -- the CDS and the LinEOS (the dashed and the solid black lines, respectively) -- characterized by the same echo timescale.
In the right panel of Fig.~\ref{fig:fecho} the red contour marks maximally compact LinEoS stars (labeled on the x-axes by the value of $\omega$) and the corresponding CDS (labeled on the y-axes by the value of their compactness $\mathcal{C}_{\rm CDS}$) with the same $\tau_{\rm echo}$. For these two configurations the effective radial potential in Eq.~(\ref{eq:PotConstantDensityStar}) is very similar (with differences in the star interior of order $\sim 1\%$ ($\sim 5\%$) for $\mathcal{C}_{\rm CDS} \simeq 0.444$  ($\mathcal{C}_{\rm CDS} \simeq 0.438$)). This is because the difference in pressure and energy density between the CDS and the LinEoS star is sizable only at small radial distance from the center, where the potential in Eq.~(\ref{eq:PotConstantDensityStar}) is overwhelmingly  dominated by the centrifugal barrier. Consequently one finds, with similar numerical accuracy, the same spectrum of QNMs -- thus, in general, ultracompact stars with the same echo timescale will tend to have a similar QNM spectrum.  In our particular case, this means that one can map the QNMs of a CDS computed in Fig.~\ref{fig:QNM} into those of a LinEoS star using the relation in Fig.~\ref{fig:fecho} with a relatively decent accuracy.
  
Spherically-symmetric static configurations are idealized objects. In reality, the unavoidable presence of rotation breaks spherical symmetry giving rise to a configuration that is axisymmetric.  For this reason we shall next turn our attention to the role of rotation.

\section{On the role of rotation} 
\label{sec:rotation}

It is important to assess the validity of our results including the effects of rotation. For slowly rotating stars this task can be accomplished by perturbing the static metric, as was pioneered in~\cite{Hartle:1967he,Hartle:1968si}. The resulting slow-rotation expansion is well motivated for old NSs. The fastest-spinning known millisecond pulsar, PSR J1748-2446ad~\cite{Hessels:2006ze}, spins with a rate of nearly 716 revolutions per second corresponding to a period of 1.396 ms. Assuming $M\simeq 2\,M_{\odot}$ and $R\simeq 13$ km, one finds (using the Newtonian approximation for the moment of inertia, $I = 2/5 MR^2$, and indicating with $J$ the total angular momentum of the star) a dimensionless spin parameter 
$\tilde{a} \equiv J/M^2 \simeq 0.35$, small enough to justify an expansion up to order $\tilde{a}^2$~\cite{Hartle:1967he,Hartle:1968si}.

The setup we are interested in, however, is different, and one cannot exclude the formation of a high-spinning remnant in the final state of the merger. For this reason, we opt for a general analysis valid for arbitrary angular velocities. The only limiting condition is that the angular velocity $\Omega$ of the rotating star must not exceed the Keplerian angular velocity $\Omega_K$ (a.k.a. the mass-shedding limit). The value of $\Omega_K$ is defined by the angular velocity of  a test particle in equilibrium  at the equatorial radius of the star, kept  bound  only by  the  balance  between  gravitational  and  centrifugal  force. In the static case the orbital angular velocity for a test particle on the surface $R_{J = 0}$ of a NS with mass $M_{J = 0}$ is $\Omega_{K}^{J = 0} = \sqrt{M_{J = 0}/R^3_{J = 0}}$.

To model the rotating star, we consider the stationary and axisymmetric spacetime in the so-called quasi-isotropic coordinates~\cite{Bardeen:1971eba,Gourgoulhon:2010ju}
\be\label{eq:QuasiIsotropic}
	\td s^2 = -N^2 \td t^2 + A^2(\td r^2 + r^2 \td \theta^2) + B^2 r^2 \sin^2\theta\left(\td \phi - \omega \td t\right)^2~,
\ee
where $N$, $A$, $B$ and $\omega$ are all functions of $(r,\theta)$.\footnote
{
The coordinate $r$ is not the same as the one used in Eq.~(\ref{eq:LineElement}). In the static, spherically symmetric limit the metric in Eq.~(\ref{eq:QuasiIsotropic}) reduces to 
\be\label{eq:LimitSpherical}
	\td s^2 
	= 	-N(r)^2 \td t^2 + A(r)^2\left[
		\td r^2 + r^2\left(\td \theta^2 + \sin^2\theta \td \phi^2\right)
	\right]~,
\ee
which is different compared to Eq.~(\ref{eq:LineElement}). In Eq.~(\ref{eq:LimitSpherical}) $r$ is the isotropic coordinate while in  Eq.~(\ref{eq:LineElement}) $r$ is the areal radius. The relation between the two radial coordinates is particularly simple in the Schwarzschild region outside the star where isotropic ($r_{\rm iso}$) and areal ($r_{\rm ar}$) radial coordinates are related by  
\be\label{eq:RadialCoo}
r_{\rm ar} = r_{\rm iso}\left(
1 + \frac{M}{2r_{\rm iso}}
\right)^2~.
\ee
The position of the light ring at $r_{\rm ar} = 3M$, therefore, corresponds to $r_{\rm iso}/M \simeq 1.866$ in isotropic coordinates.
}
Note that $\omega$ has dimension of an inverse time (frequency). It represents the dragging of inertial frames, that is the angular velocity of a zero angular momentum particle falling from infinity to a coordinate location $(r,\theta)$. In our analysis we assume that the star is rotating with an uniform angular velocity $\Omega$. For later use, it is important to remark that the form of the line element in Eq.~(\ref{eq:QuasiIsotropic}) follows from the assumption that {\it i)} spacetime is stationary and axisymmetric, 
 {\it ii)} the two corresponding Killing vectors commute and there is an isometry of the spacetime that simultaneously reverses $t \to -t$ and $\phi \to -\phi$,\footnote{Although reflection of time $t \to -t$ changes the direction of rotation, the metric is invariant under the joint reflection of $t$ and $\phi$.} 
  {\it iii)} spacetime is asymptotically flat and circular. 

For static NSs -- as for any other spherical, non-rotating, gravitating body -- the exterior spacetime is Schwarzschild. As a consequence, the light ring of ultracompact stars is always located at $r = 3M$. From the discussion in section~\ref{sec:intro}, this is not possible for ordinary static NSs as they are not able to support the required compactness. The presence of rotation substantially complicates the analysis since there is no equivalent to Birkhoff's theorem for axisymmetric rotating bodies, so the generic metric outside a rotating star is not the Kerr metric. It is thus necessary to compute the position of the light rings without relying on BH results.

In the equatorial plane $\theta = \pi/2$, the equations describing null geodesics take the form
\be\label{eq:PotentialRot}
	\dot{t} = \frac{E -\bar{\omega} L}{\bar{N}^2}~,~~~
	\dot{\phi} = \frac{L}{r^2 \bar{B}^2} + \frac{\bar{\omega}(E-\bar{\omega} L)}{\bar{N}^2}~,~~~
	\dot{r}^2 + \underbrace{\frac{1}{\bar{A}^2}\left[
	-\frac{(E-\bar{\omega} L)^2}{\bar{N}^2} + \frac{L^2}{r^2 \bar{B}^2}
	\right]}_{\equiv V_{\rm eff}(r)} = 0~,
\ee
where -- due to stationarity and axial symmetry -- there are two conserved quantities: the energy of the particle, $E$, measured at infinity and the angular momentum of the particle, $L$, evaluated with respect to the rotation axis. The dot represents the derivative w.r.t. an affine parameter $\tau$. In Eq.~(\ref{eq:PotentialRot}) we used the notation $\bar{N}(r) \equiv N(r,\theta = \pi/2)$ with similar definitions for the other metric functions. Circular null orbits (light rings) are defined by the conditions $\td r/\td \tau=0$, $\td^2r/\td \tau^2 = 0$ or, equivalently, by $V_{\rm eff} = 0$, $\td V_{\rm eff}/\td r = 0$. Explicitly, we have
\be\label{eq:Geodesics}
\bbox{
	\frac{(E-\bar{\omega} L)^2}{\bar{N}^2} = \frac{L^2}{r^2\bar{B}^2}~,~~~~~~
	\mathcal{I}(r) \equiv -\left(
	\frac{\epsilon}{\bar{N}\bar{B}}\frac{\partial \bar{\omega}}{\partial r}
	\right)r^2 + \left(
	\frac{1}{\bar{B}^3}\frac{\partial \bar{B}}{\partial r} -
	\frac{1}{\bar{N}\bar{B}^2}\frac{\partial \bar{N}}{\partial r}
	\right)r + \frac{1}{\bar{B}^2} = 0}
\ee
where $\epsilon = \pm 1$. The roots of $\mathcal{I}(r)$ define the positions of the light rings in the equatorial plane. Once the position $r_{\star}$ of a light ring is known, the first condition in Eq.~(\ref{eq:Geodesics}), evaluated at $r = r_{\star}$, gives the value of the impact parameter $D\equiv L/E$ for the corresponding circular null geodesic.

Let us start discussing the situation for ordinary NSs. We used the numerical code \verb"rns"~\cite{Stergioulas:1994ea,Nozawa:1998ak} to numerically solve the Einstein's equations for arbitrary angular velocity.
\begin{figure}[!htb!]
\begin{center}
$$\includegraphics[width=.45\textwidth]{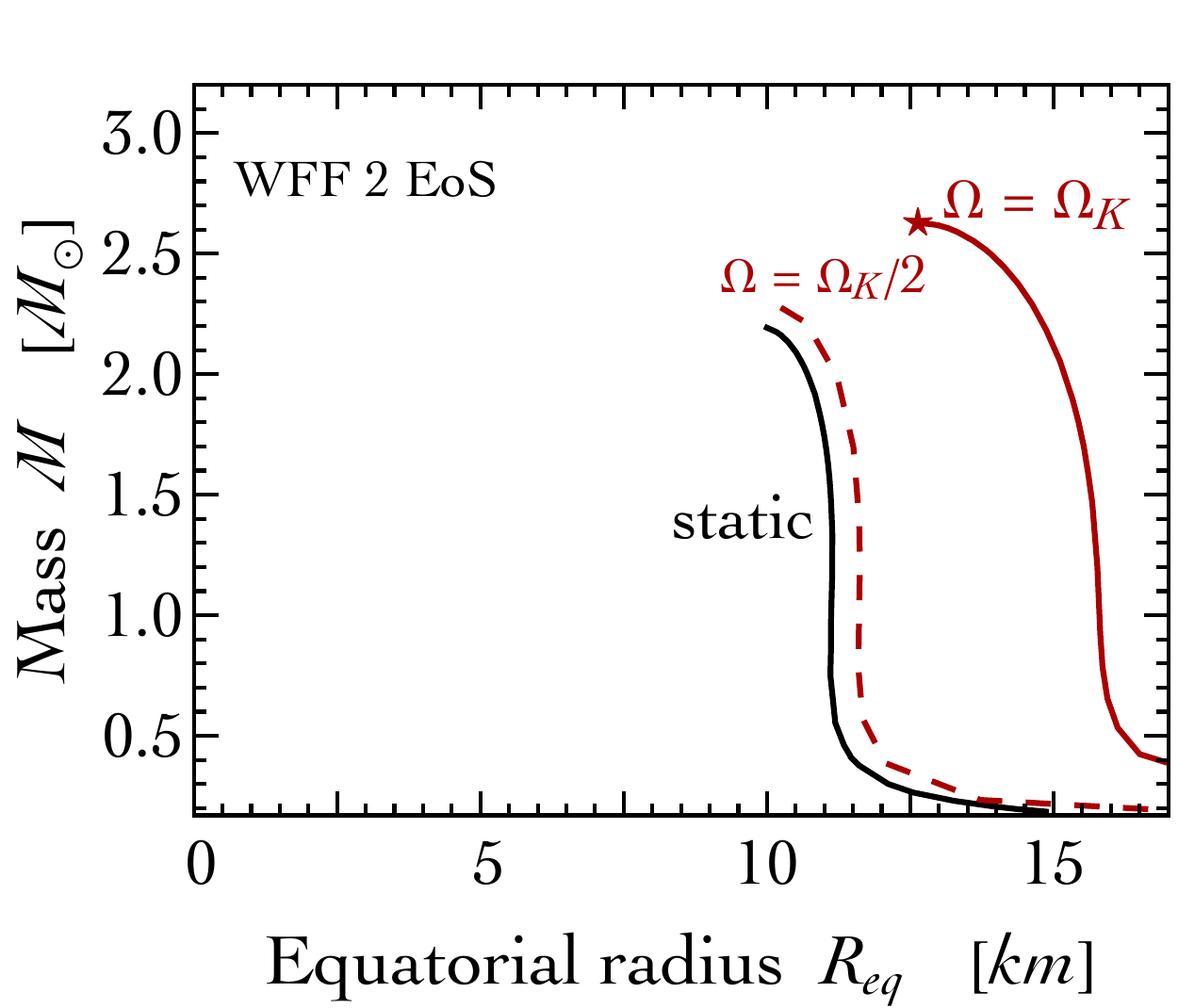}
\qquad\includegraphics[width=.465\textwidth]{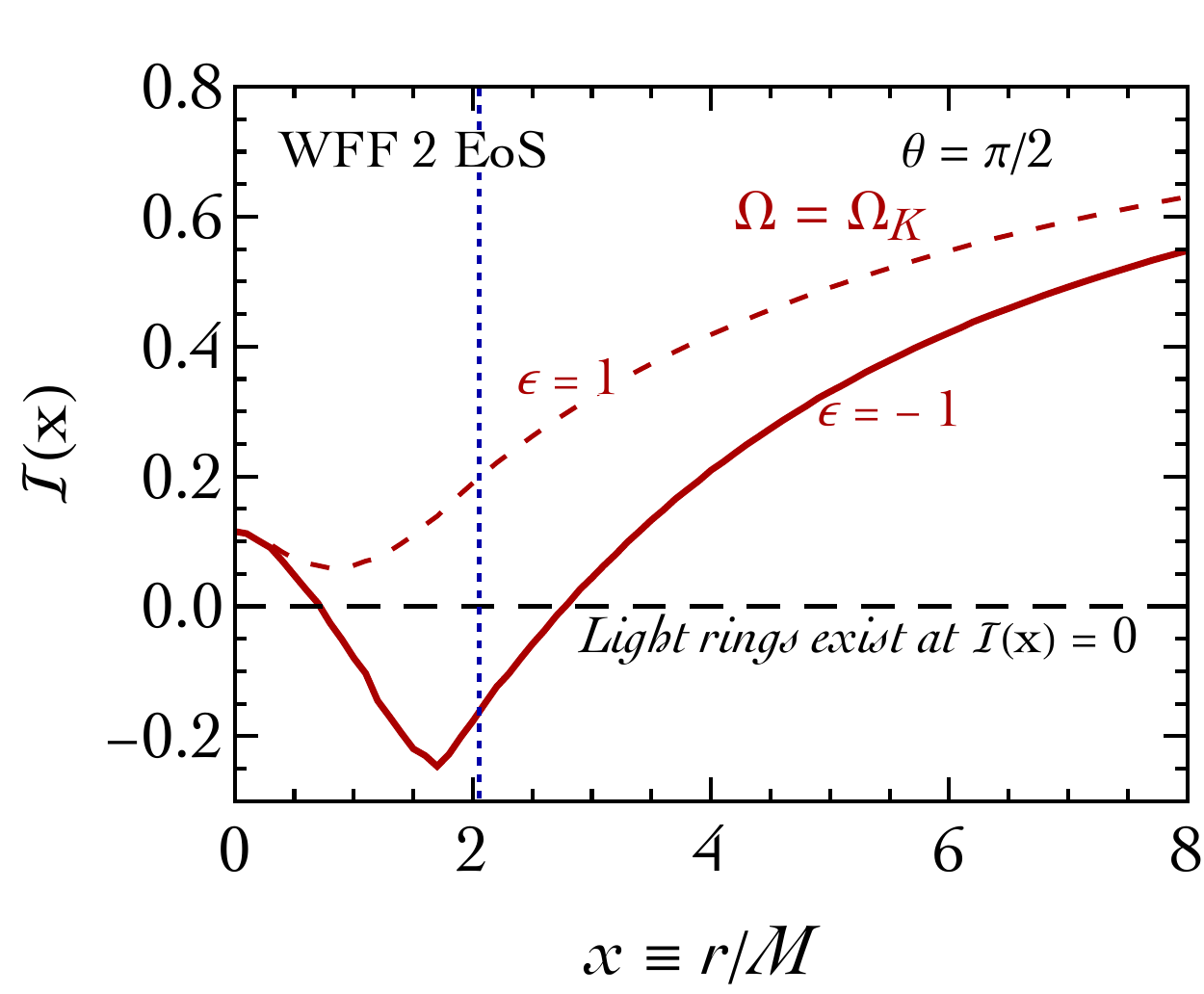}$$
\caption{\label{fig:RotationNS}\em  
Left panel. Impact of rotation on the M-R relation for NSs using the WWF\,2 EoS. In the presence of rotation, we identify the radius of the star with the equatorial circumferential radius. We show two different values for the angular velocity: the Keplerian limit $\Omega_K$ and $\Omega_K/2$. Right panel. Function $\mathcal{I}(r)$ in Eq.~(\ref{eq:Geodesics}) in the equatorial plane for the maximally compact NS with Keplerian velocity identified with a star in the left plot. The vertical dotted blue line marks the position of the surface of the star, $r = r_{\rm eq}$. For $\epsilon = -1$, the roots of $\mathcal{I}(r) = 0$ define the position of the light rings.
}
\end{center}
\end{figure}  
We cross-checked the numerical results using the Hartle-Thorne approximation in the limit of slow rotation (see appendix~\ref{app:HartleThorne}). For definiteness, we concentrate on one specific EoS among those that are allowed by the bounds in Fig.~\ref{fig:MassRadiusEOS}, and we consider the WWF\,2 EoS. The left panel of Fig.~\ref{fig:RotationNS} depicts the impact of rotation in the M-R plane. When including rotation, spherical symmetry is lost and the star becomes an oblate spheroid with a polar and an equatorial radius. We use the latter in the M-R plane.\footnote{The equator of the star is the closed line defined by $t= const$ and $\theta = \pi/2$  at the position of the surface of the star where $P=0$. It has a constant value of the coordinate $r$ introduced by the metric in Eq.~(\ref{eq:QuasiIsotropic}), $r = r_{\rm eq}$. However, this is not a good definition for the equatorial radius of the star because it is a coordinate-dependent quantity. A coordinate-independent characterization of the stellar equator is the equatorial circumferential radius $R_{\rm eq} \equiv \mathrm{C}/2\pi$, where $\mathrm{C}$ is the circumference of the star in the equatorial plane. From the metric~\eqref{eq:QuasiIsotropic} in quasi-isotropic coordinates we obtain the operative definition 
\be \label{eq:Req}
	R_{\rm eq} = \frac{1}{2\pi}\oint_{r = r_{\rm eq},\,\theta = \pi/2} \td s 
	= \frac{1}{2\pi}\oint_{r = r_{\rm eq},\,\theta = \pi/2}  \sqrt{
	B^2 r^2 \sin^2\theta \td \phi^2} = B(r_{\rm eq},\pi/2)r_{\rm eq}~.
\ee
As the definition of the mass $M$ in the rotating case we use the gravitational mass defined as the common value of the Komar and ADM mass~\cite{Gourgoulhon:2010ju}.
 }
 \begin{figure}[!htb!]
\begin{center}
$$\includegraphics[width=.45\textwidth]{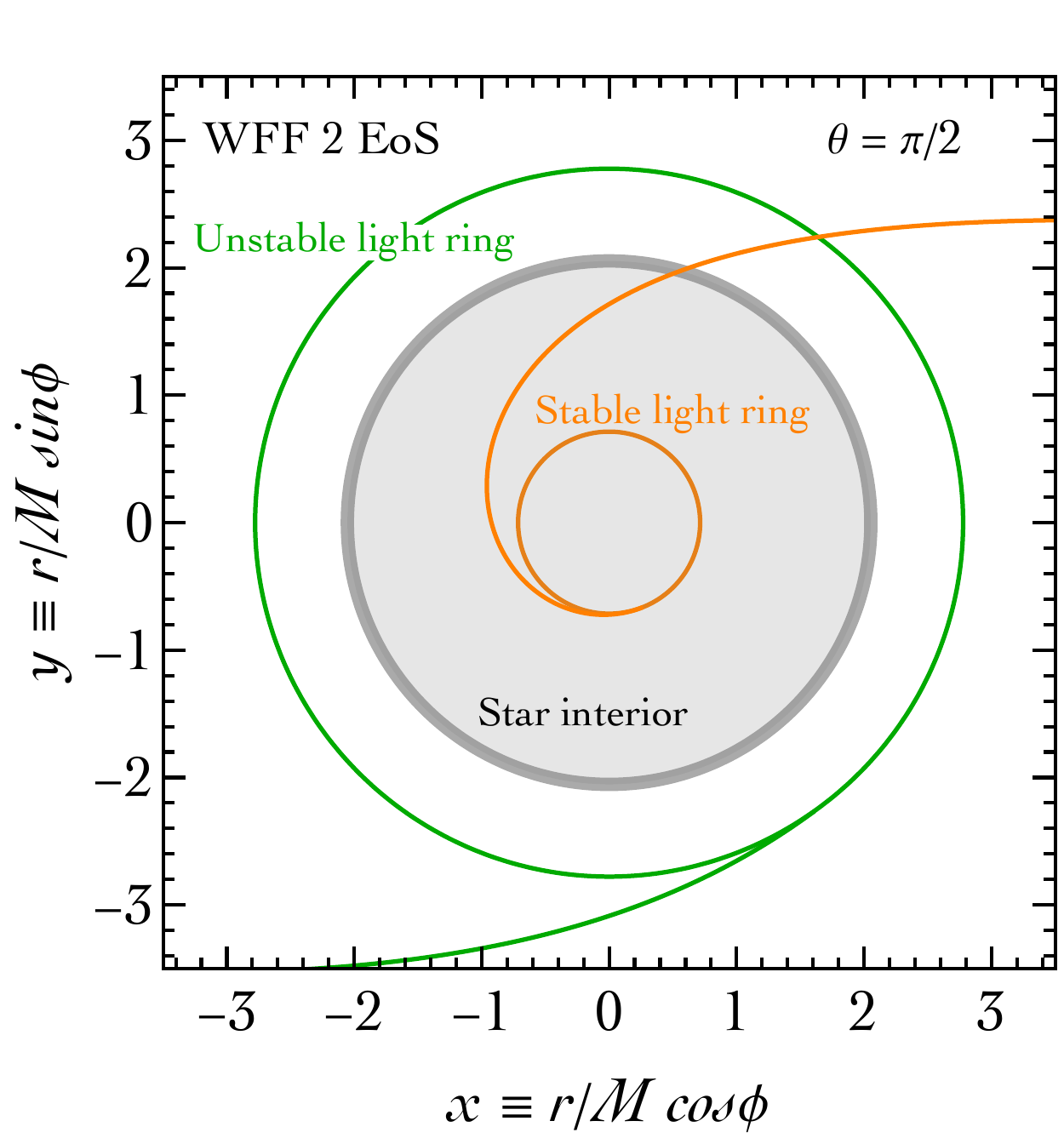}
\qquad\includegraphics[width=.45\textwidth]{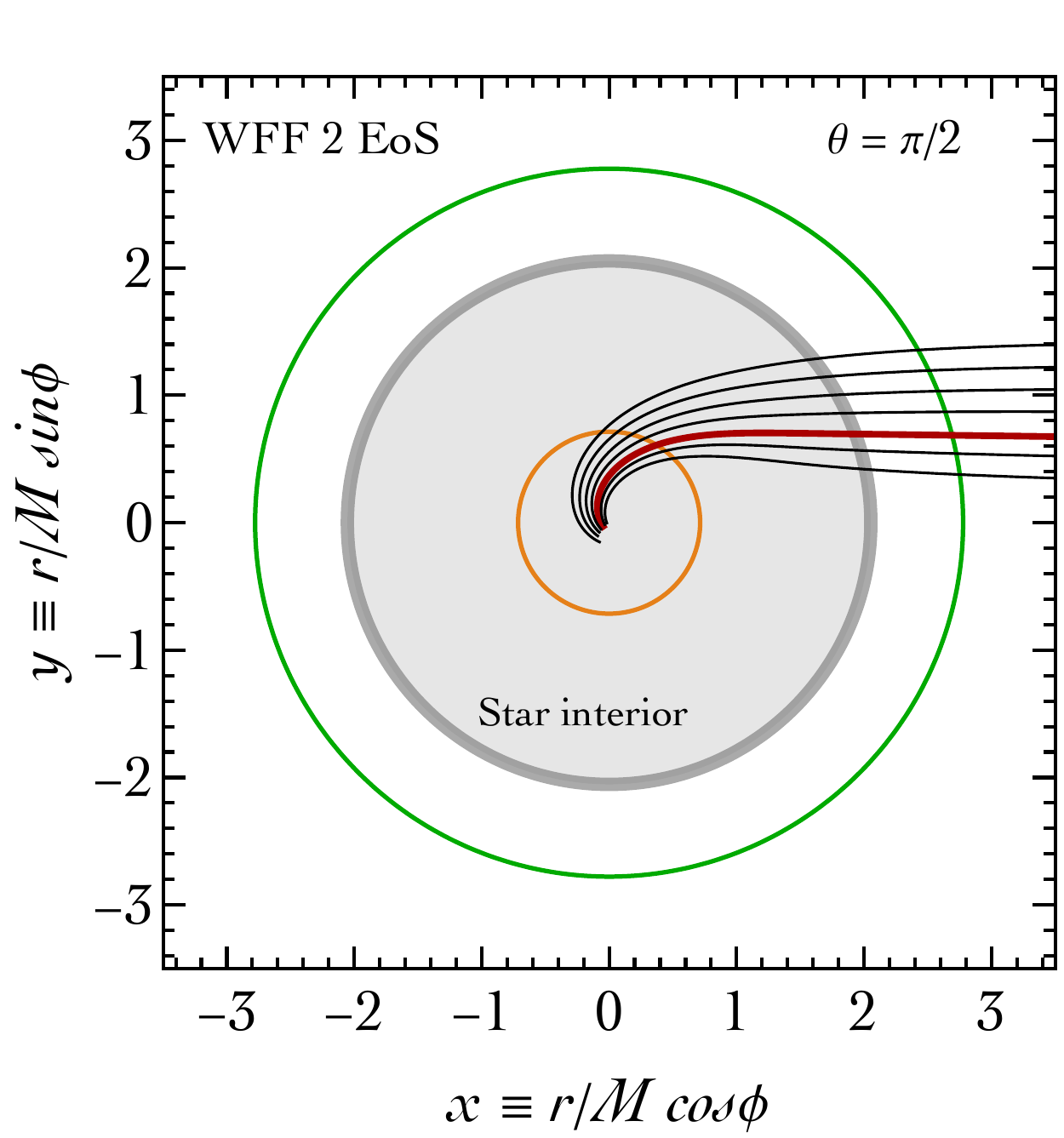}$$
\caption{\em \label{fig:geodesics} 
Left panel. Light rings for the maximally compact NS analyzed in the right panel of Fig.~\ref{fig:RotationNS} (see caption for details).
Right panel. Null geodesics pointing towards the center of the star for different values of the impact parameter.
We mark in red the principal null geodesic, i.e. the trajectory with impact parameter equal to the spin of the star.
 }
\end{center}
\end{figure} 
To illustrate the impact of rotation we consider two NS, one rotating at the Keplerian limit $\Omega = \Omega_K$ and one with an intermediate angular velocity $\Omega = \Omega_K/2$. We recover the known result according to which rotation leads, in the Keplerian limit, to a 20\% (25\%) increase of the mass (radius) of the star. Na\"{\i}vely, this means that rotation does not significantly increase the compactness of the NS (defined now as $\mathcal{C} = M/R_{\rm eq}$) compared to the static case, and one finds $\mathcal{C} < 1/3$ in all realistic situations. However, as anticipated before, to investigate the possible existence of a photon sphere we need to solve Eq.~\eqref{eq:Geodesics} to estimate the positions of null geodesics. 

In the right panel of Fig.~\ref{fig:RotationNS} we show the behavior of $\mathcal{I}(r)$ as a function of the radial coordinate $r$. We consider the Keplerian limit of the most compact static configuration (corresponding to the red star in the left panel of Fig.~\ref{fig:RotationNS}). The choice  $\epsilon = +1$ always corresponds to strictly positive values of $\mathcal{I}(r)$ (dashed line in the right panel of Fig.~\ref{fig:RotationNS}). In the case $\epsilon = -1$, on the contrary, the function $\mathcal{I}(r)$ has the possibility to vanish. This is indeed the case in Fig.~\ref{fig:RotationNS} where there are two roots corresponding to two distinct light rings. Furthermore, from Fig.~\ref{fig:RotationNS} it is evident that the ring located inside the star is stable while the outer one is unstable. The surface of the star in the equatorial plane, $r = r_{\rm eq}$, is marked with a vertical dotted blue line. We note that the existence of this light rings structure follows from a general theorem recently proved in~\cite{Cunha:2017qtt}. The theorem states that every axisymmetric, stationary, ultracompact solution to the Einstein field equations that was formed by gravitational collapse of matter obeying the null energy condition must possess at least two light rings, one of them being stable.
 
In the left panel Fig.~\ref{fig:geodesics} we show the two light rings in the equatorial plane of the NS expressed in terms of  polar coordinates.  As customary, to visualize the light rings we solved the geodetic equation using the critical value of the impact parameters computed from the first condition in Eq.~\eqref{eq:Geodesics}. As an important result we find that, \emph{even if a static star is not sufficiently compact to possess a light ring it still has the possibility to feature an unstable external light ring once rotation is taken into account.} This applies, for instance, to the specific NS we are considering in the present example. Consequently, it is not unreasonable, to speculate about the possible existence of gravitational echoes. 

The frequency of these echoes is estimated as the coordinate time needed to travel from the external light ring to the center of the star and back along a null geodesic, as was done in section~\ref{sec:GravEchoes}. A representative sample of these trajectories, each one characterized by a different impact parameter, is displayed in the right panel of Fig.~\ref{fig:geodesics}. Among them, we compute $f_{\rm echo}$ along the so-called principal null geodesic (the one marked in red in Fig.~\ref{fig:geodesics}). In analogy with Kerr BHs, they are the geodesics whose impact parameter is equal to the angular momentum of the star, $D/M = \tilde{a}$. For a NS with mass $M = 2.5\,M_{\odot}$ we find $f_{\rm echo} \simeq 16$ kHz. This is clearly too high to have phenomenological relevance.

 \begin{figure}[!htb!]
\begin{center}
$$\includegraphics[width=.465\textwidth]{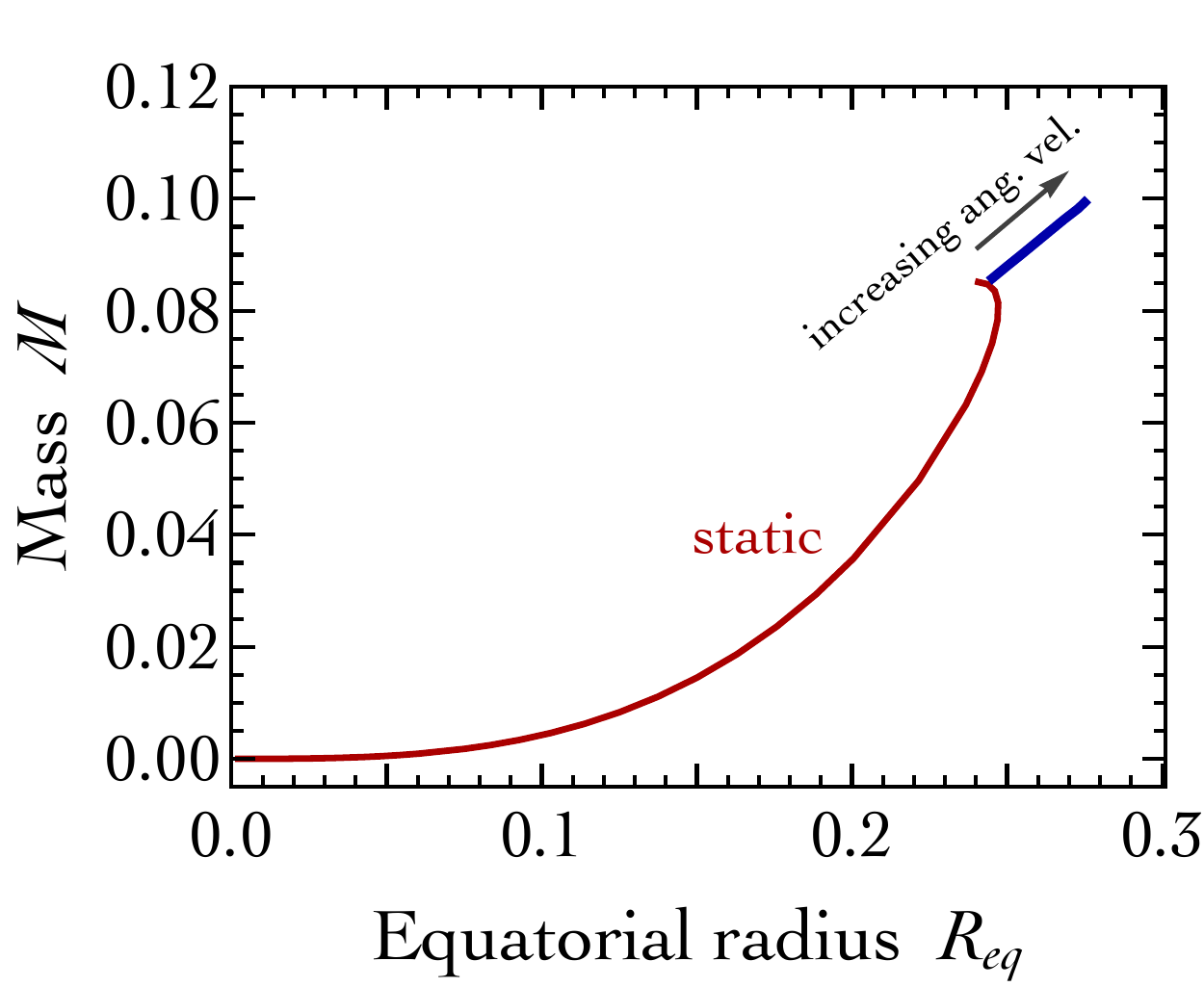}
\qquad\includegraphics[width=.45\textwidth]{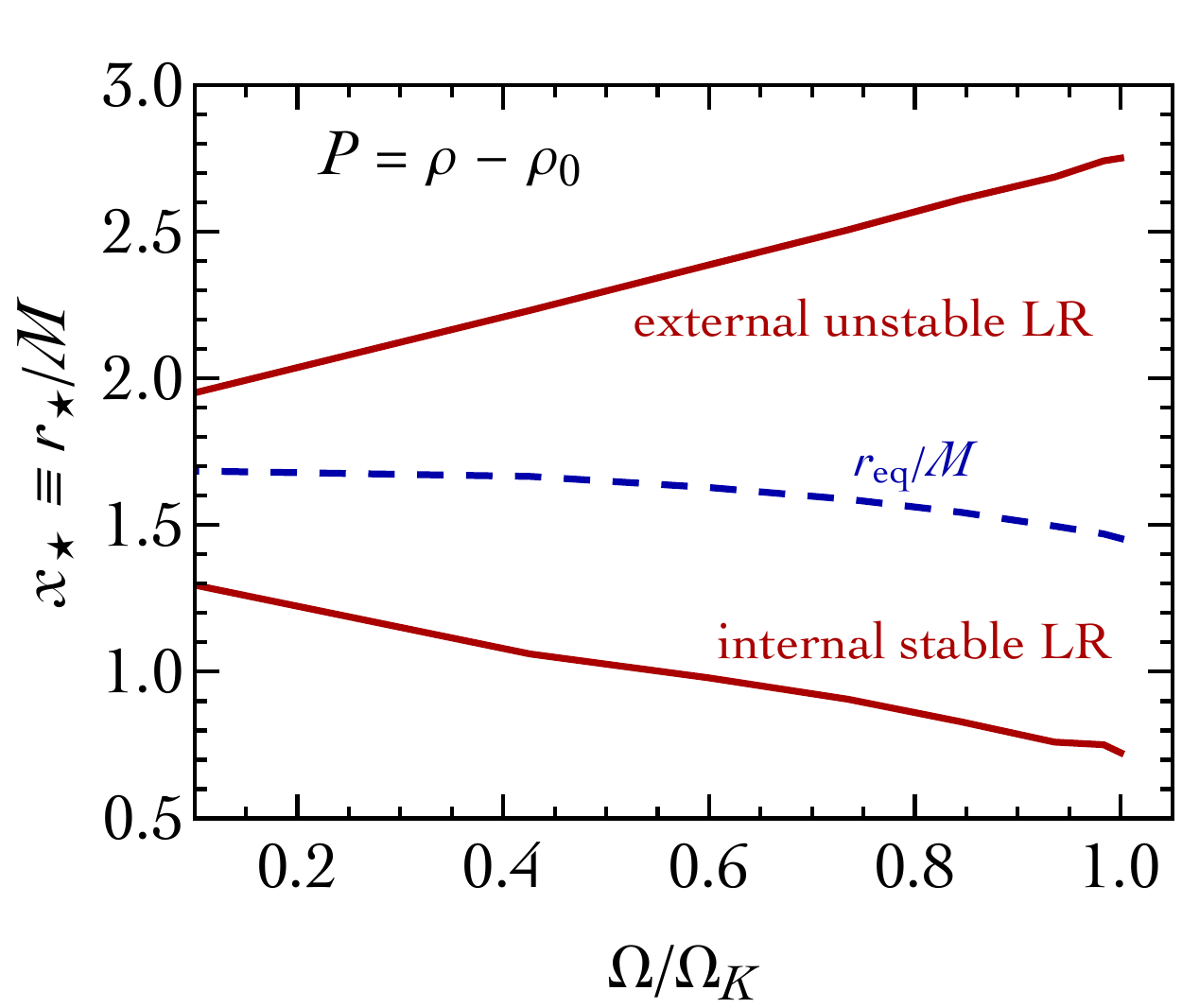}$$
\caption{\label{fig:LightRingsNS}\em 
Left panel. We show in blue the impact of non-zero angular velocity on the M-R relation for the maximally compact star with LinEoS.
In the presence of rotation, the radius of the star is defined by the equatorial circumferential radius (see Eq.~(\ref{eq:Req})).
Right panel. Position of the internal stable and external unstable light rings for the maximally compact star with LinEoS and $\omega =1$, as a function of the 
angular velocity. The blue dashed line indicates the position of the surface of the star at radial distance $r=r_{\rm eq}$.
}
\end{center}
\end{figure}
\begin{figure}[!htb!]
\begin{center}
$$\includegraphics[width=.45\textwidth]{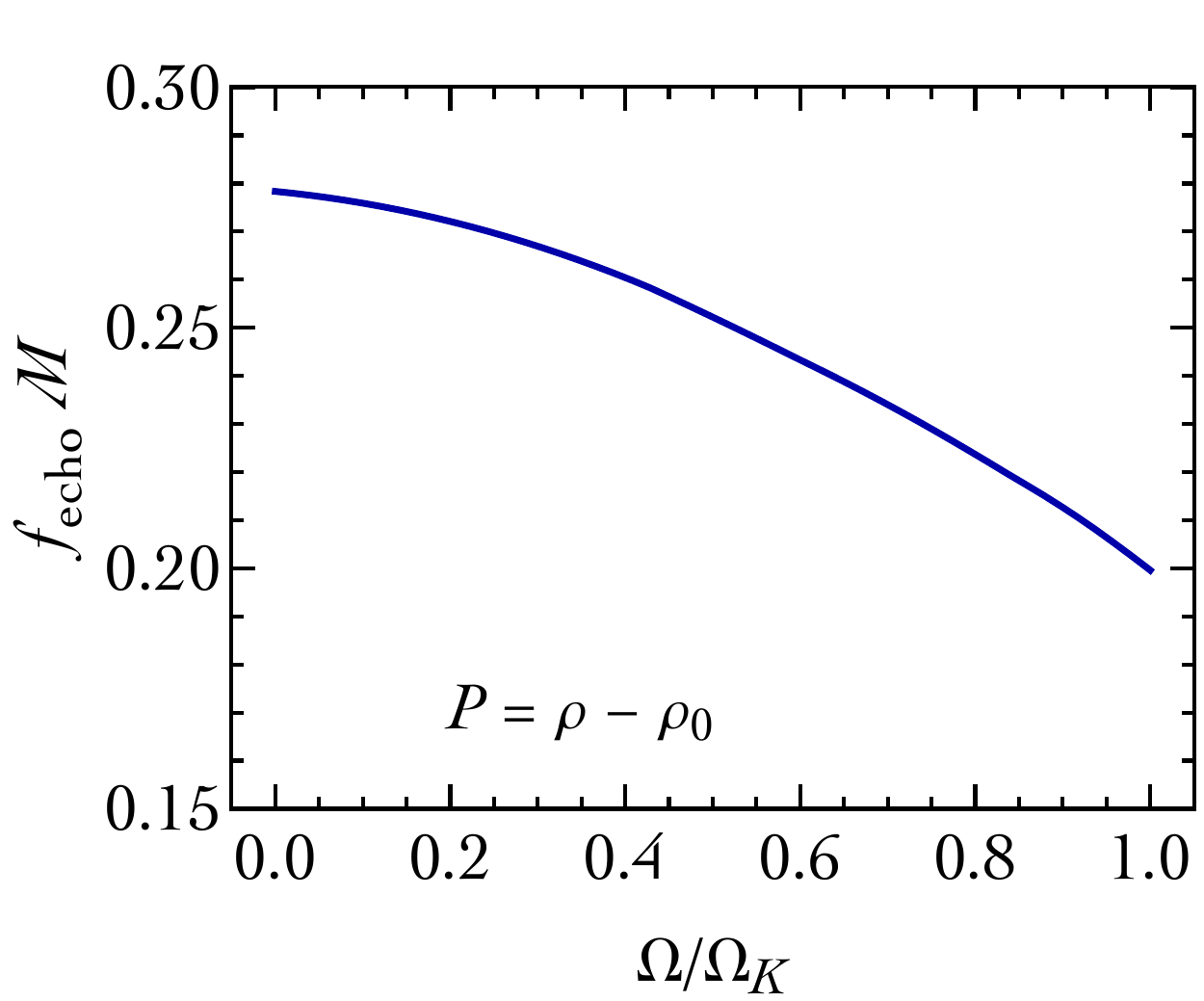}
\qquad\includegraphics[width=.45\textwidth]{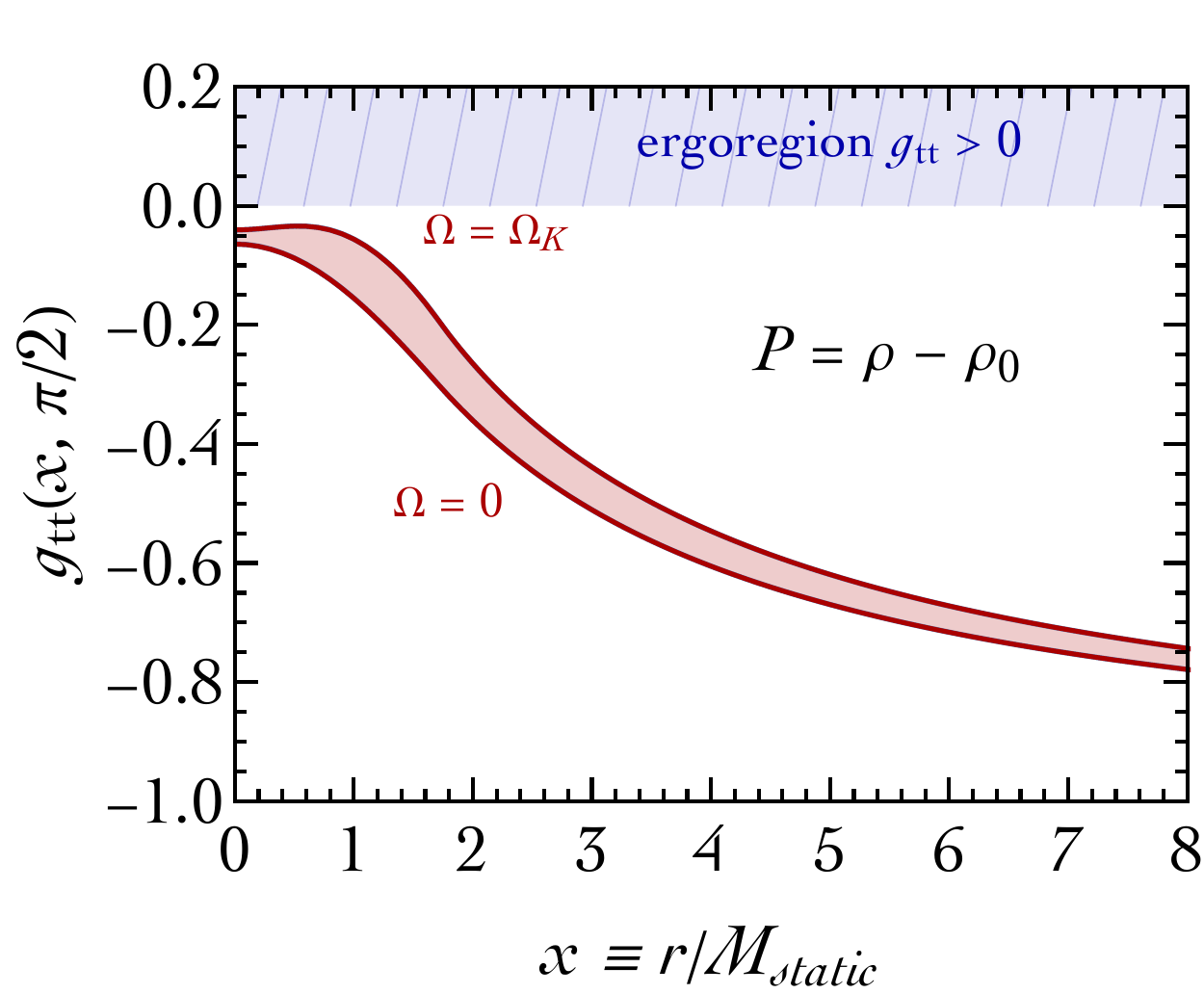}$$
\caption{\label{fig:RotEchoes}\em 
Left panel. Echo frequency for the maximally compact star with LinEoS as a function of the angular velocity (from the static configuration to the Keplerian limit).
Right panel. Metric function $g_{tt}$ in the equatorial plane for different angular velocity, from the static configuration to the Keplerian limit. Since $g_{tt} < 0$ the star never develops an ergoregion.
}
\end{center}
\end{figure}

 
A more interesting case of ultracompact exotic stars is the LinEoS that saturates the causality bound, $P = \rho-\rho_0$. The impact of rotation on their M-R relation is illustrated in the left panel of Fig.~\ref{fig:LightRingsNS}. The blue line represents the maximally compact configurations with the angular velocity varying from 0 to the mass-shedding limit. The right panel of Fig.~\ref{fig:LightRingsNS} depicts the position of the two light rings for each of the rotating solutions. 
In agreement with the theorem stated in Ref.~\cite{Cunha:2017qtt}, we find an unstable external light ring and a stable internal light ring. The left panel of Fig.~\ref{fig:RotEchoes} displays the echo frequency of the maximally compact LinEoS. It was estimated as the inverse of the coordinate time needed to travel along the principal null geodesics from the external light ring to the center of the star and back. 
The echo frequency decreases with the spin at most by $\sim 30\%$ reaching its minimum at the Keplerian limit. These results are in agreement with the findings of Ref.~\cite{Pani:2018flj} who considered CDS in the slow-rotating approximation.

BH mimickers, that are usually considered as prototypes of compact objects capable of producing GW echoes, are often plagued by the ergoregion instability~\cite{Cardoso:2014sna}.
This means that, unless the ergoregion instability operates on timescales much longer than a Hubble time, BH mimickers  are not viable candidates for GW echoes. The reason is that the GW emission associated with the ergoregion instability would produce a strong stochastic background that would already be in tension with the first observation run of aLIGO~\cite{Barausse:2018vdb}. A possible solution to this issue, as discussed in~\cite{Maggio:2017ivp}, is to  construct the BH mimicker by replacing the horizon with a reflective surface that has a non-zero absorption coefficient. A $\sim 0.5\%$ absorption rate at the surface seems to be enough to quench the instability completely. However, one should not be misled by the smallness of this number: an absorption rate at the percent level is indeed several orders of magnitude larger than the one for the NSs. It is unclear what mechanism could cause such a big effect.
 
In the following we will show that, despite being ultracompact, LinEoS satisfying the causality bound do not possess an ergoregion 
and, therefore, do not suffer from the ergoregion instability.

\subsection{Ultracompact objects {\it without} an ergoregion instability}

The ergoregion instability is turned on only after the star has developed an ergosphere~\cite{Kokkotas:2002sf}. The ergosphere is defined as the surface enclosing the ergoregion -- the region where the Killing vector $k^{\mu} = (1,0,0,0)$ becomes space-like, $k^{\mu}k^{\nu}g_{\mu\nu} = g_{tt} > 0$. A striking consequence of this is that inside the ergoregion a stationary observer 
is forced to co-rotate with the spinning compact object. Moreover, physical states with negative energy can exist within the ergoregion. Consequently, an instability  occurs because, roughly speaking, it is energetically favorable to cascade down toward even more negative energy states. 

From the line element in Eq.~\eqref{eq:QuasiIsotropic} we obtain that
\be
	g_{tt}(r,\theta) 
	= -N(r,\theta)^2 + r^2 B(r,\theta)^2\omega(r,\theta)^2\sin^2\theta 
     \, \leqslant g_{tt}(r,\pi/2) \equiv \bar{g}_{tt}
	~.
\ee
In the right panel of Fig.~\ref{fig:RotEchoes} we show the metric function $\bar{g}_{tt}$ of the maximally compact star with LinEoS saturating the causality bound for different values of the angular velocity, from the static configuration to the Keplerian limit. As $\bar{g}_{tt} < 0$ it is seen that an ergoregion never forms, not even for extreme values of angular velocity. We confirmed numerically that $g_{tt} < 0$ remains true also outside the equatorial plane.

In conclusion, we have explicitly constructed an example of a physically viable (i.e. with an EoS obeying physical assumptions) ultracompact object that 
is not affected by any ergoregion instability.

\section{Discussion and Conclusions} 
\label{sec:conclusions}

Are gravitational echoes phenomenologically relevant?  Our analysis 
suggests that a neat echo signal as the one expected from 
BH mimickers belongs to the realm of unphysical ultracompact objects, like CDS or stars with LinEoS strongly violating ($\omega \gg 1$) the causality bound. This is, in particular, illustrated by the left panel of Fig.~\ref{fig:fecho}. 

However, in light of the result presented in section~\ref{sec:rotation}, it is still intriguing to ask whether physically motivated exotic ultracompact objects supported by the maximally stiff LinEoS may leave some imprint in GW interferometers. Let us elaborate more on this point, bearing in mind that we shall keep the following discussion at a more speculative level.

\subsection{Are physically motivated gravitational echoes detectable?}

In~\cite{Abbott:2017dke}, the LIGO scientific collaboration and the Virgo collaboration searched for GWs associated to the remnant of the binary NS merger GW170817 over a frequency range $f = [1024,\,4096]$ Hz. This analysis shows that this frequency band is a good target for the detection of a post-merger signal.
 \begin{figure}[!htb!]
\begin{center}
\includegraphics[width=.55\textwidth]{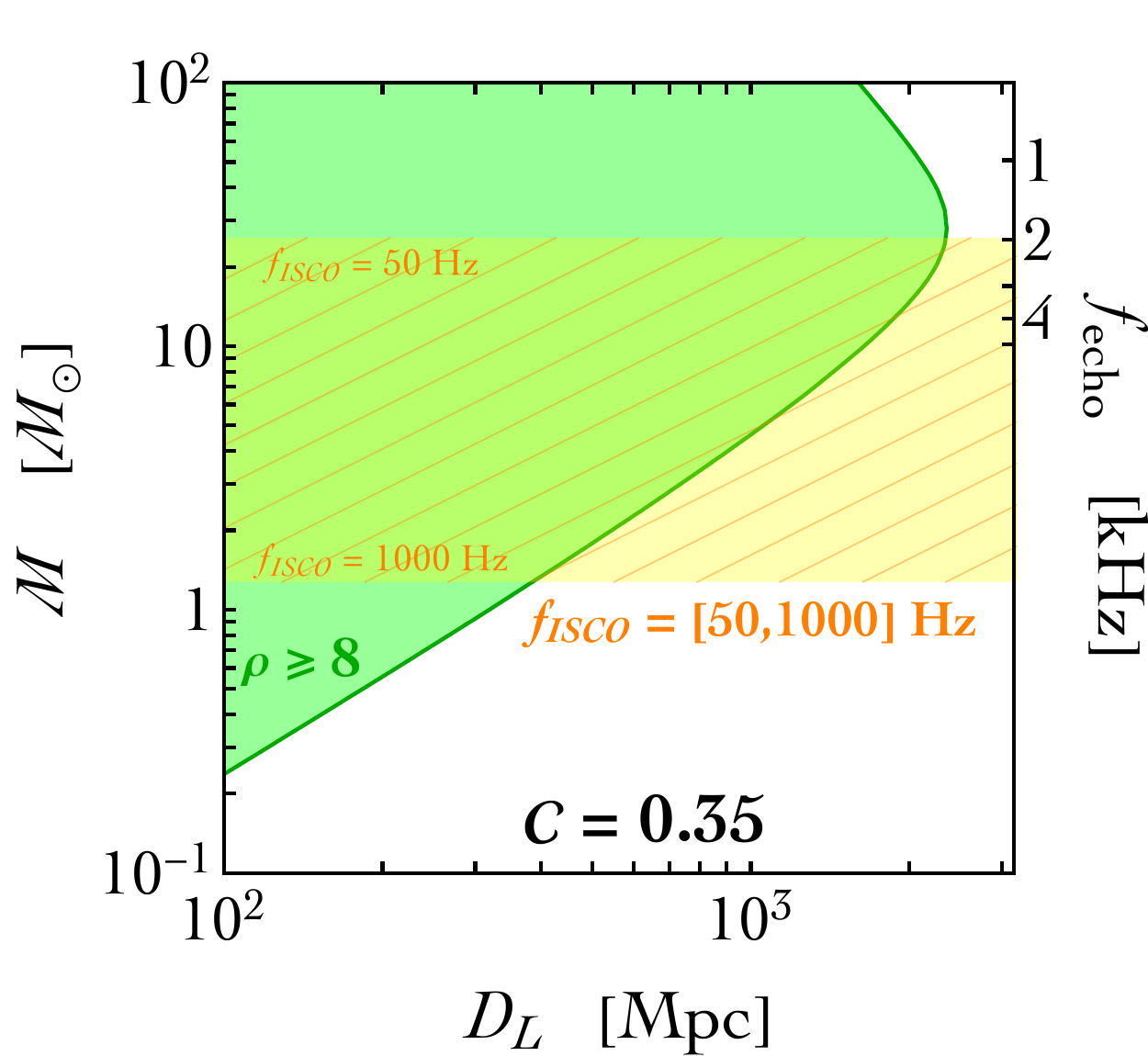}
\caption{\em \label{fig:LIGO} 
The aLIGO best sensitivity range for the merger of two exotic stars supported by the LinEoS 
as a function of their equal mass $M$ (left y-axis)
 and the luminosity distance of the event (x-axis).
We consider objects with maximal compactness $\mathcal{C} = 0.35$. 
The green regions corresponds to a signal-to-noise ratio above the nominal threshold $\rho \geqslant 8$.
The  yellow  band  corresponds  to  the  GW  frequency  range $f_{\rm ISCO}= [50,\,1000]$ Hz.
On the right y-axis we show the typical echo frequency assuming the formation of a maximally compact exotic object 
with LinEoS and mass $M$ in the final state. Frequencies in the interval $f = [1024,\,4096]$ Hz are typically used 
in the search for post-merger signals.
}
\end{center}
\end{figure} 
We are, therefore, in the position to speculate about potential detection prospects. The scenario we have in mind is the following: Consider the merger of two ultracompact exotic objects with  LinEoS $P = \rho -\rho_0$ discussed in this paper. In the initial phase of the merger, at large separation distance, the two bodies are effectively point-like, and evolve according to Kepler's third law. GWs  are emitted with characteristic frequency that is twice  the  orbital  frequency of the inspiral motion. As the two objects get closer and closer, their orbital distance decreases and their speed increases. The typical frequency that marks the end of the inspiral phase is approximately $f_{\rm ISCO} = \mathcal{C}^{3/2}/3^{3/2}\pi M_{\rm tot}$, where $M_{\rm tot}$ is the total mass of the system, and we assume the same  compactness $\mathcal{C}$ for the two merging bodies~\cite{Giudice:2016zpa}. The merger and the ringdown phase are observable if the characteristic frequency $f_{\rm ISCO}$ falls within the aLIGO sensitivity range, namely $f_{\rm ISCO} = [50,\,1000]$ Hz. Furthermore, a detection is possible only if the signal-to-noise ratio $\rho$ is larger than a certain threshold, usually fixed to be $\rho \geqslant 8$. For a GW with strain $h(t)$, and Fourier transform $\tilde{h}(f)$, the signal-to-noise ratio can e estimated by the following integral in the frequency domain
 \be
 \rho^2 = 4\int_0^{\infty}\frac{\left|\tilde{h}(f)\right|^2}{S_n(f)}df~,~~~~~~~~~~~
 \tilde{h}(f) \approx \frac{\sqrt{5/24}}{\pi^{2/3}D_L}M_C^{5/6}f^{-7/6}~,
 \ee 
 where $S_n(f)$ is the noise power spectral density~\cite{Ajith:2011ec}, $M_C$ the chirp mass, $D_L$ the luminosity distance, and where we  used the quadrupole approximation truncated at the Newtonian order for $\tilde{h}(f)$. To give an idea, in Fig.~\ref{fig:LIGO} we consider fixed compactness $\mathcal{C} = 0.35$, and we show in green the region of the plane $(M,\,D_L)$ in which $\rho \geqslant 8$. For simplicity, we consider equal masses $M_1 = M_2 = M$. We superimpose, in yellow with diagonal meshes, the region where $f_{\rm ISCO} = [50,\,1000]$ Hz. We conclude that aLIGO and VIRGO are sensitive to the merger of exotic compact objects with $M \sim \mathcal{O}(10)\,M_\odot$. The frequency $f_{\rm ISCO}$ for such masses is close to the lower end of the sensitivity interval, and it should be possible to detect and reconstruct the waveform corresponding to the post-merger phase. We expect sizable deviations w.r.t. the merger of two BHs with the same mass, because 
 of finite-size effects due to different compactness and tidal deformations~\cite{Giudice:2016zpa}. However, we are not aware of any numerical simulation done in our setup, and it is not simple to predict how the extreme stiffness of the EoS will influence the shape of the waveform. The only point that we want to make in the present discussion is that, once the mass $M$ of the final state object originating from the merger is specified, it is possible to compute the echo frequency as illustrated in section~\ref{sec:echoes}. We are implicitly assuming here that an exotic ultracompact object with a maximally stiff LinEoS is created in the final state. On the right-$y$ axis in Fig.~\ref{fig:LIGO} we show the echo frequency $f_{\rm echo}$ corresponding to $M$ (where now $M$ is the mass of the final state object). Interestingly, an hypothetical signal lies in the frequency interval $f = [1024,\,4096]$ Hz quoted by the analysis championed in~\cite{Abbott:2017dke}. 

We argue that it is in principle possible to detect a signal associated to the presence of 
the unstable light ring encircling an ultracompact exotic object supported by the maximally stiff 
EoS $P = \rho - \rho_0$. 
Of course, the characteristic shape of such 
``echo'' signal is not simple to predict if the object is not asymptotically close in compactness to the Buchdahl limit, 
and  only a numerical simulation can  reveal 
whether the presence of the photon sphere in a maximally stiff object affects or not the post-merger dynamics.
A study in this direction using publicly available codes~\cite{einsteintoolkit} is underway.

\subsection{Summary}

In the following, we summarize our findings.

\begin{itemize}

\item [$\circ$] In section~\ref{sec:stars} we considered static, spherically symmetric. We argue that perfect fluid stars are the most likely candidates for ultracompact (i.e. with $R< 3M$) objects, and, under a number of physical assumptions listed in section~\ref{sec:stars}, we identified the EoS $P=\omega(\rho-\rho_0)$ as the one that is able to support the most compact fluid stars. We stressed the role of the causality constraint that imposes the condition $\omega \leqslant 1$. This result is per se not new in the literature. However, we presented the material in a fairly original way, making use of simple analytical considerations to support our conclusions and drawing parallels between perfect fluid stars, boson stars and exotic objects that do not respect the physical assumptions on which our analysis is based. 
  
\item [$\circ$] In section~\ref{sec:echoes} we analyzed the solutions discussed in the previous point having in mind the possibility that such ultracompact objects, if involved in the final state of a merger process, generate gravitational echoes. The most original part of our material is concentrated in section~\ref{sec:rotation}, where we studied the effect of non-zero rotation. We found that, even for angular velocity close to the Keplerian limit, the ultracompact objects discussed in sections~\ref{sec:stars},\,\ref{sec:echoes} -- despite being characterized by the presence of a stable internal light ring -- do not develop an ergoregion, and are not affected by ergoregion instabilities.

\end{itemize}

Our main conclusion is that ultracompact objects supported by {\it physical} EoS are not able to generate gravitational echoes like those that characterize the relaxation phase of a BH mimicker. This result is in agreement with the findings of Ref.~\cite{Mannarelli:2018pjb}, that we complemented here with a comprehensive analysis of rotation.

However, the message we would like to convey is not entirely negative. The age of GW astronomy has just started and, in light of the tremendous amount of data that is expected in the next years, we must leave no stone unturned in our quest for new phenomena beyond the current knowledge of GR, particle physics and nuclear physics. In this respect, we believe that the ultracompact solutions found in this paper deserve further study. The extreme stiffness of the EoS $P = \rho - \rho_0$, together with the existence of an external unstable light ring and the absence of ergoregion instabilities, may cause distinctive signatures in a merger event  if compared with ordinary NSs or BHs. Furthermore, it would be interesting to understand under which conditions, if any, standard (or exotic) matter is able to support such stiff EoS.

\section*{Acknowledgments}

We thank Aleksi Kurkela for discussions  and Pedro Cunha and Carlos Herdeiro for their useful comments about the tentative instability related to stable light rings. This work was supported by the grants IUT23-6, PUT799, and by EU through the ERDF CoE program grant TK133 and by the Estonian Research Council via the Mobilitas Plus grant MOBTT5.

\appendix

\section{The Buchdahl limit and related results}
\label{app:Buchdahl}

The Buchdahl limit \eqref{eq:Buch} can be stated as an upper bound on the compactness, $\mathcal{C} \leq 4/9$, of non-rotating fluid stars. It holds under the mild assumptions on the EoS laid out in section~\ref{sec:stars} and the requirement that no region of a spherically symmetric star can be inside its gravitational radius, $r > 2 m$. In this appendix we will derive an inequality that allows us to compare fluid stars with different mass profiles. The Buchdahl limit is then obtained by comparing viable fluid stars with constant density stars. Finally we list some inequalities relating central pressure and compactness.

Throughout this section we assume that $\rho, e^{\nu}, e^{\lambda} \geqslant 0$ everywhere. First we prove the following statement comparing two stars with known mass and anisotropy profiles:

\begin{theorem}
Take two stars with the same mass $M$ and radius $R$. If their mass and pressure anisotropy profiles satisfy
\be\label{eq:lemma_assump}
	m(r) \geqslant \bar m(r)~, \qquad
	e^{\lambda(r)/2} \left[\left(\frac{m(r)}{r^3}\right)^{\prime} - \frac{8\pi \Delta(r)}{ r}\right] \leqslant e^{\bar \lambda(r)/2}  \left[\left(\frac{\bar m(r)}{r^3}\right)^{\prime} - \frac{8\pi \bar \Delta(r)}{r}\right]~
\ee
and $\nu'(r) \bar \nu^{\prime}(r) \geqslant 0$, where the unbarred quantities correspond to the first and the barred quantities to the second star, respectively, then the components of the metric satisfy
\be\label{eq:lemma_result}
	e^{\lambda(r)} \geqslant e^{\bar \lambda(r)}~, \qquad
	e^{\nu(r)} \leqslant e^{\bar \nu(r)}~.
\ee
The inequality \eqref{eq:lemma_result} is saturated if and only if \eqref{eq:lemma_assump} is saturated.
\end{theorem}

The first inequality in \eqref{eq:lemma_result} is trivially equivalent to the first inequality of \eqref{eq:lemma_assump} by the relation between $e^{\lambda}$ and $m$.  
To prove the second inequality we start by eliminating pressure from the Einstein and continuity equations~(\ref{eq:tt}-\ref{eq:cont}). Taking the first derivative of the $rr$ component of the Einstein equations and then using the $tt$ and continuity equations we obtain
\be\label{eq:TOVreprocessed}
	\left(\frac{m}{r^3}\right)^{\prime} - \frac{8\pi \Delta}{r}
	= \frac{B}{A}\left(\frac{B A^{\prime}}{r}\right)^{\prime}~,
\ee
where we denoted $A \equiv e^{\nu/2}$, $B \equiv e^{-\lambda/2} \equiv \sqrt{1-2 m/r}$ for the sake of compactness. By \eqref{eq:TOVreprocessed} the second inequality in \eqref{eq:lemma_assump} is equivalent to
$
	\left(B A^{\prime}/r\right)^{\prime}/A \leqslant \left(\bar B \bar A^{\prime}/r\right)^{\prime}/\bar A.
$
It can be recast as 
$
	\left[\left(B \bar A A^{\prime} - \bar B A \bar A^{\prime}\right)/r\right]^{\prime} 
	\leqslant A^{\prime}\bar A^{\prime} \left(B  - \bar B\right)/r 
	\leqslant 0.
$
The last inequality holds because $B \leqslant \bar B$ and the assumption $\nu'(r) \bar \nu^{\prime}(r) \geqslant 0$. Integrating in the range $[r,R]$ then gives
$
	(B \bar A A^{\prime} - \bar B A \bar A^{\prime})/r \geqslant 0, 
$
where we used that, by Birkhoff's theorem, the exterior solutions match, i.e. $B(R) = A(R) = \bar B(R) = \bar A(R)$, $B^{\prime}(R) = A^{\prime}(R) = \bar B^{\prime}(R) = \bar A^{\prime}(R)$. Again, using $B \leqslant \bar B$ we obtain $A^{\prime}/A \geqslant \bar A^{\prime}/\bar A$. After, again, integrating in $[r,R]$ and using Birkhoff's theorem we get $\bar A \geqslant A$, which proves \eqref{eq:lemma_result}. Since for a given $m(r)$ and $\Delta(r)$ the metric is determined uniquely, saturation of \eqref{eq:lemma_assump} implies the saturation \eqref{eq:lemma_result} and vice versa.

Consider now stars with a monotonously decreasing density profile and a non-negative pressure anisotropy,
\be\label{eq:lemma2_assump}
	\rho'(r) \leqslant 0~, \qquad
	\Delta(r) \geqslant 0~,
\ee
These conditions are saturated by a CDS.  Note that the condition $\rho^{\prime} \leqslant 0$ is equivalent to $(m/r^3)^{\prime} \leqslant 0$: Integrating the latter in $[r,R]$ gives $m/r^3 \geqslant M/R^3$. So, Eq. \eqref{eq:lemma2_assump} implies \eqref{eq:lemma_assump} with $\bar m = m_{\rm CDS} \equiv M r^3/R^3$ the CDS mass profile and $\bar \Delta = 0$. Finally, \eqref{eq:lemma2_assump} also implies that the condition $\nu'(r) \bar \nu'(r) \geqslant 0$ is satisfied: Since $\bar \nu = \nu_{\rm CDS}$ we have $\bar \nu^{\prime} > 0$ and, by using the notation introduced above, \eqref{eq:TOVreprocessed} now implies that $\left(B A^{\prime}/r\right)^{\prime} \leqslant 0$ which after integration in $[r,R]$ and applying the exterior at the boundary solution gives $M/R^3 \leqslant B(r) A^{\prime}(r)/r$ and thus $\nu^{\prime}(r) \geqslant 0$ as long as $\nu^{\prime}$ is continuous at the boundary.

In conclusion, for stars satisfying the condition \eqref{eq:lemma2_assump} the following holds:
\begin{enumerate}[label=\roman{*})]
	\item The components of the metric are bounded by the corresponding components of the CDS
\bea\label{ineq:lambdanu}
	e^{\nu(r)} &\leqslant e^{\nu_{\rm CDS}(r)} = \left[ \frac{3}{2}\left(1 - 2 \mathcal{C}\right)^{1/2} - \frac{1}{2} \left(1 -  \frac{2\mathcal{C} r^2}{R^2}\right)^{1/2} \right]^{2}~, \\
	e^{\lambda(r)} &\geqslant e^{\lambda_{\rm CDS}(r)} = \left( 1 -  \frac{2\mathcal{C} r^2}{R^2} \right)^{-1}~.
\eea

	\item  They satisfy the Buchdahl limit~\eqref{eq:Buch}
\be
	\mathcal{C} \leqslant \frac{4}{9}~,
\ee
as a result of the requirement $e^{\nu(r)} > 0$.

	\item The coordinate time required for light to travel from the photon sphere to center of the star and back, i.e. the characteristic time delay for gravitational echoes, satisfies  (see Eq.~\eqref{eq:TauBound} for the explicit expression)
\be
	\tau_{\rm echo}/M \leqslant (\tau/M)_{\rm echo, CDS}~.
\ee
This follows trivially from $\td \tau/\td r \equiv \exp[(\lambda-\nu)/2]$ and Eq.~\eqref{ineq:lambdanu}.

\end{enumerate}

The condition \eqref{eq:lemma2_assump} is automatically satisfied for perfect fluid stars that obey the weak energy condition \eqref{eq:WEC} and are microscopically stable \eqref{eq:MS}: The r.h.s. of the TOV equation \eqref{eq:TOV} is negative when $\rho, P \geqslant 0$ and $r > 2m$, so, by $\td \rho/\td P \geqslant 0$, we obtain $\rho' = (\td \rho/\td P) P^{\prime} \leqslant 0$. As $\Delta = 0$ by construction, both conditions in \eqref{eq:lemma2_assump} are fulfilled. Moreover, we can replace the assumption $\Delta = 0$ with $\Delta \geqslant 0$ and the conclusions remain valid, since the the r.h.s. of the TOV equation \eqref{eq:TOV} will still guarantee $P^{\prime} \geqslant 0$.

In the case of isotropic stars we can make use of the internal solution \eqref{eq:nu_int}, which we repeat here for convenience
\be\label{eq:MetricNu}
	e^{\nu(r)/2} = e^{-\lambda(R)/2}\exp\{-H[P(r)]\}~,
	~~~~ r\leqslant R~.
\ee
For isotropic stars obeying the weak energy condition and microscopic stability the central pseudoenthalpy thus satisfies
\be\label{eq:BuchPressure}
	H(P_c) 
	\geqslant -\ln\left[\frac{3}{2}\left( 1 - 2 \mathcal{C}\right)^{1/2} - \frac{1}{2}\right]~,
\ee
where $P_c$ stands for the pressure in the center of the star. This inequality follows from the internal solution \eqref{eq:nu_int}. Importantly, the r.h.s. diverges when $\mathcal C \to 4/9$, i.e. as the as the compactness approaches the Buchdahl limit~\eqref{eq:Buch}. Because $H(P)$ is a monotonously growing function of pressure, the inequality \eqref{eq:BuchPressure} implies a lower bound on pressure, or equivalently, an upper bound on compactness when the central pressure is known. 

A more specific relation to pressure can be obtained by imposing causality, or more generally, the condition $\partial P/\partial \rho \leqslant \omega$. Integrating in the range $P \in [0,P_c]$ then gives $\rho \geqslant \rho_0 + P/\omega$, where $\rho_0 \equiv \rho_c - P_c/\omega >0$ and $\rho_c$ is the central density. Thus $H(P) \leqslant \int \td P/(\rho_0 + P(1+1/\omega))$ which together with \eqref{eq:BuchPressure} yields 
\be
	P_c/\rho_0 \geqslant \frac{\omega}{\omega+1}\left[ \left(\frac{3}{2}\left( 1 - 2 \mathcal{C}\right)^{1/2} - \frac{1}{2}\right)^{-\frac{\omega + 1}{\omega}} - 1\right]~.
\ee
In particular, the pressure diverges when compactness approaches the Buchdahl limit.

\begin{figure}[!tb!]
\centering
$$
\includegraphics[width=.43 \linewidth]{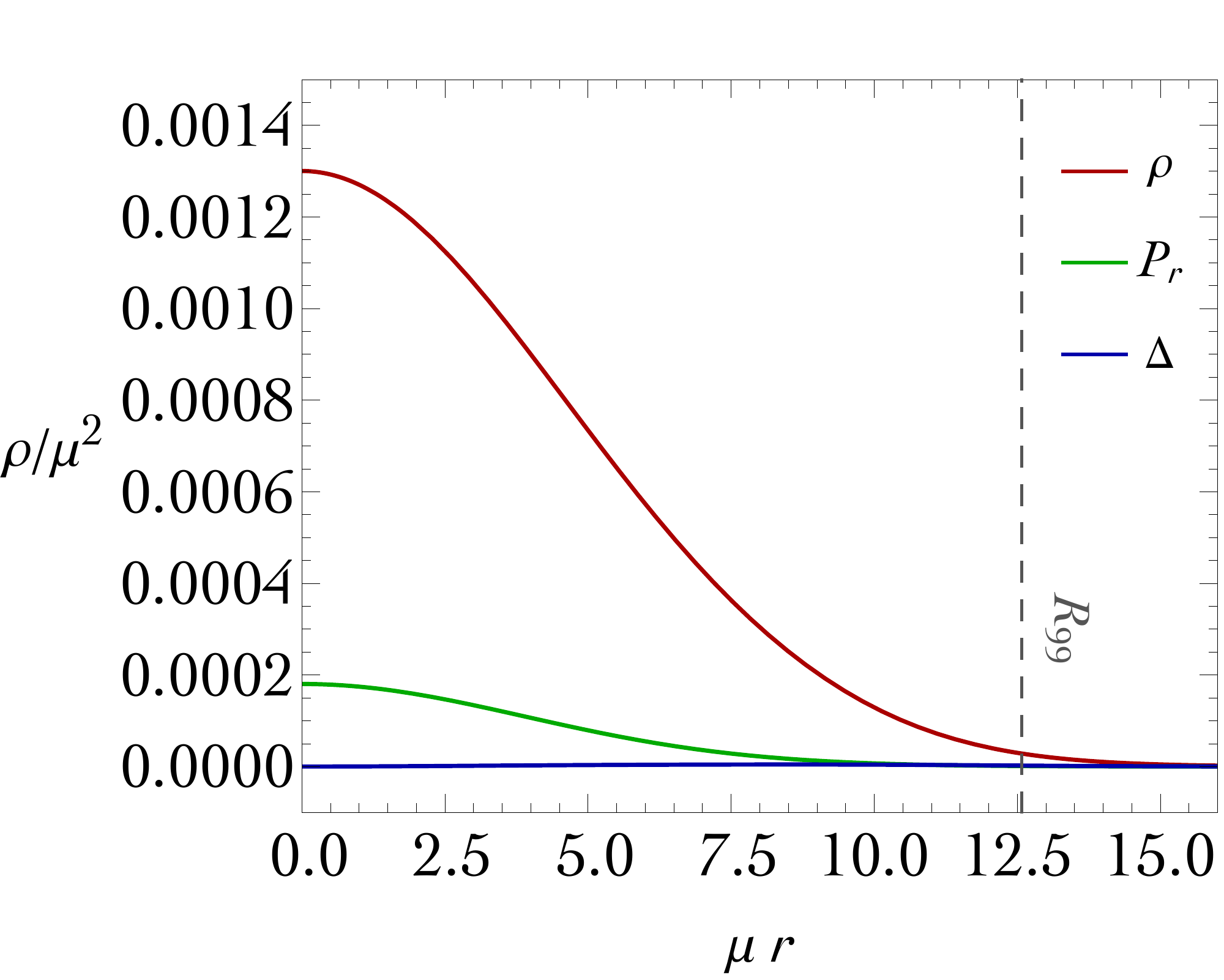}
\qquad\quad
\includegraphics[width=.43 \linewidth]{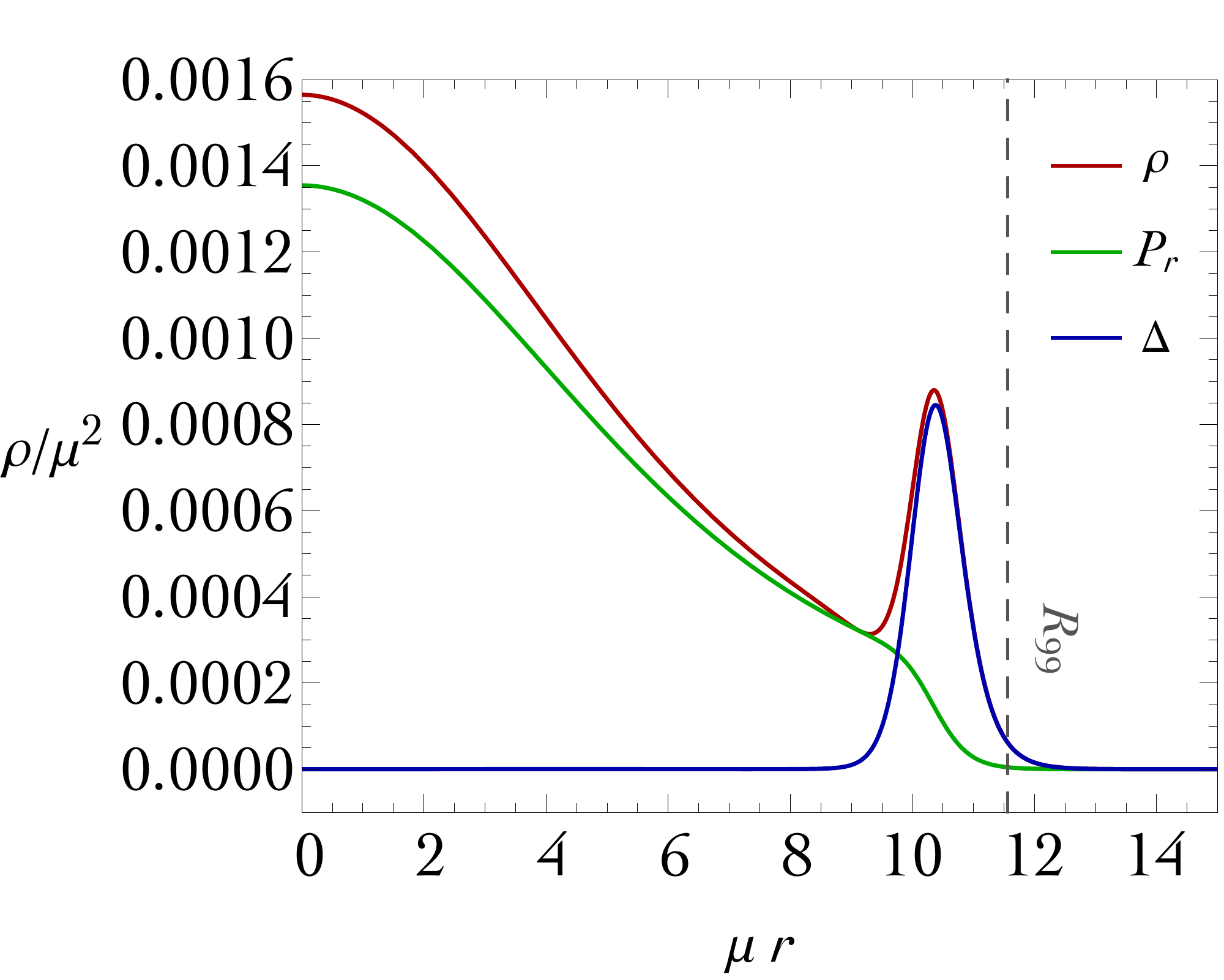}
$$
\caption{\label{fig:BS}\em 
The density profiles of a compact massive boson star ($\mathcal C = 0.156$) with $\lambda = 1000$ and $\phi_c = 0.02 \sigma_0$ (left panel) and compact solitonic boson star ($\mathcal C = 0.30$) with $\sigma_0 = 0.05$ and $\phi_c = 1.09 \sigma_0$ (right panel). Both configurations are the heaviest allowed by their potentials. The results are presented in units where $\mu = 1$. The vertical line denotes the sphere containing $99\%$ of the stars mass.
}
\end{figure} 

\section{Boson stars}\label{app:BosonStars}

Consider boson stars consisting of a complex scalar field $\Phi$. The field equations are solved with the stationary ansatz, $\Phi = \phi e^{-i\omega t}$, obey the equations
\bea
	m^{\prime} &= 4\pi r^2 \rho~, 
	\\
	\frac{\nu^{\prime} }{2r e^{\lambda}}  &= \frac{m}{r^3} + 4\pi P_{\rm r}~, 
	\\
	\phi'' &= \left(\frac{\lambda^{\prime}-\nu^{\prime}}{2} - \frac{2}{r} \right) \phi' + e^{\lambda}\left(-e^{-\nu}\omega^2 \phi + V'/2\right)~,
	\label{eq:phi}
\eea
where $V(|\Phi|) \equiv V(\phi)$ is the potential of the scalar field and $e^{-\lambda} \equiv 1 - 2m/r$. The energy density, radial pressure and tangential pressure are
\bea
	\rho = e^{-\nu}\omega^2 \phi^2 + e^{-\lambda} (\phi')^2 + V~, \\
	P_{\rm r} = e^{-\nu}\omega^2 \phi^2 + e^{-\lambda} (\phi')^2 - V~, \\
	P_{\rm t} = e^{-\nu}\omega^2 \phi^2 - e^{-\lambda} (\phi')^2 - V~,
\eea
respectively. The pressure anisotropy therefore reads $\Delta = 2e^{-\lambda} (\phi')^2$. These equations provide a closed system and imply the continuity \eqref{eq:cont} and the TOV equation \eqref{eq:TOV}. The boundary conditions at the origin read $m(0) = 0$, $\nu(0) = \nu_c$, $\phi(0) = \phi_c$, $\phi'(0) = 0$, while at the spatial infinity we require that the field decays exponentially, $\phi \propto e^{-r \sqrt{\mu^2-\omega^2}}/r$. The ground state frequency is the smallest frequency $\omega$ for which the latter condition is satisfied.

Density profiles of a massive boson star, $V(\Phi) = \mu^2 |\Phi|^2 + \lambda |\Phi|^4$, and a solitonic boson star, $V(\Phi) = \mu^2 |\Phi|^2 (1- |\Phi|^4/\sigma_0^2)^2$,  are given in Fig.~\ref{fig:BS}. Both solutions are obtained by numerically solving the  represent the system of equations \eqref{eq:phi} and using the shooting method to find $\omega$. They represent the heaviest stars possible with $\lambda = 1000$ and $\sigma_0 = 0.05$, respectively, and have a compactness close to the maximal compactness allowed in their respective class of boson stars. The violation of condition \eqref{eq:Buch_cond} (or \eqref{eq:lemma2_assump}) for the Buchdahl bound is clearly seen in the solitonic boson star but not in the massive boson star.

We remark that it is possible to construct boson stars that possess light rings even if $\mathcal C < 1/3$. In such cases the exterior unstable light rings are created by the matter distribution, instead of being inherited from the exterior Schwarzschild metric, as is the case for ultracompact objects. For example, solitonic boson stars can acquire a light ring pair due to their high density shell depicted in Fig.~\ref{fig:BS}~\cite{Macedo:2013jja}. Mini boson stars provide another example as they can develop a light ring due to a central density peak, although sufficiently peaked configurations are only found in the unstable branch~\cite{Cunha:2017wao}.

\section{Comparison with the Hartle-Thorne approximation}
\label{app:HartleThorne}

The metric best-suited for the slow-rotation expansion~\cite{Hartle:1967he,Hartle:1968si} is based on the 
quasi-Schwarzschild coordinates $(t,r,\theta,\phi)$ describing -- in full generality -- the line element 
\be
	\td s^2 = g_{\mu\nu}dx^{\mu}dx^{\nu} = -H(r,\theta)^2 \td t^2 + Q(r,\theta)^2 \td r^2 + r^2 K(r,\theta)^2
	\left\{\td \theta^2 + \sin^2\theta\left[ \td \phi - \omega(r,\theta)\td t \right]^2
	\right\}~.
\ee
Expanding up to the second order in the angular velocity of the star $\Omega$, the spacetime geometry 
 takes the form
\begin{eqnarray}\label{eq:HartleThorne}
 	\td s^2 &=& -\left\{
 	e^{\nu}\left[
	1 + 2h_0 + 2h_2\mathcal{P}_2(\cos\theta)\right] - \omega^2 r^2 \sin^2\theta\right\} \td t^2 
	-2\omega r^2\sin^2\theta \td t \td \phi  \\
	&+& e^{\lambda}\left\{
	1 + \frac{2\left[m_0 + m_2\mathcal{P}_2(\cos\theta)\right]}{r - m}
	\right\}\td r^2 
	+ r^2
	\left[1 + 2\left(v_2- h_2\right)\mathcal{P}_2(\cos\theta)\right]
	\left(\td \theta^2 + \sin^2\theta \td \phi^2\right)~,\nonumber
\end{eqnarray}
 where $\nu = \nu(r)$, $\lambda = \lambda(r)$, $m = m(r)$ are the metric functions and mass of the corresponding static solution, and  $\mathcal{P}_2(\cos\theta)$ is the Legendre polynomial of second order.
 The angular velocity of local inertial frames $\omega(r)$ is proportional to $\Omega$ 
while the functions $h_0(r)$, $h_2(r)$, $m_0(r)$, $m_2(r)$, $v_2(r)$ are of order $\Omega^2$, 
and must be calculated from the Einstein equations. We follow the procedure outlined in~\cite{Hartle:1967he,Hartle:1968si}, and we refer the interest reader to these original references for the 
details of the computation.

We implement the Hartle-Thorne approximation for the equation of state $P = \omega(\rho - \rho_0)$. 
To validate our solutions, we start discussing the I-Love-Q relations~\cite{Yagi:2013bca,Yagi:2013awa}.
The I-Love-Q are universal relations between the moment of inertia $I$, the tidal Love number $\lambda$ and the quadrupole moment $Q$ 
that are independent of the EoS.\footnote{We define, as customary, the dimensionless quantities $\bar{I} \equiv I/M^3$, 
$\bar{Q} \equiv Q/M^3\tilde{a}^2$, $\bar{\lambda} = 2k_{2}^{\rm tid}/3\mathcal{C}^5$, 
where $k_{2}^{\rm tid}$ is the tidal apsidal constant. Notice that we use the value of the mass computed in the 
 static limit ($M_{\rm static}$ in the figures of this appedix).} 
 We follow the standard computation for $I$, $Q$ and $\lambda$~\cite{Yagi:2013awa}. The only subtle point arises in the computation of the Love number because we are dealing with a case in which the density does not vanish at the surface of the star.
 As a consequence, one needs to include the effect of the discontinuity 
 with the external spacetime in the computation of $\lambda$~\cite{Postnikov:2010yn}. 
The true origin of the universality dictated by the I-Love-Q relations is still unknown\footnote{Ref.~\cite{Yagi:2014qua}
 suggests that universality arises as an emergent approximate symmetry related to the iso-density contours of the  star.} but it is  
 conjectured that they are valid to approximately 1\%
 accuracy for
any physically
reasonable EoS provided that {\it i)} the stars do not rotate maximally, {\it ii)} magnetic fields are not extremely strong, and 
 {\it iii)}  GR is valid. The I-Love-Q relations, therefore, must apply also to the maximally stiff EoS $P = \omega(\rho - \rho_0)$.
Our results complement the ones presented in~\cite{Silva:2017uov} in which the I-Love-Q relations were tested for a 
piecewise EoS that matches a tabulated NS EoS at low densities, while matching a stiff EoS in the high-density region.
 \begin{figure}[!htb!]
 \centering
  \includegraphics[width=.48\linewidth]{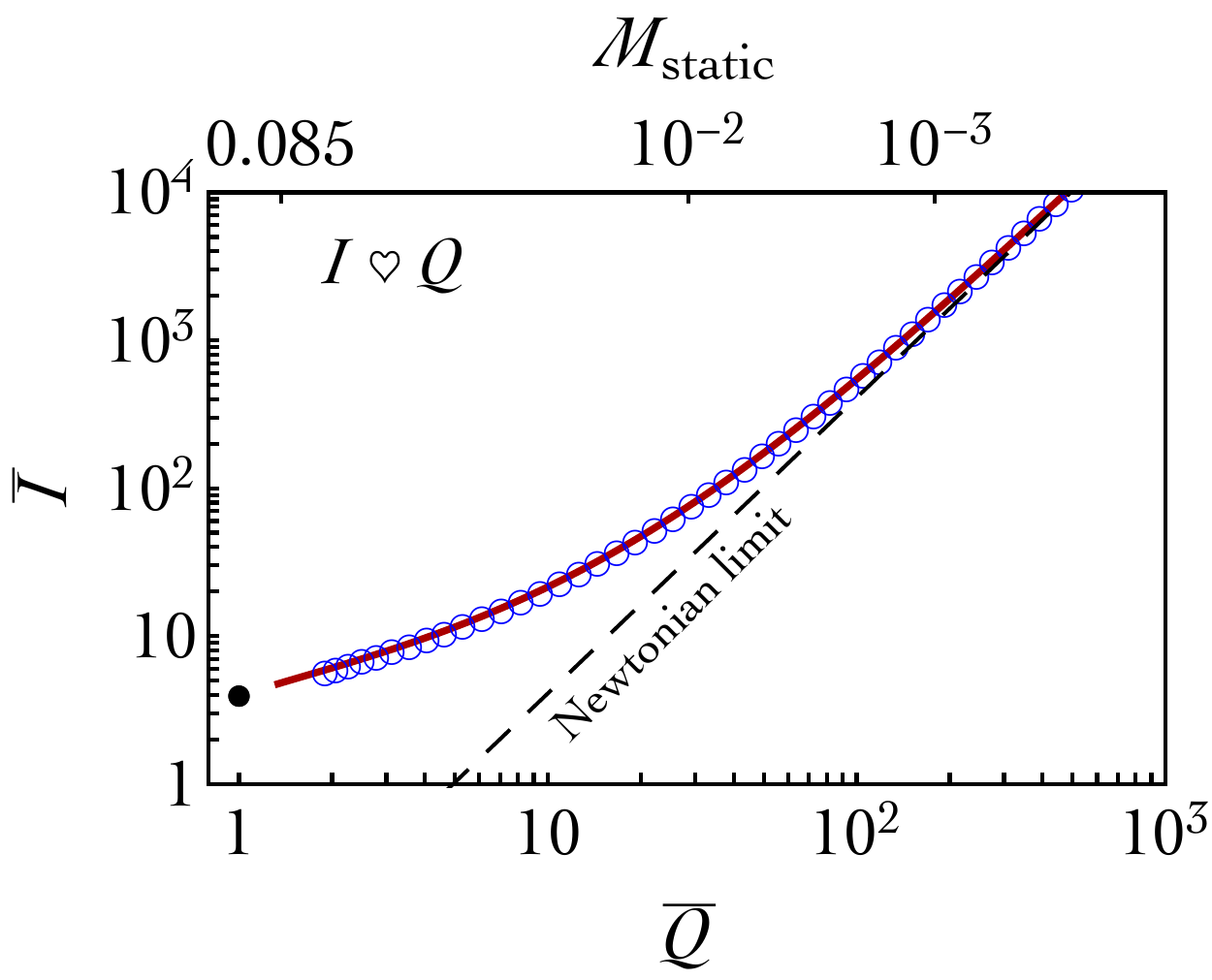}
\minipage{0.5\textwidth}
\centering
  \includegraphics[width=.95\linewidth]{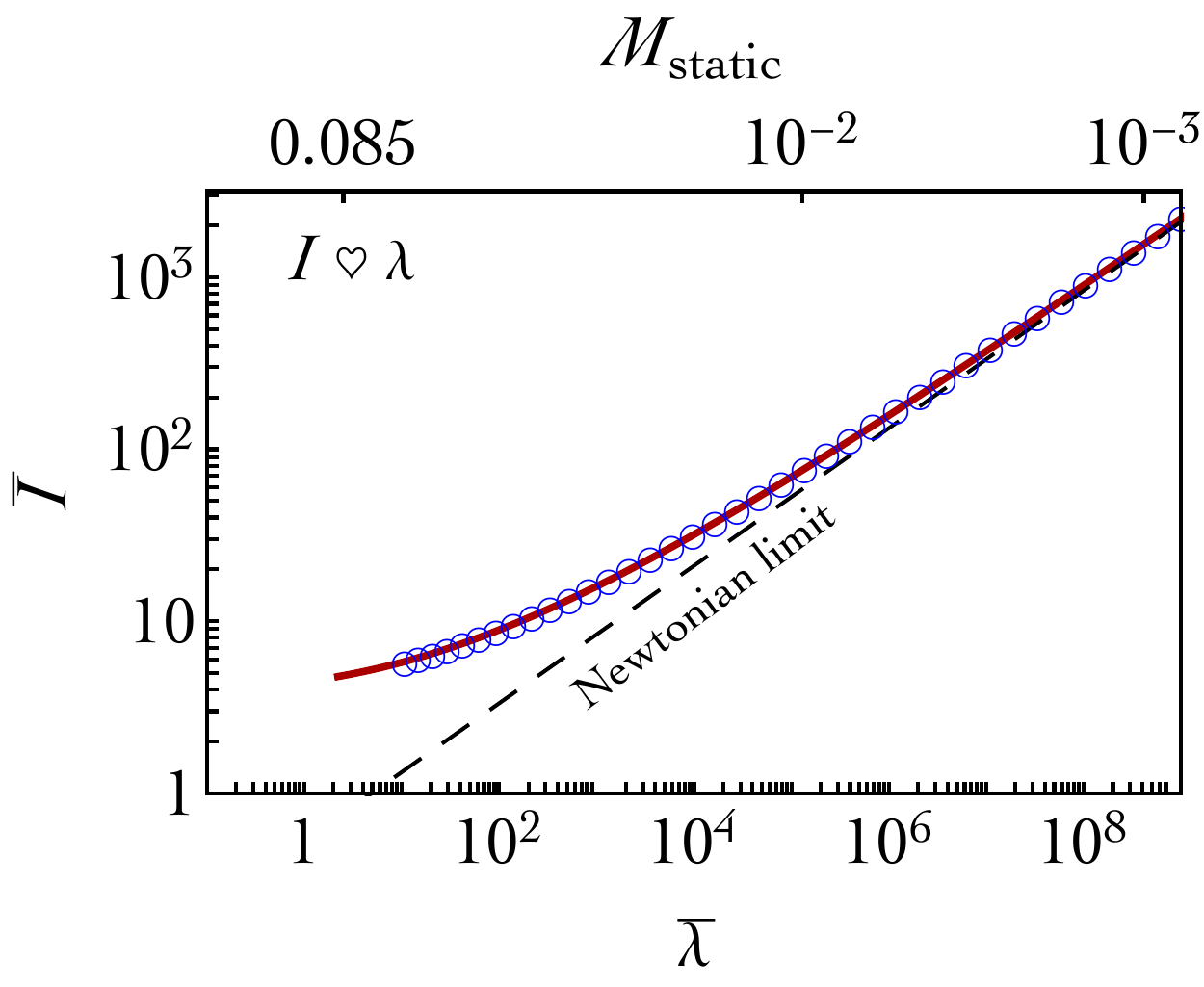}
\endminipage 
\minipage{0.5\textwidth}
  \includegraphics[width=.95\linewidth]{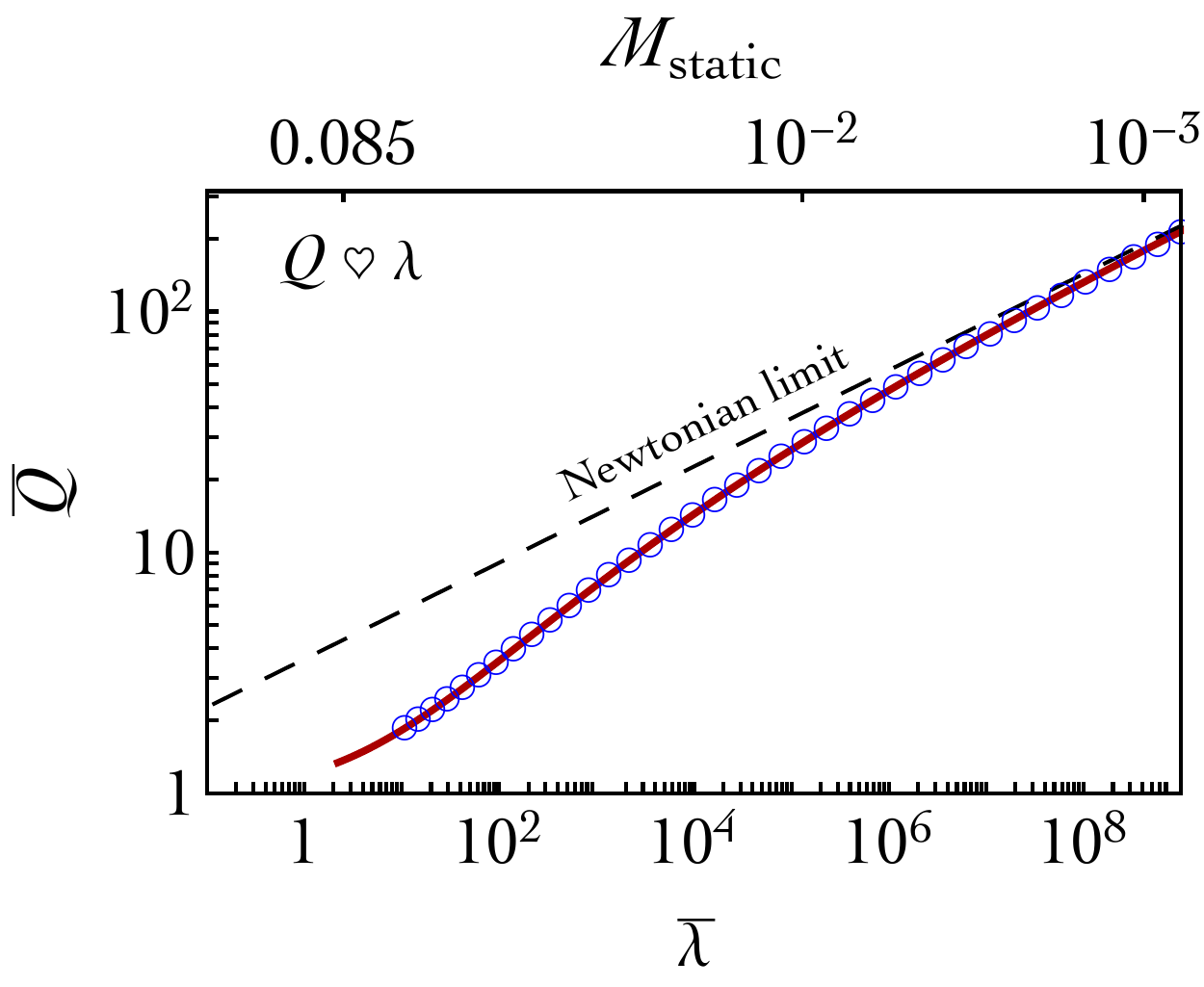}
\endminipage \\
\vspace{-0 cm}
\caption{\label{fig:ILoveQ}\em 
I-Love-Q relations for the maximally stiff equation of state $P = \rho -\rho_0$ (red solid lines) computed by means of the 
Hartle-Thorne approximation.
For comparison, the empty blue circles represent the I-Love-Q relations for a polytrope with $n=1$. 
}
\end{figure} 
In Fig.~\ref{fig:ILoveQ} we show the three I-Love-Q relations in the case saturating the causality condition $\omega = 1$. For comparison, we also show the  
result for  a polytrope with index $n = 1$. As expected, the I-Love-Q relations are valid for our maximally stiff LinEoS, and  we find an excellent agreement with the universal fitting function 
proposed in~\cite{Yagi:2013awa}.

We now move to discuss the main point of this appendix, that is the comparison with the numerical results presented 
in section~\ref{sec:rotation}.
 \begin{figure}[!htb!]
\minipage{0.5\textwidth}
\centering
  \includegraphics[width=.95\linewidth]{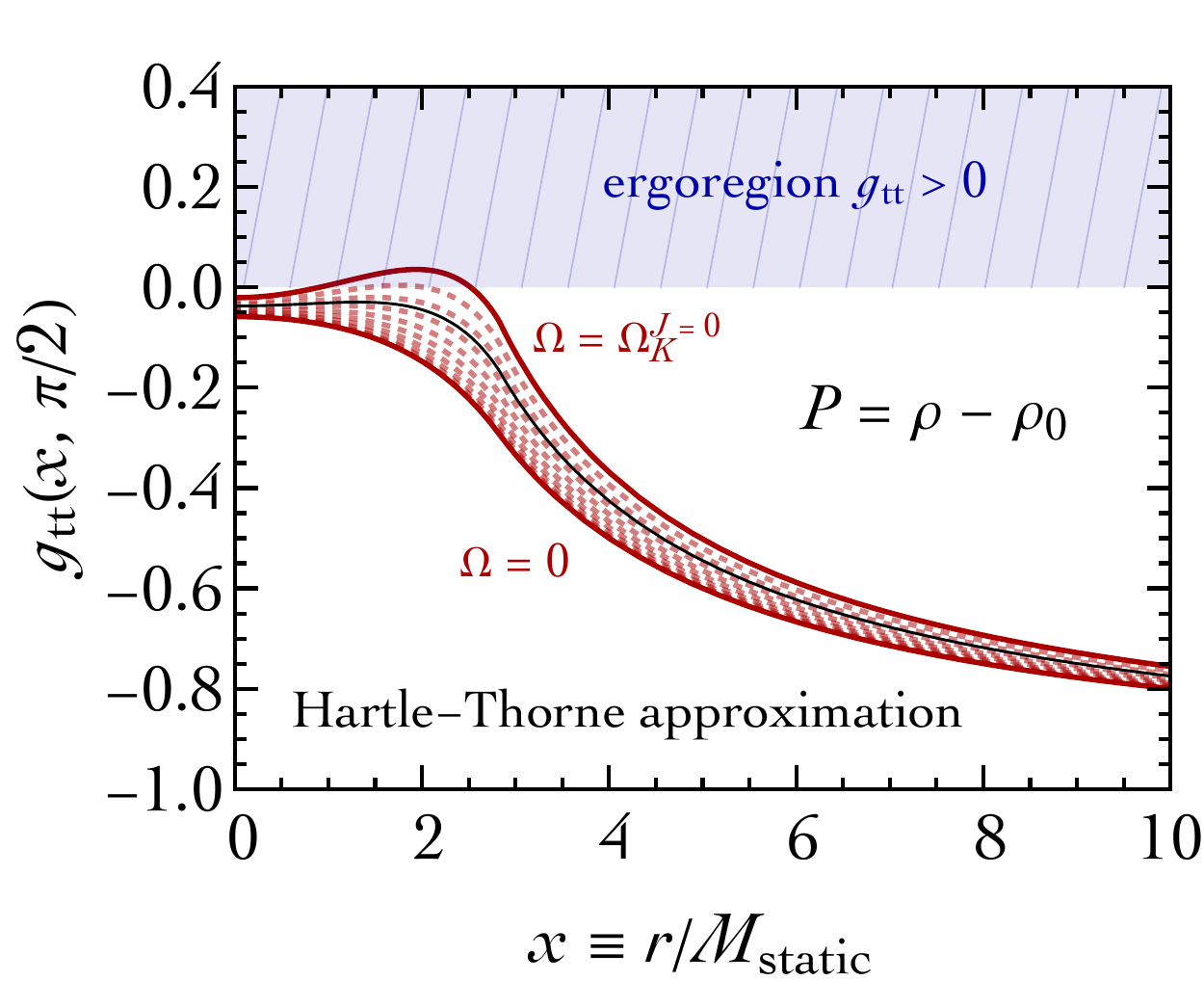}
\endminipage 
\minipage{0.5\textwidth}
  \includegraphics[width=.95\linewidth]{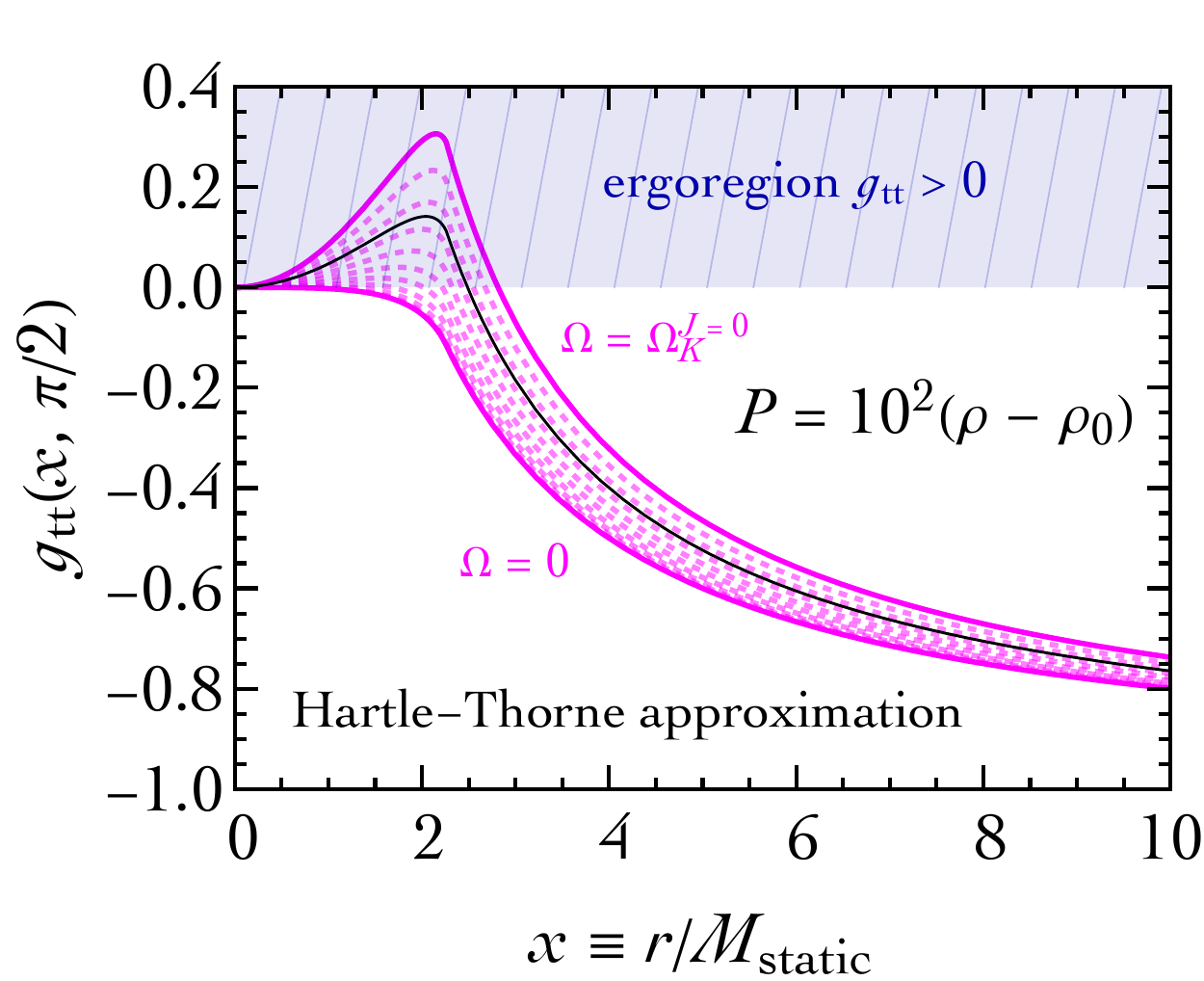}
\endminipage \\
\vspace{-0 cm}
\caption{\label{fig:HartleThorne}\em  Metric component $g_{tt}$ in Eq.~(\ref{eq:HartleThorne}) in the equatorial plane 
as a function of the radial Hartle-Thorne coordinate $r$ normalized to the static mass for different 
values of the angular velocity.  
We use the LinEoS $P = \omega(\rho - \rho_0)$ with $\omega = 1$ (left panel) and $\omega = 10^2$ (right panel). 
In both cases we perturb the static configuration with the maximal allowed compactness (see Fig.~\ref{fig:MassRadiusPlot}).  
}
\end{figure} 
In particular, we shall investigate the formation of the ergoregion in order to validate the result presented in the right panel of Fig.~\ref{fig:LightRingsNS}.
In Fig.~\ref{fig:HartleThorne} we show the metric function $g_{tt}$ in the equatorial plane computed by means of the 
Hartle-Thorne approximation for increasing values of the angular velocity, from $\Omega = 0$ to the mass shedding limit 
$\Omega = \Omega_{K}^{J = 0}$. 
  In the left panel, we focus on the case with $\omega = 1$, and we consider a configuration that in the static limit is close to the maximal allowed compactness. We find that it is difficult to form an ergoregion even for 
  extreme values of the angular velocity. In the comparison with Fig.~\ref{fig:LightRingsNS}, 
  the reader should keep in mind the Keplerian limit $\Omega_K$ used in section~\ref{sec:rotation} does not 
  coincide  with the corresponding value in the static case. 
  On the contrary,  since the equatorial radius satisfies $R_{\rm eq} > R$, at mass-shedding we have in general
  $\Omega_K < \Omega_K^{J = 0}$. Numerical studies suggest that $\Omega_K \approx  0.75\, \Omega_K^{J = 0}$, and one
   should trust the computation in Fig.~\ref{fig:HartleThorne} at most up to this value (black solid lines) thus 
    validating the result in Fig.~\ref{fig:LightRingsNS} according to which there is no ergoregion even in the Keplerian limit.
For completeness, in the right panel of Fig.~\ref{fig:HartleThorne} we show the same computation for the non-physical case 
with $\omega = 10^2$. In this case the star can be much more compact (see Fig.~\ref{fig:MassRadiusPlot}), 
and an ergoregion is present even for moderately small values of the angular velocity  thus leading to the presence of a potentially dangerous ergoregion instability that usually plagues ultracompact BH mimickers.
Finally, we compute the light rings in the Hartle-Thorne spacetime.
 \begin{figure}[!htb!]
\begin{center}
\includegraphics[width=.45\textwidth]{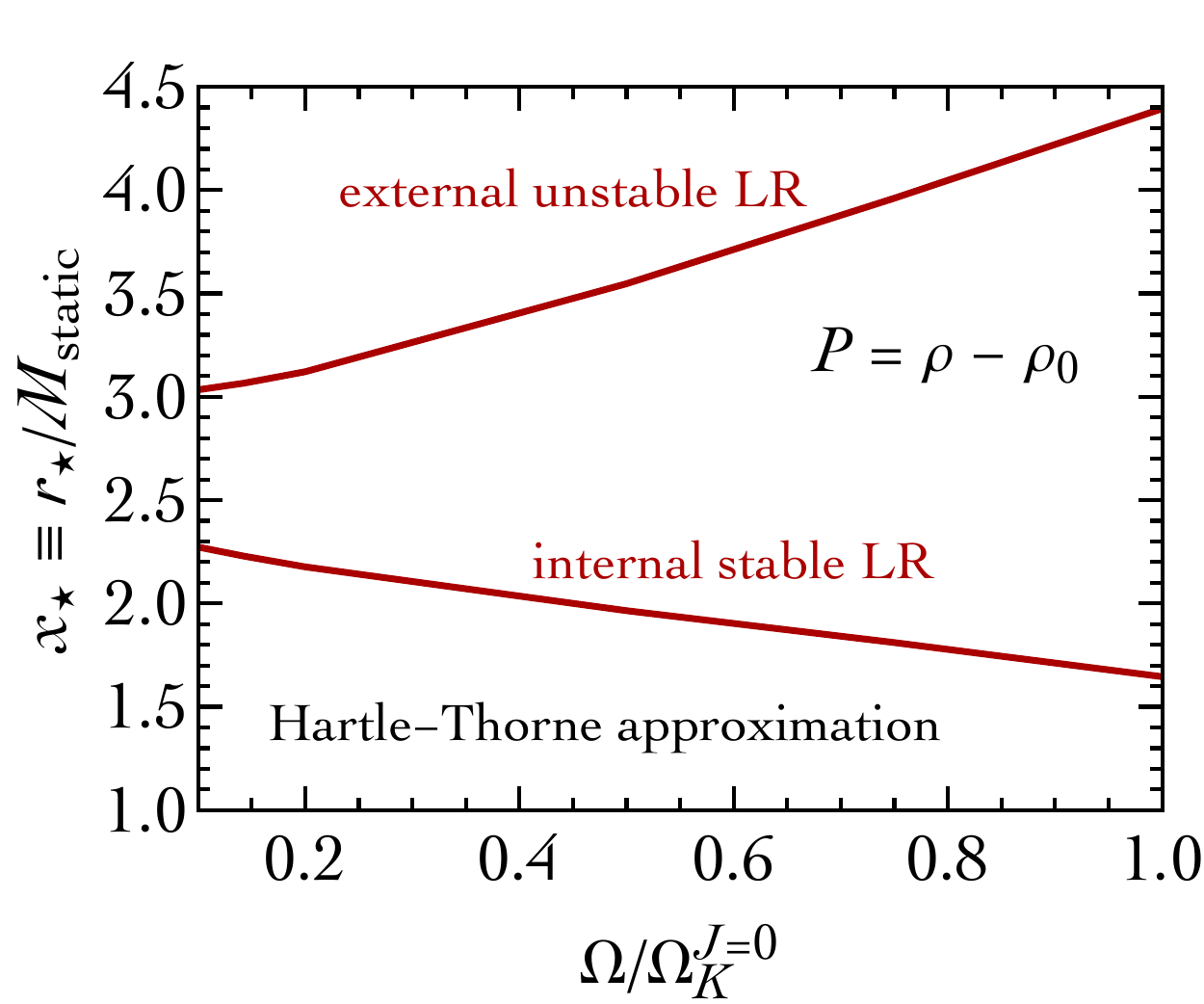}
\caption{\em \label{fig:HartleThorneLR} 
Position of the light rings in the equatorial plane as a function of the angular velocity.
 We perturb the maximally compact star with LinEoS using the Hartle-Thorne approximation. 
 }
\end{center}
\end{figure}
As done in the right panel of Fig.~\ref{fig:LightRingsNS}, we consider the maximally compact star with LinEoS 
and $\omega = 1$.
In the equatorial plane, the equation for null geodesics takes the form
\be
\dot{r}^2 + \underbrace{\left[
\frac{E^2 g_{\phi\phi} + 2EL g_{t\phi} + L^2 g_{tt}}
{g_{rr}(g_{tt}g_{\phi\phi} - g_{t\phi}^2)}
\right]}_{V_{\rm eff}^{\rm HT}(r)}= 0~,
\ee
where the metric functions follow from Eq.~(\ref{eq:HartleThorne}) with $\theta = \pi/2$.
The position of the light rings can be obtained by imposing the conditions 
$V_{\rm eff}^{\rm HT}(r) = 0$, $\td V_{\rm eff}^{\rm HT}(r)/\td r = 0$.
In practice, the light rings are defined by the roots of the equation
\be
D^2 g_{tt}^{\prime} + 2Dg_{t\phi}^{\prime} + g_{\phi\phi}^{\prime} = 0~,~~~~~~~
D = \frac{-g_{t\phi} + \epsilon\sqrt{g_{t\phi}^2-g_{tt}g_{\phi\phi}}}{g_{tt}}~,
\ee
where $D\equiv L/E$ is the impact parameter, $\epsilon= \pm 1$, and $^\prime$ indicates the derivative w.r.t. $r$.
We show our result in Fig.~\ref{fig:HartleThorneLR} where we plot the position of both the internal stable and 
external unstable light rings as a function of the angular velocity.
Notice that, by construction, outside the star the Hartle-Thorne metric falls back into the Schwarzschild metric in the 
limit of zero angular velocity.  
It follows that, in this limit, the position of the external light ring 
must converge towards the value $r = 3M$, as numerically verified in our computation.

\newpage

\vspace{-5mm}
\bibliographystyle{JHEP}
\bibliography{echoes}{}

\end{document}